\documentclass[twocolumn,english,amsmath,amssymb,superscriptaddress,nofootinbib]{revtex4-1}
\usepackage{mathptmx}
\usepackage[T1]{fontenc}
\usepackage[latin9]{inputenc}
\usepackage{amsmath}
\usepackage{amssymb}
\usepackage{graphicx}
\usepackage{esint}

\makeatletter

\providecommand{\tabularnewline}{\\}

\usepackage{babel}

\usepackage{xr}
\makeatother

\usepackage{babel}
\begin{document}

\title{Analysis of a multi-mode plasmonic nano-laser with a inhomogeneous distribution of molecular emitters}

\author{Yuan Zhang}
\email{yzhang@phys.au.dk}

\address{Department of Physics and Astronomy, Aarhus University, Ny Munkegade
120, DK-8000 Aarhus C, Denmark}

\author{Klaus M{\o}lmer}
\email{moelmer@phys.au.dk}

\address{Department of Physics and Astronomy, Aarhus University, Ny Munkegade
120, DK-8000 Aarhus C, Denmark}
\begin{abstract}
We extend Lamb's reduced density matrix laser theory to analyze the inhomogeneous molecular couplings and the mode-correlation in a plasmonic nano-laser consisting of a gold sphere and many dye molecules interacting with a driving optical field and with the quantized plasmon modes. The molecular inhomogeneity is accounted for by simulating their random distribution around the sphere. Our analysis shows that in order to obtain lasing we must employ a large number of strongly driven molecules to compensate strong damping of the plasmon modes. The compact molecular arrangement, however, can lead to molecular energy-shifts and thus reduce the excitation of the plasmon modes and ultimately suggests a maximum limit for the plasmon excitation for any specific system.
\end{abstract}
\maketitle

\section{Introduction}

The interaction between metals and light has been investigated
for more than a century with Maxwell's electromagnetic theory.
One essential insight obtained is that the electromagnetic (EM) field is enhanced
and localized around metal nano-particles (MNP) and on the interfaces between
metallic films and dielectrics \cite{MPelton} due to the excitation of surface plasmons 
involving collective oscillations of conductance electrons in the metal. The enhancement
 boosts the interaction between quantum emitters and the EM field \cite{MSTame,RMMa-0,MPelton,PBerini} 
 and thus leads to enhanced absorption \cite{NICade,YZelinskyy}, emission \cite{PAnger,YZhang-4} and Raman scattering \cite{MFleischmann,PJohansson}.
This can be utilized to improve the sensibility of spectroscopic
instruments \cite{SYDing} and the efficiency of LEDs \cite{XFGu,NGao}
and solar cells \cite{HAAtwater,LJWu}.

The localization introduces EM modes with mode volumes that are not limited by the 
wavelength of free-space light \cite{YYin,PBerini}. These modes can be excited if 
externally pumped quantum emitters are placed near MNPs or metallic films. Under suitable
conditions, the energy loss of those modes can be even compensated and the system can achieve lasing. 
This phenomenon known as SPASER, was proposed by Bergman and Stockman \cite{Bergman} and verified firstly by Noginov, et. al. \cite{MANoginov} with an experiment involving a gold nano-sphere
and many dye molecules. Since then many experimental demonstrations have been reported with structures like semiconductor wires \cite{JHo,RFOulton,CYWu,YJLu-1,YJLu,YHou,QZhang,TPPHSi,YHChou,BTChou}/squares
\cite{RMMa,RMMa-1} on metallic films, semiconductor pillars \cite{MTHill,MPNezhad,SHkwon,JHLee,MKha,KDing}/dots
\cite{AMatsudira,CYLu}/wires \cite{CYLu-1,SWChang} inside metallic
cavities as well as dye molecules in periodically arranged MNP arrays
\cite{JYSuh,WZhou,AYang,AYang-1,AHSchokker}.

\begin{figure}[!htb]
\begin{centering}
\includegraphics[scale=0.3]{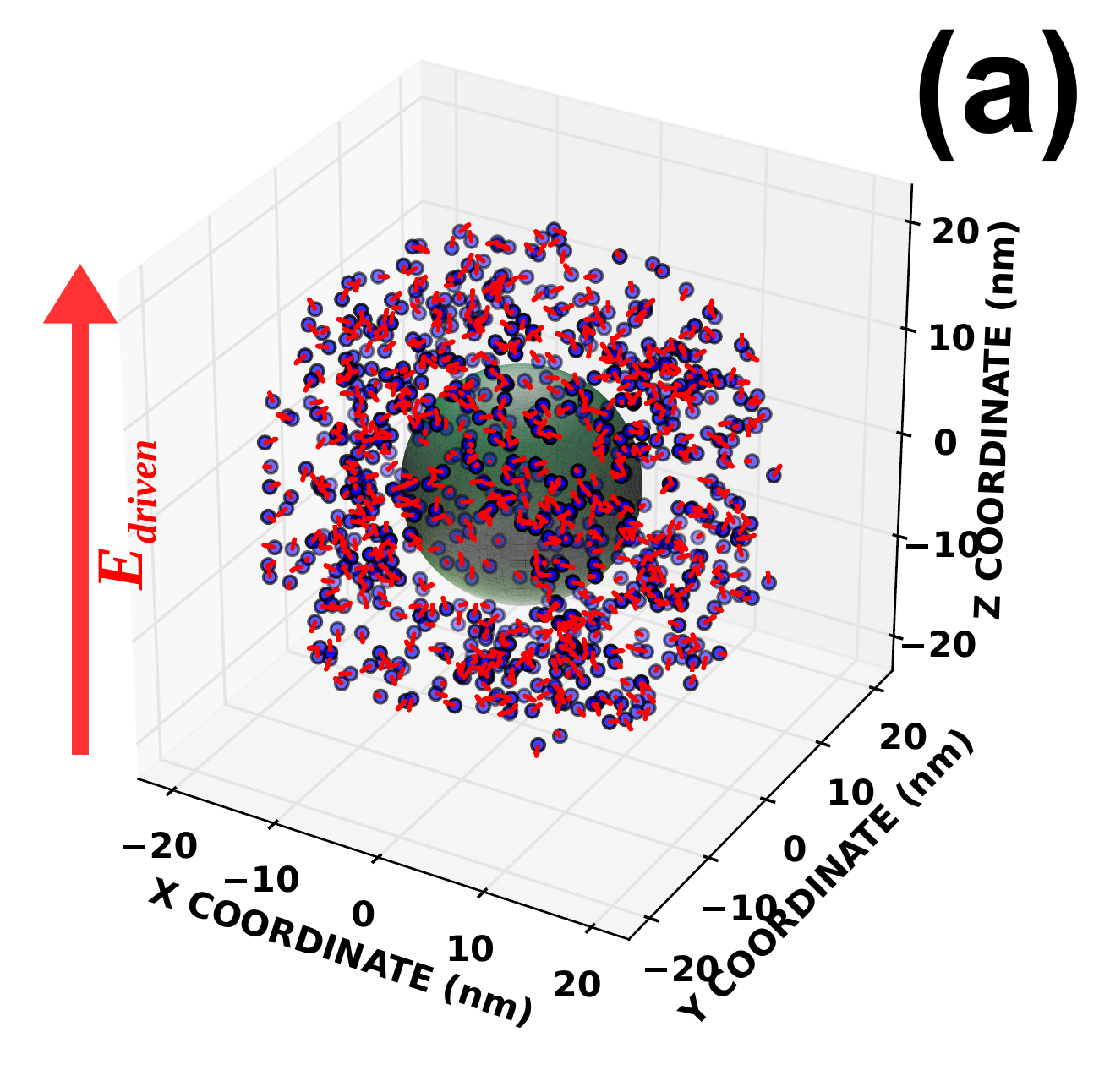}\includegraphics[scale=0.8]{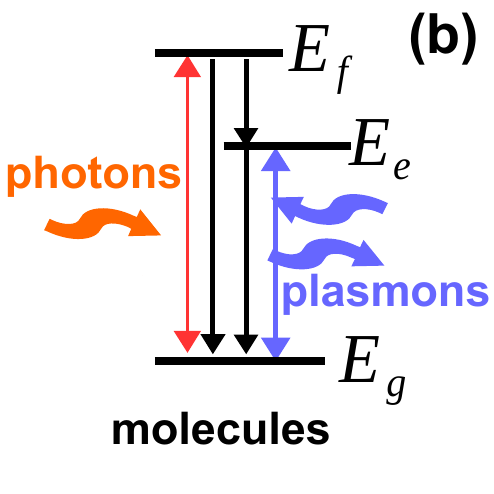}
\par\end{centering}
\begin{centering}
\includegraphics[scale=0.66]{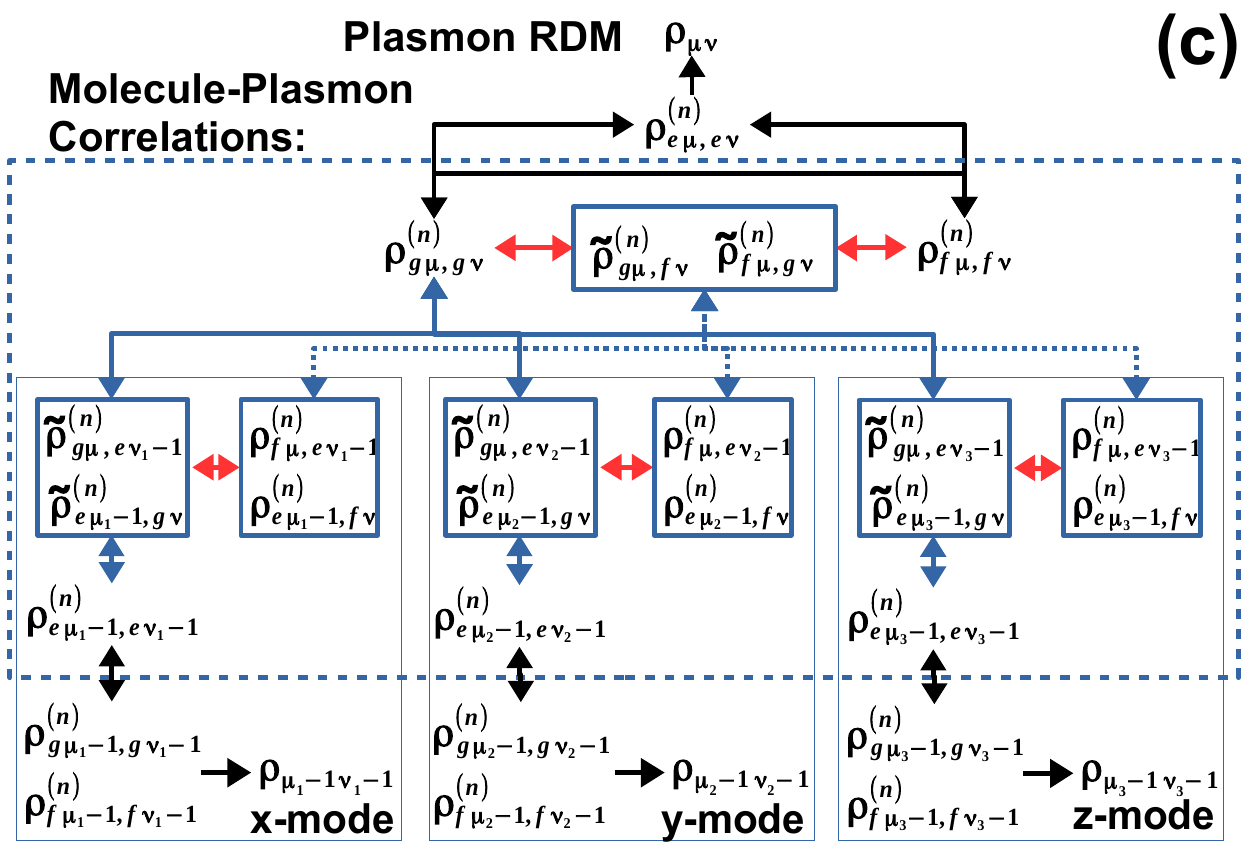}
\par\end{centering}
\caption{\label{fig:scheme-three-level}  Panel (a) illustrates our system, composed of a gold nano-sphere ($10$ nm radius) surrounded
by a layer ($12.5$ nm inner- and $22.5$ nm outer-radius) of $800$ randomly
distributed dye molecules (blue dots) with randomly oriented transition
dipole moments (the red arrows); the driving field is polarized along
the $z$-axis. Panel (b) shows the effective three-level ($E_{g}$, $E_{e}$
and $E_{f}$) molecules interacting coherently with the driving field (the red
arrow) and with the plasmons (the blue arrows) and experiencing
dissipation (the black arrows). Panel (c) shows how the reduced density matrix elements of the plasmon modes
$\rho_{\mu\nu}$ with $\mu\equiv\left\{ \mu_{x}\mu_{y}\mu_{z}\right\} $ depend on the
molecule-plasmon correlations $\rho_{a\mu,b\nu}^{\left(n\right)}$. The thin solid blue boxes indicate the correlations related to $x,y$, and $z$ modes (from left to right), and the colors of the arrows indicate the physical origin of the dependence, cf., panel (b).
Approximate analytical expressions for the correlations are obtained to derive a closed set of  equations for
$\rho_{\mu\nu}$ (for more details see text). }
\end{figure}

In order to theoretically describe these systems, we have to determine the lasing modes and consider how
the gain medium transfers energy to these modes. The modes can be analyzed by solving Maxwell\textquoteright s
equations analytically \cite{YHChou,SWChang-1} or numerically \cite{JHo,RFOulton,CYWu,YJLu-1,YJLu,YHou,QZhang,TPPHSi,YHChou,BTChou,MTHill,MPNezhad,
SHkwon,JHLee,MKha,KDing,JYSuh,WZhou,AYang,AYang-1,AHSchokker,NLi}.
The energy transfer requires us to model the gain medium as a random spatial distribution of multi-level emitters.
The multi-level model allows us to couple some levels with external driving fields or electron reservoirs to describe pumping mechanism and couple other levels with the lasing modes. It means that we have to deal with a complex theoretical problem involving many emitters, many levels and many modes. To reduce the complexity, semi-classical theories have been developed and utilized, for example, rate equations \cite{AMatsudira,CYLu,CYLu-1} and Maxwell-Bloch equations \cite{WZhou,AYang,AYang-1}. Because of the mean-field approximations involved, these theories, however, yield no statistical information about the  the lasing modes and thus advanced full quantum laser theories are needed. Unfortunately, so far those quantum theories treat the emitters as identical two-level systems \cite{MRichter,VMPar} and are thus incapable of dealing with randomly distributed emitters.  

Most existing theories can be viewed as effective descriptions, where the emitters are treated only in an average sense. They can reproduce main characteristics measured in experiments because the inhomogeneity of the emitters becomes irrelevant if huge amounts of them are involved. However, this may not be the case in the plasmonic nano-laser. Because of the strong inhomogeneous subwavelength distribution of the near-field in the nano-laser, the spatial distribution of emitters can significantly affect how they convert the pumping energy into the plasmon energy and even determine whether the systems can achieve lasing or not. By analyzing this influence, we can achieve more insights about the systems and more importantly understand how to improve the system performance by engineering the spatial distribution.   

In this article, we provide a systematic analysis of the inhomogeneity of molecular emitters in a nano-laser of Fig. \ref{fig:scheme-three-level}(a), which resembles the one studied in \cite{MANoginov}. As the basis for our analysis, we will firstly describe our theoretical model in Sec. \ref{sec:model}. To account for multi-molecules and multi-modes, we have extended the density matrix laser theory of Lamb \cite{MSargent} in our model. By following the procedure developed in  \cite{YZhang,YZhang-1}, we first establish a reduced density operator equation for the entire system and then derive a quantum master equation for the reduced density matrix (RDM) of the plasmon modes. Our extended theory allows us to analyze how the molecular distribution affects the plasmon statistics and the molecule-induced mode-correlations. This analysis is presented in Sec. \ref{sec:Calculations}.  In the end, we summarize our findings and present an outlook for future work.

\section{Theoretical Model\label{sec:model}}
As indicated by Fig.\ref{fig:scheme-three-level} (a), we consider a random arrangement of molecular emitters separated by more than $2.5$ nm from the surface of a gold nano-sphere of 10 nm radius. The separation guarantees that electron tunneling is suppressed \cite{KJSavage} and the molecule-MNP coupling is dominated by Coulomb  coupling. The molecules are assumed to be resonant with the dipole plasmons of the sphere.  The higher multipole plasmons have minor influence on the system \cite{YZhang-3} and contribute only as an off-resonant reservoir to the excited-state decay rate of the molecules \cite{JGersten}.

\subsection{Reduced Density Operator Equation\label{subsec:RDOE}}
For a nano-sphere, there are three degenerate dipole plasmon modes with transition
dipole moments pointing along three orthogonal axes. Therefore, we can label them by $j=x,y,z$ or $j=1,2,3$. These modes can be described as quantum harmonic oscillators with Hamiltonian $H_{\mathrm{pl}}=\sum_{j}\hbar\omega_{j}C_{j}^{+}C_{j}$, where $C_{j}^{+}$ and $C_{j}$ are creation and annihilation operators  and  $\hbar\omega_{j}=\hbar\omega_{\mathrm{pl}}$ is their excitation energy \cite{Gweick}. We describe the molecules as three-level systems with the internal energy level scheme and transitions shown in Fig.1(b). The molecular Hamiltonian reads $H_{{\rm e}}=\sum_{n=1}^{N_{{\rm e}}}\sum_{a_{n}}E_{na}\left|a_{n}\right\rangle \left\langle a_{n}\right|$ where ground states $\left|a_{n}=g_{n}\right\rangle $, first $\left|a_{n}=e_{n}\right\rangle $
and second $\left|a_{n}=f_{n}\right\rangle $ excited states have the energies $E_{ng}$,$E_{ne}$,$E_{nf}$, respectively \cite{ThreeLevels}. We assume that the plasmon modes are resonant with the ground-to-first excited state transition, cf. the blue arrow in Fig.\ref{fig:scheme-three-level}(b), and introduces
the coupling Hamiltonian $V_{{\rm pl-e}}=\hbar\sum_{n=1}^{N_{{\rm e}}}v_{ge}^{\left(jn\right)}\left(C_{j}^{+}\left|g_{n}\right\rangle \left\langle e_{n}\right|+{\rm h.c.}\right)$ 
in the \emph{rotating wave approximation}. Here, the coefficient 
$\hbar v_{ge}^{\left(jn\right)}=\left[\mathbf{d}_{ge}^{\left( n\right)}\cdot\mathbf{d}_{j}-3\left(\mathbf{d}_{ge}^{\left(n\right)}\cdot\hat{\mathbf{x}}_{n}\right)\left(\mathbf{d}_{j}\cdot\hat{\mathbf{x}}_{n}\right)\right]/\left|\mathbf{X}_{n}\right|^{3}$  is determined by the transition dipole moment $\mathbf{d}_{ge}^{\left(n\right)}$ of the molecules, $\mathrm{\mathbf{d}}_{j}=d_{\mathrm{pl}}\mathrm{\mathbf{e}}_{j}$
of the plasmon modes as well as the distances $X_n$ and directional unit vectors $\hat{\mathbf{x}}_{n}$
connecting the $n^{\rm th}$ molecule and the sphere-center. We assume that a classical driving field is resonant with the ground-to-second excited state transition, cf. the red arrow in Fig.\ref{fig:scheme-three-level}(b), and introduces the coupling Hamiltonian $V_{{\rm e}}\left(t\right)=\hbar\sum_{n=1}^{N_{{\rm e}}}v_{gf}^{\left(n\right)}\left(e^{i\omega_{0}t}\left|g_{n}\right\rangle \left\langle f_{n}\right|+{\rm h.c.}\right)$
in the \emph{rotating wave approximation}. Here, the coefficient $\hbar v_{gf}^{\left(n\right)}=\mathbf{d}_{gf}^{\left(n\right)}\cdot\mathbf{n}E_{0}$ is determined by another molecular transition dipole moment $\mathbf{d}_{gf}^{\left(n\right)}$ and the driving field is specified by a frequency $\omega_{0}$, a
polarization vector $\mathbf{n}$ and an amplitude $E_{0}$ \cite{NFO}. Here, we consider continuous optical excitation and thus $E_{0}$ is time-independent.

The density operator $\hat{\rho}$ for the quantized plasmon modes and the molecular emitters obeys the following quantum master equation
\begin{equation}
\frac{\partial}{\partial t}\hat{\rho}=-\frac{i}{\hbar}\left[H_{{\rm pl}}+H_{{\rm e}}+V_{{\rm pl-e}}+V_{{\rm e}}\left(t\right),\hat{\rho}\right]-\mathcal{D}\left[\hat{\rho}\right],\label{eq:RDO}
\end{equation}
where the system dissipation is accounted for by the \emph{Lindblad} terms:
\begin{equation}
\mathcal{D}\left[\hat{\rho}\right]=\left(1/2\right)\sum_{u}k_{u}\left(\left[\hat{L}_{u}^{+}\hat{L}_{u},\hat{\rho}\right]_{+}-2\hat{L}_{u}\hat{\rho}\hat{L}_{u}^{+}\right).\label{eq:Dissipation}
\end{equation}
The damping of the plasmon modes is included by terms with $k_{u}=\gamma_{j}=\gamma_{\mathrm{pl}}$,
$\hat{L}_{u}=C_{j}$  for each mode $j$. The decay
processes of the molecules are included by terms with $k_{u}=k_{a\to b}^{\left(n\right)}$, $\hat{L}_{u}=\left|b_{n}\right\rangle \left\langle a_{n}\right|$
for $E_{na}>E_{nb}$ for each molecule, cf. the black arrows in Fig.\ref{fig:scheme-three-level}(b). For the sake of simplicity,
we ignore pure molecular dephasing.

\subsection{Plasmon Reduced Density Matrix Equation\label{subsec:PRDME}}

The main goal of our analysis is to determine the plasmon state populations (probabilities) and correlations as quantified by the reduced density matrix (RDM) with elements
$\rho_{\mu\nu}\equiv\text{tr}_{\text{S}}\left\{ \hat{\rho}\left(t\right)\left|\nu\right\rangle \left\langle \mu\right|\right\} $, where $\text{tr}_{\text{S}}$ denotes the trace over the system and  $\left|\mu\right\rangle \equiv\left|\mu_{x}\right\rangle \left|\mu_{y}\right\rangle \left|\mu_{z}\right\rangle $ and $\left|\nu\right\rangle \equiv\left|\nu_{x}\right\rangle \left|\nu_{y}\right\rangle \left|\nu_{z}\right\rangle $ denote product states of the plasmon occupation number Fock states.
From Eq. \eqref{eq:RDO}, we observe that $\rho_{\mu\nu}$ depends on the molecule-plasmon correlations $\rho_{e\mu_{j}-1,g\nu}^{\left(n\right)}$,
$\rho_{g\mu,e\nu_{j}-1}^{\left(n\right)}$ and $\rho_{e\mu,g\nu_{j}+1}^{\left(n\right)}$,
$\rho_{g\mu_{j}+1,e\nu}^{\left(n\right)}$ with a general definition $\rho_{a\mu,b\nu}^{\left(n\right)}\equiv\text{tr}_{\text{S}}\left\{ \hat{\rho}\left(t\right)\left|b_{n}\right\rangle \left\langle a_{n}\right|\times\left|\nu\right\rangle \left\langle \mu\right|\right\} $, cf. Appendix \ref{sec:derviation-plasmon-rdm}. 

The equations for the correlations also follow from Eq. \eqref{eq:RDO}. 
These equations result in dependence between the plasmon RDM and the correlations, shown in Fig. \ref{fig:scheme-three-level}(c), which is caused by the couplings and the dissipation rates in the master equation (\ref{eq:RDO}).  This dependence also indicates our procedure to solve those inter-dependent equations: Because both molecular and plasmonic dissipation rates contribute to the
decay of the correlations, they must decay faster and thus may adiabatically \cite{MOScully} follow
the plasmon RDM elements which are only affected by the plasmon damping. Because of the molecular dissipation, the correlations
represented within the blue dashed box of Fig. \ref{fig:scheme-three-level}(c) depend on the correlations
outside the box. Fortunately, they all can be expressed as functions of the plasmon RDM because of the symmetry hidden in the coupling Hamiltonians. 
Finally, we back substitute these expressions and obtain closed dynamic equations for the plasmon RDM, where the molecules
contribute by several coefficients, cf. Eq.\eqref{eq:equation-plasmon-rdm} in Appendix \ref{sec:derviation-plasmon-rdm}.

The diagonal elements $P_{\mu}\equiv\rho_{\mu\mu}$ are
the populations (the probabilities) of the plasmon number states $\left|\mu\right\rangle $
while the off-diagonal elements $\rho_{\mu\nu}$ ($\mu\neq\nu$)  represent the 
coherence of the plasmons. Here, we focus on the populations by
solving the equations for those diagonal elements:
\begin{align}
 & \frac{\partial}{\partial t}P_{\mu}=-\sum_{j=1}^{3} \left[ \left(\gamma_{j}\mu_{j}+\kappa{}_{\mu}^{\left(j\right)}\right)P_{\mu} - \sum_{k=1}^{3}\eta_{\mu}^{\left(jk\right)}P_{\mu_{k}-1} \right ]\nonumber \\
 & +\sum_{j=1}^{3} \left[ \left(\gamma_{j}\left(\mu_{j}+1\right)+\kappa_{\mu_{j}+1}^{\left(j\right)}\right)P_{\mu_{j}+1}-\sum_{k=1}^{3}\eta_{\mu_{j}+1}^{\left(jk\right)}P_{\mu_{j}+1\mu_{k}-1} \right], \label{eq:EPSP}
\end{align}
where the rates $\kappa_{\mu}^{\left(j\right)}\equiv-\sum_{n=1}^{N_{\mathrm{e}}}\alpha{}_{\mu\mu}^{\left(jn\right)}$ and $\eta_{\mu}^{\left(jk\right)}\equiv-\sum_{n=1}^{N_{\mathrm{e}}}\beta_{\mu\mu}^{\left(jkn\right)}$
include contributions from individual molecule $\alpha{}_{\mu\mu}^{\left(jn\right)}$
and $\beta_{\mu\mu}^{\left(jkn\right)}$ respectively, cf. Eqs. \eqref{eq:alpha}
and \eqref{eq:beta} in the Appendix \ref{sec:derviation-plasmon-rdm}. Here, $\mu_j \pm 1$ for $j=x$ indicates 
$(\mu_x \pm 1, \mu_y, \mu_z)$ and $\mu_j +1 \mu_k -1$ for $j=x$ and $k=y$ denotes $(\mu_x + 1, \mu_y -1, \mu_z)$.
Since the former rates decrease the population of higher
plasmon states but increase that of lower states, they can be interpreted
as molecule-induced plasmon damping rates. Since the latter rates
have the opposite effect on the population, they can be interpreted as
molecule-induced plasmon pumping rates. The latter rates depend
on two plasmon mode indices and thus account for correlation
between different plasmon modes induced by the molecules. These rates
can be considered as extended Einstein's AB coefficients accounting for the multi-plasmon
modes, the molecular pumping mechanism and the molecular inhomogeneity.

At steady-state the second line of Eq. \eqref{eq:EPSP} is recovered if we replace
$\mu_{j}$ by $\mu_{j}+1$ on the right side of the first line, which suggests a recursion relation of the populations. 
We obtain such a relation by setting the first line to zero: 
\begin{equation}
P_{\mu}=\frac{\sum_{k=1}^{3}\left(\sum_{j=1}^{3}\eta_{\mu}^{\left(jk\right)}\right)P_{\mu_{k}-1}}{\sum_{j=1}^{3}\left(\gamma_{j}\mu_{j}+\kappa{}_{\mu}^{\left(j\right)}\right)}. \label{eq:final-recursion-relation}
\end{equation}
Together with the normalization condition $\sum_{\mu}P_{\mu}=1$,
the above relation can be utilized to easily calculate the populations
 according to the procedure outlined in Fig. \ref{fig:procedure} in Appendix \ref{sec:derviation-plasmon-rdm}.
Although $P_{\mu_{x}\mu_{y}\mu_{z}}$ contains all the information about the
three dipole plasmons, it is more intuitive to consider physical
quantities related to one or two dipole plasmon. We calculate them by tracing out
one mode to get  $P_{\mu_{x}\mu_{y}}=\sum_{\mu_{z}}P_{\mu_{x}\mu_{y}\mu_{z}}$,
$P_{\mu_{y}\mu_{z}}=\sum_{\mu_{x}}P_{\mu_{x}\mu_{y}\mu_{z}}$ and
$P_{\mu_{x}\mu_{z}}=\sum_{\mu_{y}}P_{\mu_{x}\mu_{y}\mu_{z}}$ (the joint population of two modes) or by tracing 
out two modes to get $P_{\mu_{x}}=\sum_{\mu_{y}\mu_{z}}P_{\mu_{x}\mu_{y}\mu_{z}}$,
$P_{\mu_{y}}=\sum_{\mu_{x}\mu_{z}}P_{\mu_{x}\mu_{y}\mu_{z}}$ and
$P_{\mu_{z}}=\sum_{\mu_{x}\mu_{y}}P_{\mu_{x}\mu_{y}\mu_{z}}$  (the reduced population of one mode). We
can also quantify the strength of plasmon excitation with the so-called plasmon
mean numbers: $N_{j}\equiv\sum_{\mu_{j}}\mu_{j}P_{\mu_{j}}$ and the
plasmon statistics with the so-called (steady-state) second order
correlation functions: $g_{j}^{\left(2\right)} (0) \equiv\sum_{\mu_{j}}\mu_{j}\left(\mu_{j}-1\right)P_{\mu_{j}}$.
To analyze how the individual molecule contributes to the plasmon
excitation, we can calculate the population of individual molecular states: $P_{g}^{\left(n\right)}\equiv\sum_{\mu}\rho_{g\mu,g\mu}^{\left(n\right)}$,
$P_{f}^{\left(n\right)}\equiv\sum_{\mu}\rho_{f\mu,f\mu}^{\left(n\right)}$
and $P_{e}^{\left(n\right)}\equiv\sum_{\mu}\rho_{e\mu,e\mu}^{\left(n\right)}$,
which are actually determined by $P_{\mu_{x}\mu_{y}\mu_{z}}$, cf.
Eqs. \eqref{eq:rhogg-4}, \eqref{eq:rhoff-4}  and \eqref{eq:rhoee-4}
in Appendix \ref{sec:collection-population}.

\section{Results\label{sec:Calculations}}
The above theoretical model provides clues about how the molecular inhomogeneity 
may affect the system performance. The molecular inhomogeneity mainly originates from their positions and orientations of 
their transition dipole moments, which leads to that all the molecules interact with the three modes simultaneously 
but with random strengths. Since this situation is too complex, we shift its discussion to the end and first consider
 a special configuration where all the molecules are located along the equator of the gold sphere, 
cf. Fig.\ref{fig:single-plasmon-mode} (a). 

\subsection{Configuration with Single Dipole Plasmon Mode}

\begin{figure}
\begin{centering}
\includegraphics[scale=0.25]{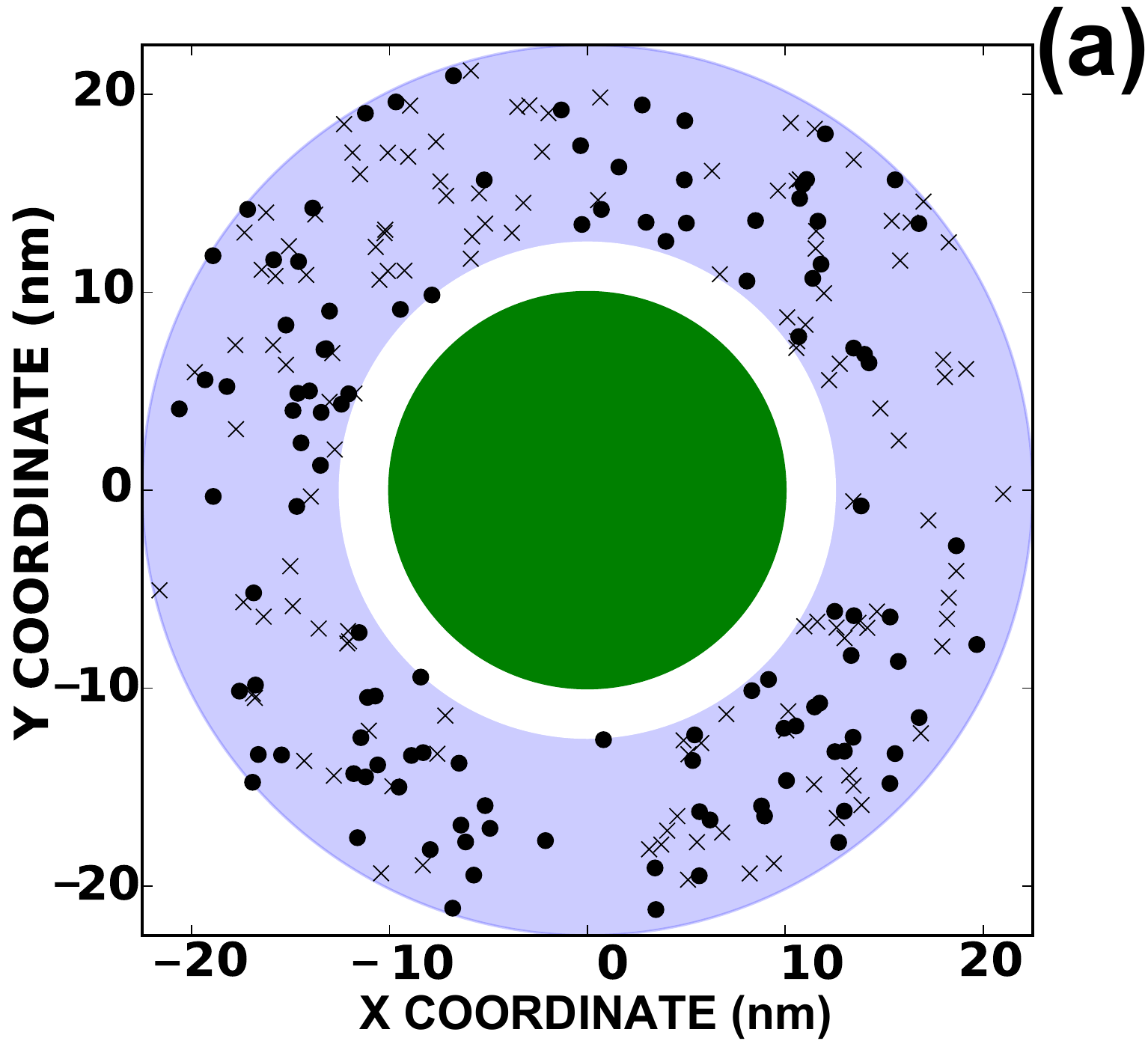}\includegraphics[scale=0.24]{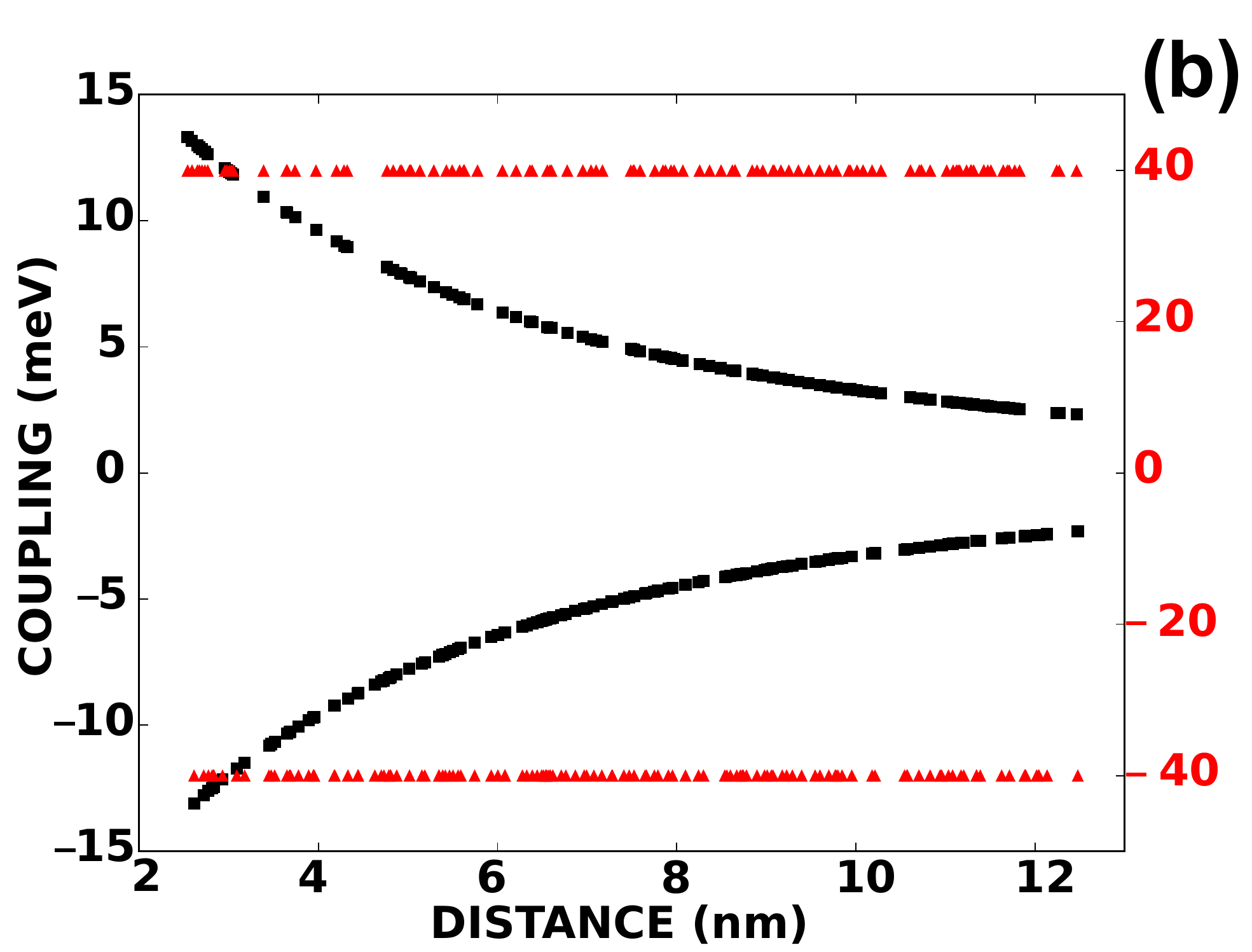}
\par\end{centering}
\begin{centering}
\includegraphics[scale=0.22]{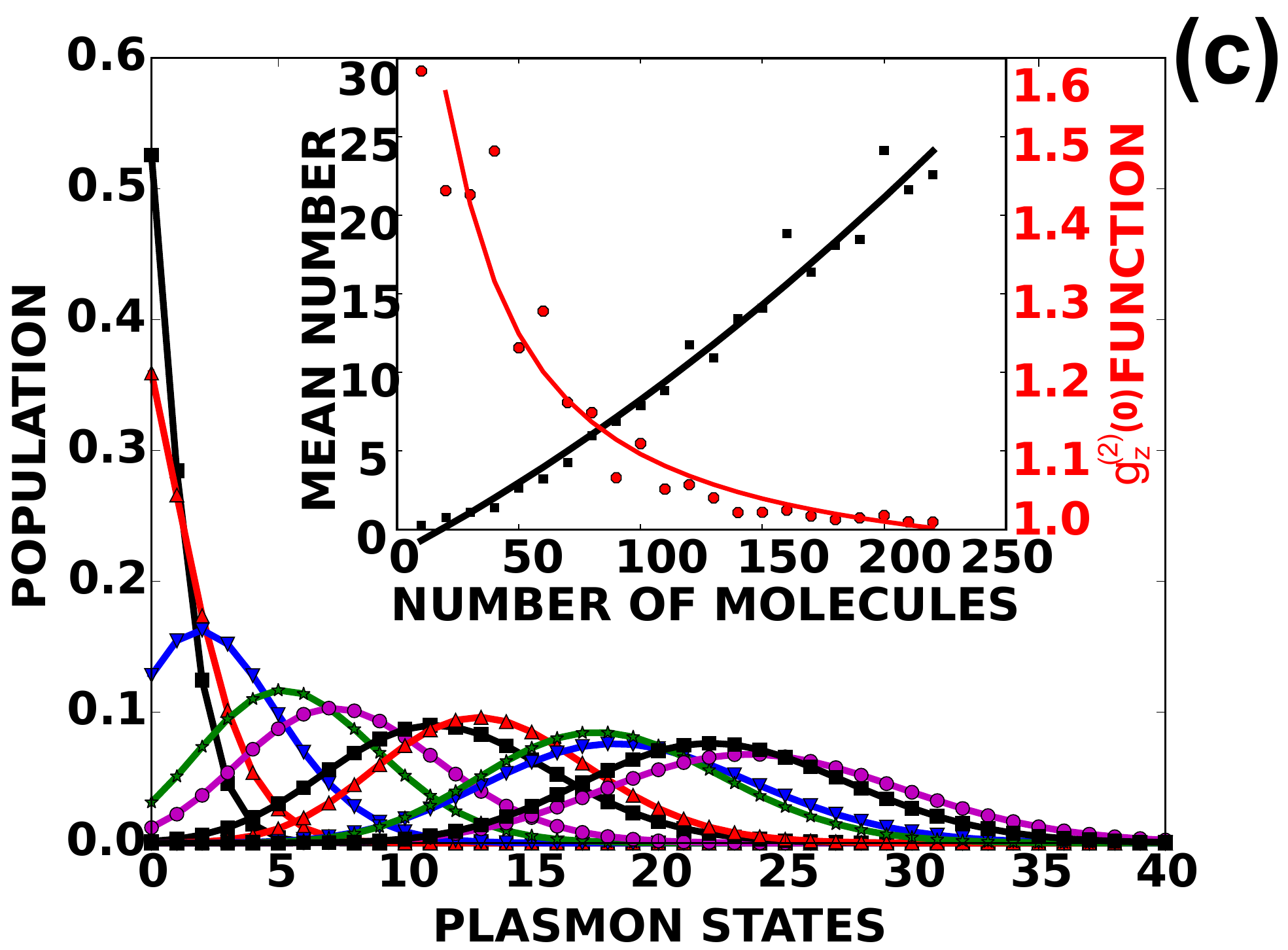}\includegraphics[scale=0.22]{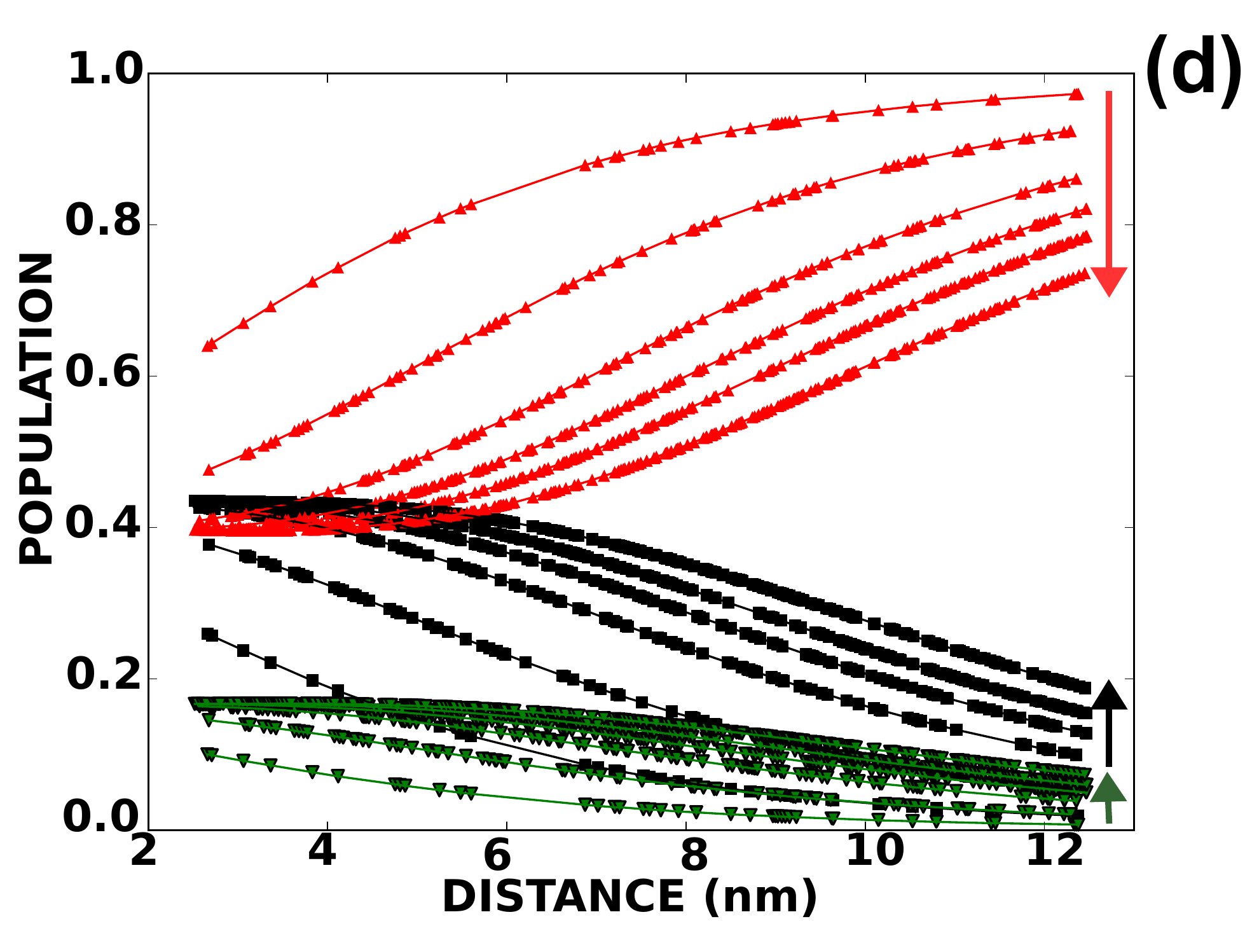}
\par\end{centering}
\begin{centering}
\includegraphics[scale=0.22]{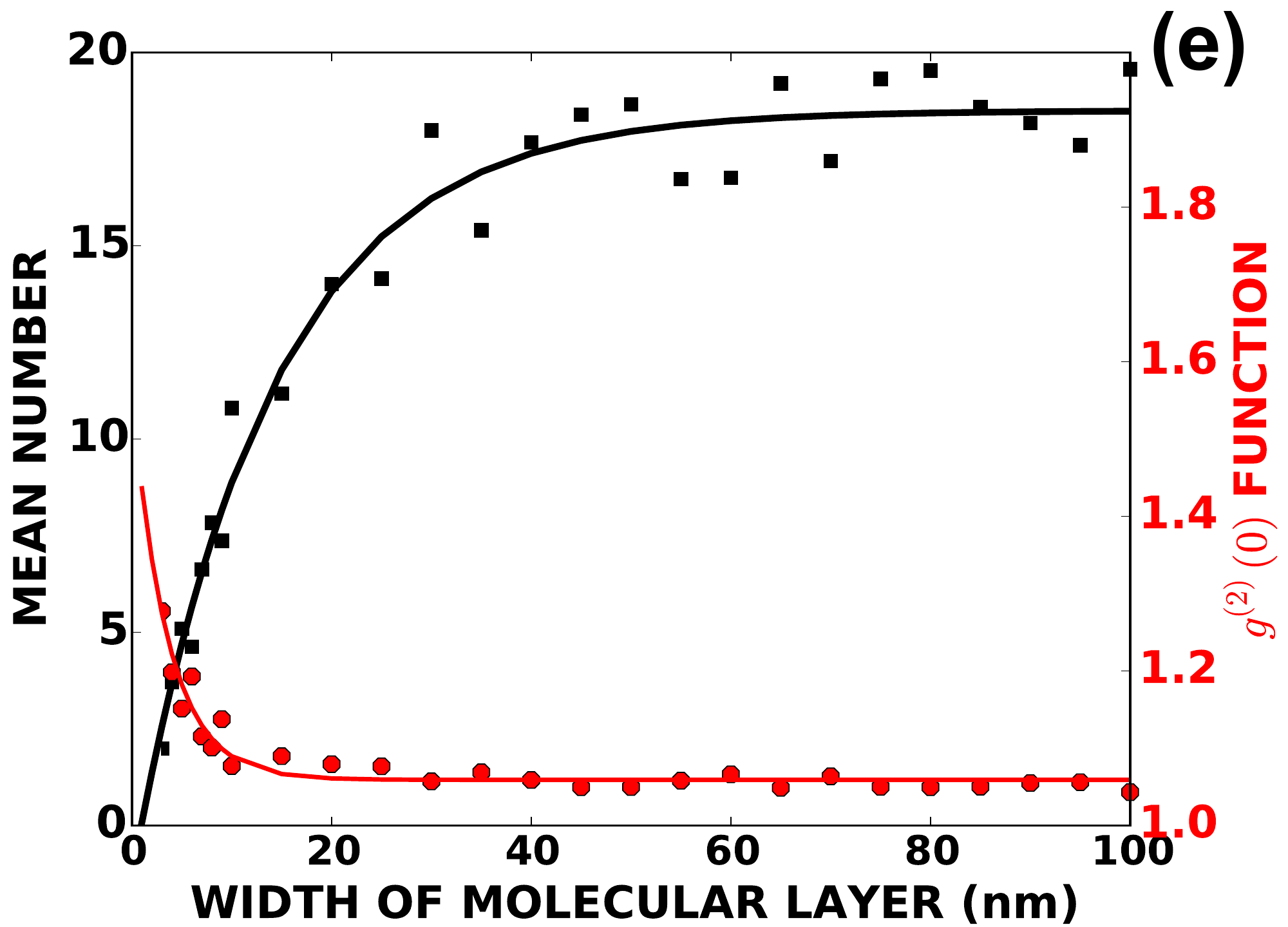}\includegraphics[scale=0.22]{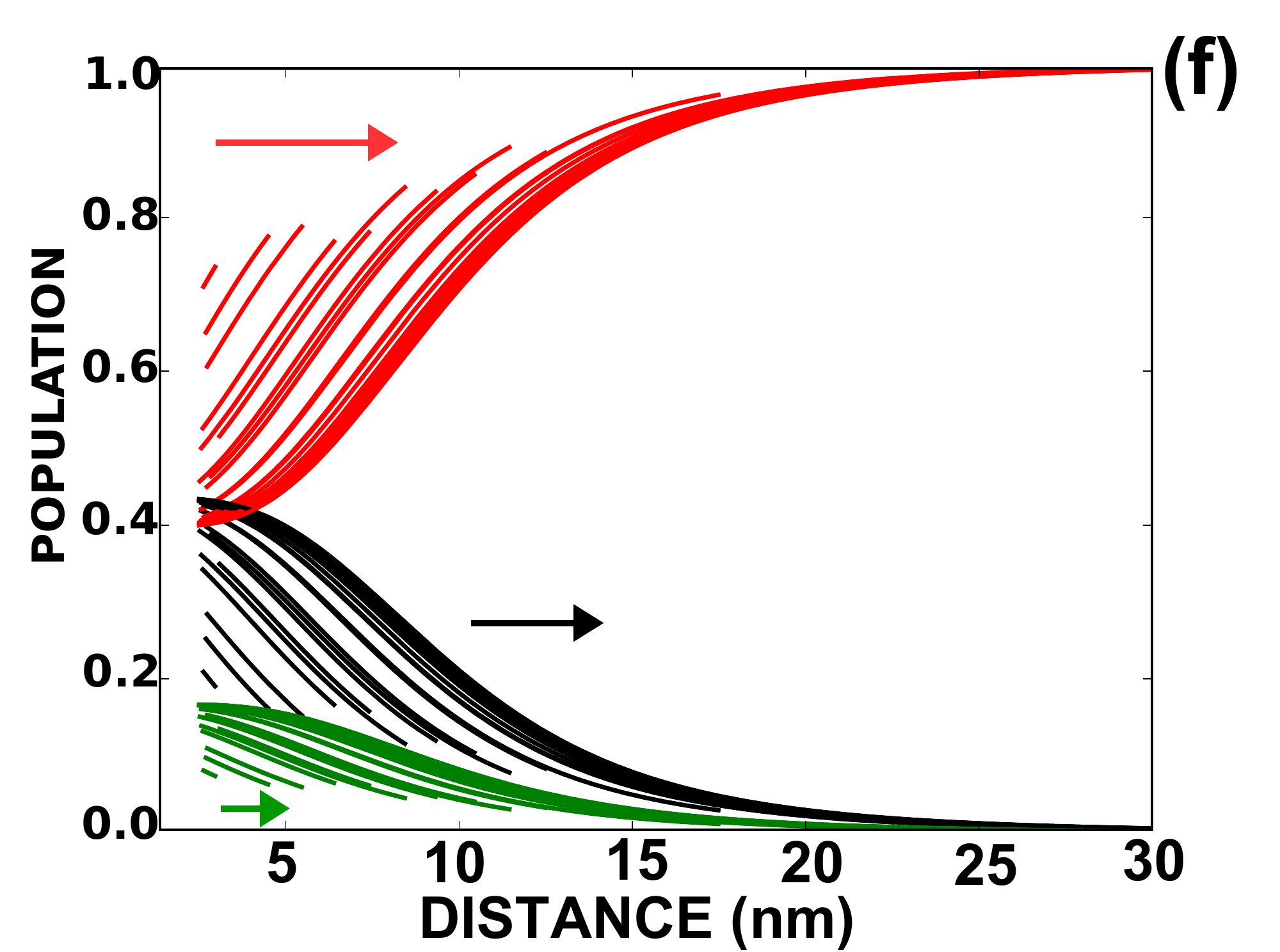}
\par\end{centering}
\caption{ \label{fig:single-plasmon-mode} Configuration with single dipole plasmon mode. Panel (a): a gold nano-sphere (the green filled circle) with
$240$ randomly distributed molecules in a layer ($r_{\mathrm{in}}=12.5$
inner- and $r_{\mathrm{out}}=22.5$ nm outer-radius); the black dots
and crosses indicate the molecular transition dipole moment along
the z-axis; the driving field is polarized
along the positive $z$-axis. Panel (b): the coupling with the dipole
plasmon (the black squares) and with the driving field (the red triangles)
for molecules with different distances $d_n$ to the
sphere-surface.  Panel (c): plasmon state population $P_{\mu_{z}}$
for systems with increasing number of molecules $N_{\mathrm{e}}$ from $20$ to $220$
in a step of $20$ (from the left to right curve); the inset shows
plasmon mean number $N_{z}$ (the black squares fitted with $-1.54 + 8.26\times10^{-2} N_{\rm e} +1.54 \times 10^{-4} N_{\rm e}^2$) and the
$g_{z}^{\left(2\right)}(0)$-functions (the red octagons fitted with $0.74 e^{-0.02 N_{\mathrm e} } +0.98 $) versus
$N_{\mathrm{e}}$. Panel (d): population of molecular states $P_{a}^{\left(n\right)}$ versus
$d_{n}$; $P_{g}^{\left(n\right)}$(the black squares
and curves), $P_{e}^{\left(n\right)}$ (the red up-triangles and curves),
$P_{f}^{\left(n\right)}$ (the green down-triangles and curves); the
arrows on the right indicate the increase of $N_{\mathrm{e}}$. Panel (e-f): systems
with different widths $W=r_{\mathrm{out}}-r_{\mathrm{in}}$ of the molecular
layer (fixed molecular density);
panel (e) shows $N_{z}$ (the black squares fitted with $-19.79 e^{-0.72 W} +18.49 $) and $g_{z}^{2}\left(0\right)$-function
(the red circles fitted with $0.50 e^{-0.28 W } +1.06 $)
; panel (f) shows $P_{g}^{\left(n\right)}$, $P_{e}^{\left(n\right)}$ and $P_{f}^{\left(n\right)}$ versus $d_n$ with $W$
from $1$ nm to $100$ nm (the zoomed area in the inset); the arrows
indicate the increase of $W$. Physical parameters are specified in Table \ref{tab:parameters} in Appendix \ref{sec:parameters}. }
\end{figure}

For the configuration in Fig. \ref{fig:single-plasmon-mode} (a), the molecule-plasmon coupling is
reduced to $\hbar v_{ge}^{\left(jn\right)}= \delta_{j,z} \left( \pm d_{ge}^{(n)} d_{\rm pl} /X_n^{3} \right)$ 
with positive (negative) sign for the molecules oriented along the positive (negative) z-axis. 
Obviously, the dipole plasmon x- and y-mode are not involved and thus can be ignored in the following analysis. 
The coupling depends inversely on the cubic of 
 the molecule-sphere center distance $X_n=a_{\rm MNP} + d_n$, cf. the black squares in Fig. \ref{fig:single-plasmon-mode} (b).
Here, $a_{\rm MNP}$ is the radius of the sphere and $d_n$ the distance to the sphere-surface. 
In contrast, the driving field coupling $\hbar v_{gf}^{\left(n\right)}$ on the molecules depends only on the molecular
orientations but not the positions, cf. the red triangles in Fig. \ref{fig:single-plasmon-mode} (b). 
If all the molecules have the same distance to the sphere-surface, they are equivalent and 
the resulting ideal system has been already analyzed in \cite{YZhang-1}. There,
we focused on the pumping mechanism and found the optimal parameters 
of the system leading to the strongest plasmon excitation, cf. Table \ref{tab:parameters}
in Appendix \ref{sec:parameters}. These parameters will be used as reference parameters for 
the following simulations.

To compensate the strong plasmon damping, the number of molecules coupled strongly
with the plasmons is an essential parameter. It was demonstrated in the experiment \cite{XGMeng} that
the system properties like emission wavelength, intensity and pumping threshold strongly depend on 
the concentration (number) of the molecules. Here, we analyze this dependence from three aspects: density of molecules, 
spatial extension of molecular layer and molecular level shift. 

As indicated by Fig.\ref{fig:single-plasmon-mode} (b), the molecules close to the sphere couple strongly with the plasmons. Therefore,
those molecules contribute more to the plasmon excitation than other 
molecules. By increasing the molecular density, we increase the number of molecules and thus the plasmon excitation. This is clearly 
reflected  in Fig. \ref{fig:single-plasmon-mode} (c) by the increased populations $P_{\mu_{z}}$ of higher plasmon excited states 
and the increased plasmon mean number $N_{z}$ (the black dots and curve in the inset)
with increasing number of molecules $N_{\mathrm{e}}$ from $10$ to $240$.
For larger  $N_{\mathrm{e}}$, $P_{\mu_{z}}$ resemble Poisson distributions indicating the formation
of a coherent state and $N_{z}$ approaches $25$, which is much larger than unity and thus indicates that the system is lasing.
 This conclusion is further confirmed by the $g_z^{\left(2\right)}(0)$-function, cf. the red dots and red curve in the inset of 
Fig. \ref{fig:single-plasmon-mode} (c),  which approaches unity for large $N_{\mathrm{e}}$, i.e. the Poisson limit.
 The fluctuation of the dots around the curves in the inset is caused by different molecular distribution 
in each simulation and may thus represent fluctuations encountered in experiments.   

To understand why the increasing molecular density can increase the plasmon excitation,
here, we analyze the state population for every molecule $P_{a}^{\left(n\right)}$,
 cf. Fig. \ref{fig:single-plasmon-mode}(d). First, we notice that the molecule-plasmon coupling  $V_{{\rm pl-e}}$ leads to
reversible processes (spontaneous emission, stimulated emission and absorption of the plasmons)
since it enters into our master equation as a coherent coupling, cf. Eq. (\ref{eq:RDO}).
These processes tend to balance the population of the molecular excited states  $P_{e}^{\left(n\right)}$
and ground states $P_{g}^{\left(n\right)}$. This leads to the reduced $P_{e}^{\left(n\right)}$,
cf. the red curves and arrow, and the increased $P_{g}^{\left(n\right)}$, cf. the black curves and arrow,
 with increasing $N_{\mathrm{e}}$. In addition, because the reduced coupling with increasing distance $d_n$ (cf. 
in Fig. \ref{fig:single-plasmon-mode}(b)) reduces the rate of the processes, 
 the $P_{e}^{\left(n\right)}$ increase and $P_{g}^{\left(n\right)}$ decrease with increasing $d_n$.
The population of the higher excited state $P_{f}^{\left(n\right)}$ is mainly determined by 
the strong decay rate from this state to the middle excited state and thus is always smaller than the other populations. 

In the following, we consider the effect of varying the spatial extent
(width $W$) of the molecular layer, cf. Fig. \ref{fig:single-plasmon-mode}(e), which 
 can also be studied in experiments like \cite{MANoginov,XGMeng} by precisely controlling 
the synthesis time of the molecular layer. In this case, the molecules are added far away from the sphere surface and
will thus contribute less to the plasmon excitation because of the reduced molecule-plasmon coupling,
cf. Fig. \ref{fig:single-plasmon-mode} (b). As a result, the plasmon mean number  $N_{z}$  and the $g_{z}^{\left(2\right)}\left(0\right)$-function
 saturate for large $W$ as displayed by Fig. \ref{fig:single-plasmon-mode}(e).
In addition, we find that the data points are close to the fitted curve
for small $W$ but fluctuate a lot for large $W$. This can be easily
understood with the change of the molecular state population $P_{a}^{\left(n\right)}$, cf.
Fig. \ref{fig:single-plasmon-mode}(f). When $W$ increases from 1
nm to $30$ nm,  $P_{g}^{\left(n\right)},P_{e}^{\left(n\right)},P_{f}^{\left(n\right)}$
change dramatically, cf. the zoomed inset, since all molecules contribute to the plasmon excitation.
Therefore, $N_{z}$ increases and the molecular inhomogeneity
 has less influence on the plasmon excitation. However,
when $W$ increases further, $P_{g}^{\left(n\right)},P_{e}^{\left(n\right)},P_{f}^{\left(n\right)}$
 change less and now the molecules distributed 
in the region near to the sphere will significantly affect $N_{z}$.

\begin{figure}
\begin{centering}
\includegraphics[scale=0.23]{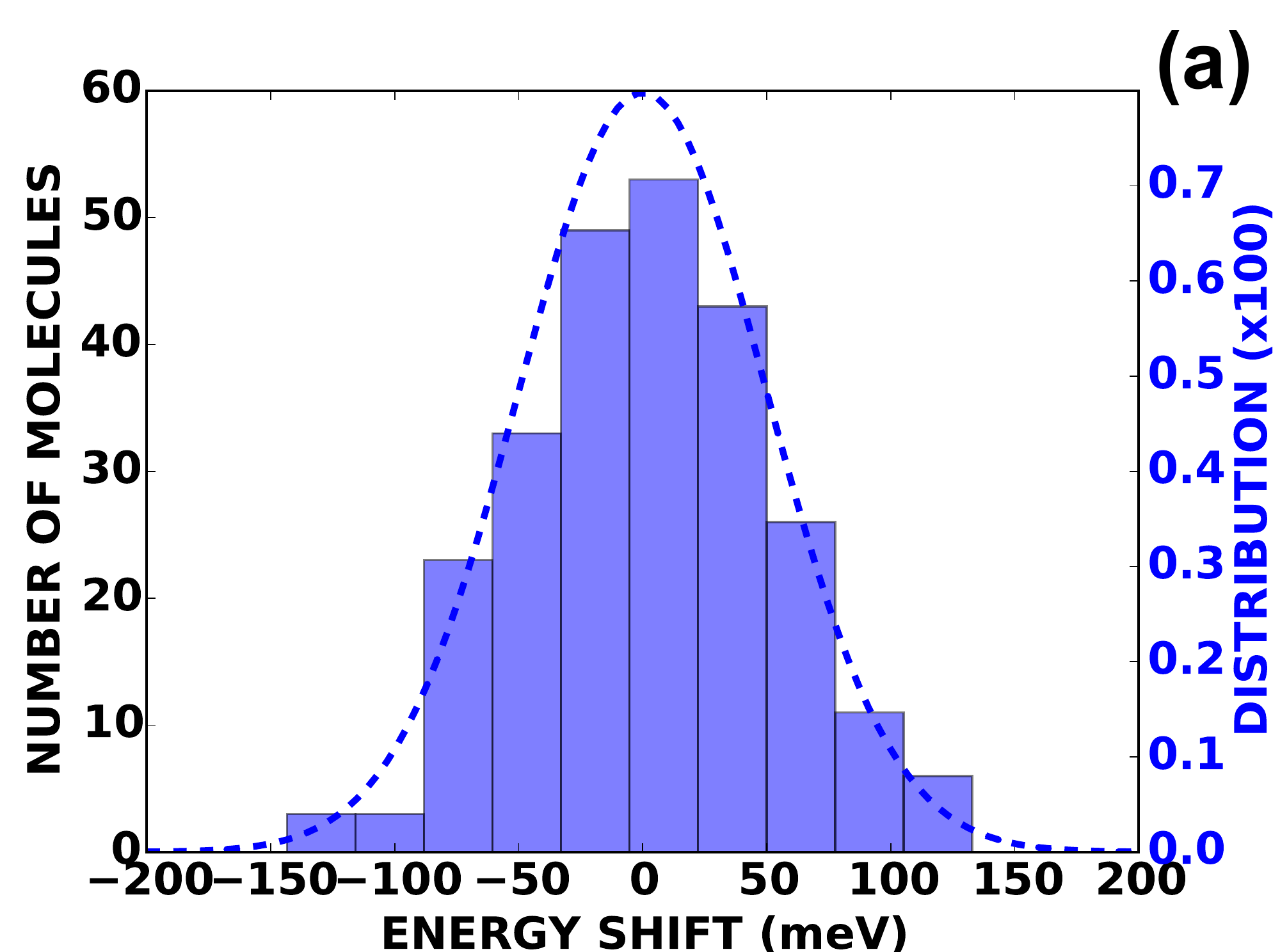}\includegraphics[scale=0.23]{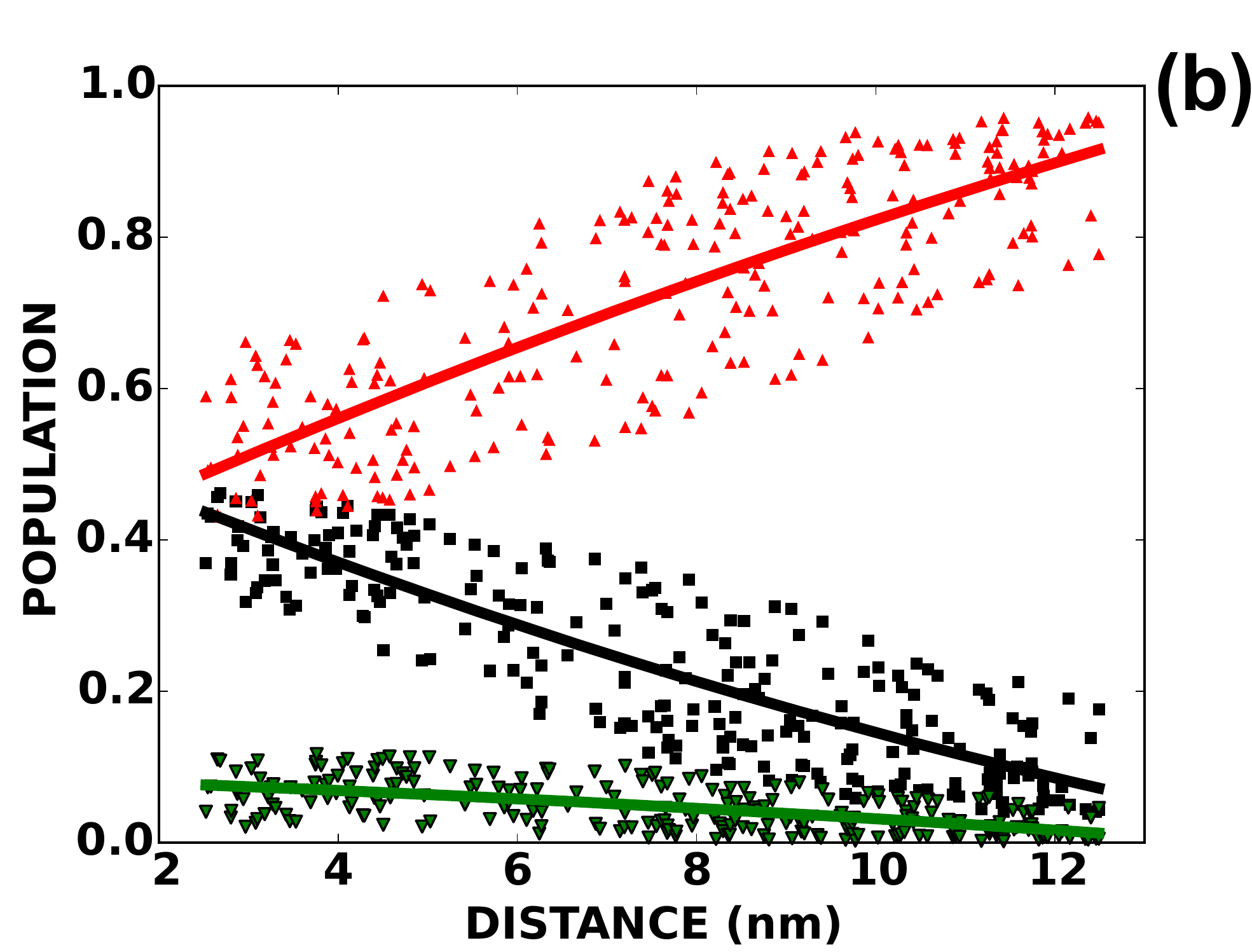}
\par\end{centering}
\begin{centering}
\includegraphics[scale=0.23]{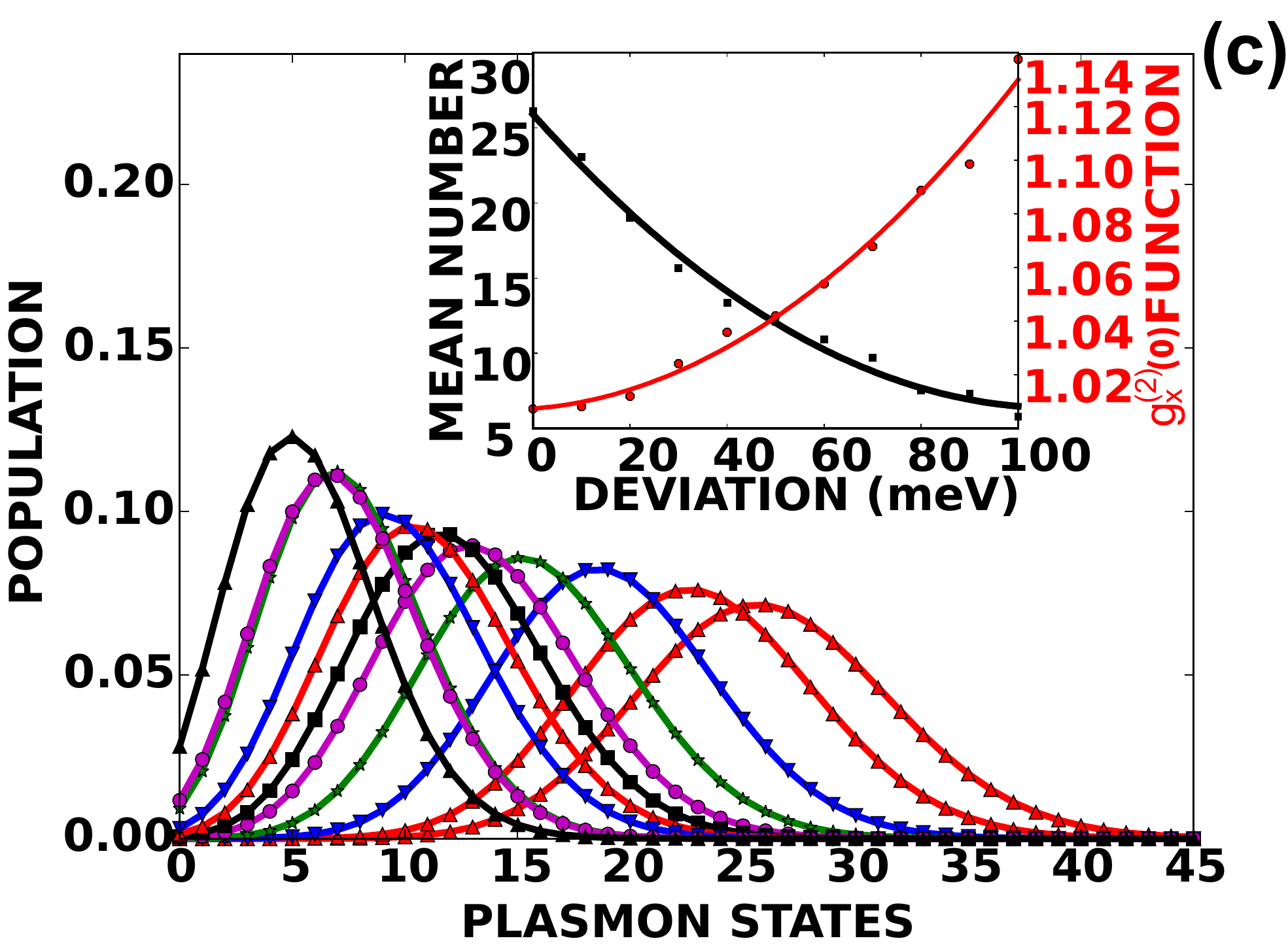}\includegraphics[scale=0.23]{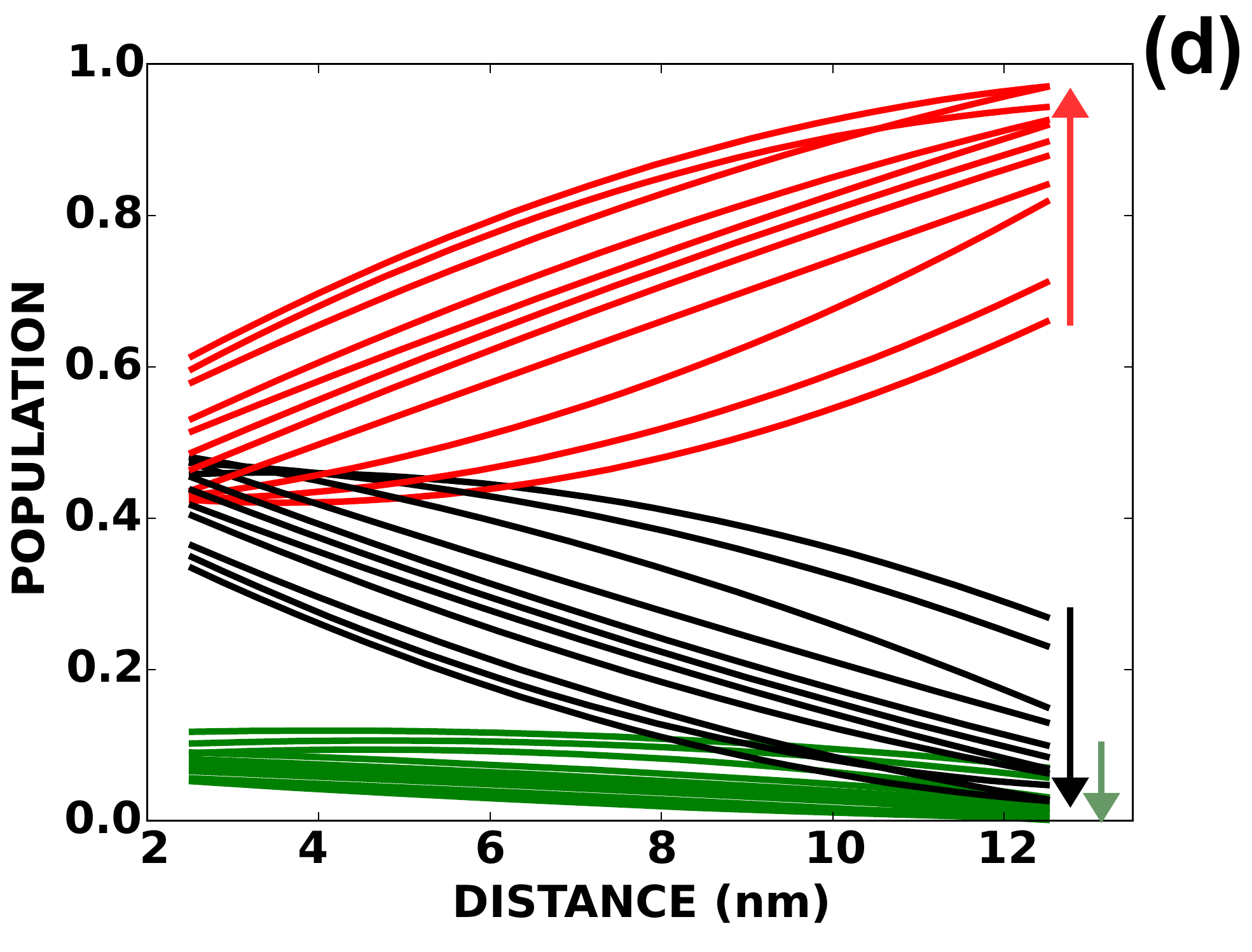}
\par\end{centering}
\caption{\label{fig:single-plasmon-mode-energy-shift}Effect of molecular energy-shift
 $\delta E^{\left(n\right)}$ for a system with $N_{\mathrm{e}}=250$ molecules.
Panel (a): histogram of $\delta E^{\left(n\right)}$ as well as the
Gaussian distribution with the deviation $\sigma=$50 meV. Panel (b):
population of molecular states $P_{a}^{\left(n\right)}$ versus the
molecular distance to the sphere-surface, $P_{g}^{\left(n\right)}$
(the black squares and curve), $P_{e}^{\left(n\right)}$ (the red
up-triangles and curve), $P_{f}^{\left(n\right)}$ (the green down-triangles
and curve); the random populations fitted by polynomial function.
Panel (c): plasmon state population $P_{\mu_{z}}$ with increasing
$\sigma$ from $0$ meV to $100$ meV in a step of $10$ meV (from
the right to left curve); the inset: plasmon mean number $N_{z}$
(the black dots and curve) and $g_{z}^{\left(2\right)}\left(0\right)$-function
(the red dots and curve). Panel (d): fitted populations of molecular states
for increasing $\sigma$ indicated by the arrows. The strength of
the driving field is $E_{0}=9\times10^{7}$ $\mathrm{V/m}$. Physical
parameters are specified in Table \ref{tab:parameters} in Appendix \ref{sec:parameters}.}
\end{figure}

The compact molecular arrangement around the sphere implies that the molecules may directly 
interact with each other through Coulomb coupling. If the molecular concentration is very high, 
 electron transfer between molecules becomes possible. Although this process may be relevant here,
  it is however beyond the scope of our theory. For not too high concentration, 
  the electron transfer can be ignored but direct energy exchange between excited molecular dipoles
  can lead to energy-shift  (inhomogeneous broadening). 
  In principle, such effect can be accurately described 
by directly incorporating the inter-molecular energy exchange coupling into the system Hamiltonian in the 
master equation (\ref{eq:RDO}). However, here, we follow  an easier, phenomenological 
 way to account for such effect by introducing random energy shift $\delta E^{\left(n\right)}$
to individual molecule with a \emph{Gaussian} distribution \cite{HHaken}, cf. Fig. \ref{fig:single-plasmon-mode-energy-shift} (a),
$p\left(\delta E\right)=\left(1/\sqrt{2\pi}\sigma\right)\exp\left\{ -\left(\delta E\right)^{2}/\left(2\sigma^{2}\right)\right\} $ (with
standard deviation $\sigma$).  It means that the transition energies are modified as 
$\hbar\omega_{eg}^{\left(n\right)}=\hbar\omega_{eg}+\delta E^{\left(n\right)}$
and $\hbar\omega_{fg}^{\left(n\right)}=\hbar\omega_{fg}+\delta E^{\left(n\right)}$, compared to the values in Table I in  Appendix A.

The consequence of energy-shift is to perturb the perfect resonant condition 
 for the molecular pumping and the molecule-plasmon energy transfer assumed previously. 
This is reflected by the irregular change of the state populations $P_{a}^{\left(n\right)}$
for the molecules at similar distances to the sphere surface, cf. the dots in  Fig. \ref{fig:single-plasmon-mode-energy-shift} (b).
However, since the majority of molecules has no or small energy shift as shown in Fig. \ref{fig:single-plasmon-mode-energy-shift} (a),
the populations $P_{a}^{\left(n\right)}$ in Fig. \ref{fig:single-plasmon-mode-energy-shift} (b) still roughly follow the same trend 
observed in Fig. \ref{fig:single-plasmon-mode} (d), cf. the solid lines. The broadening of the transition energies is also 
reflected in the shift of the plasmon state population $P_{\mu_z}$ to lower states, a reduced plasmon 
mean number $N_z$ and an increased $g_z^{(2)}(0)$-function with increasing deviation $\sigma$ of 
the energy-shift from 0 meV to 100 meV, cf. Fig. \ref{fig:single-plasmon-mode-energy-shift} (c). 
These features can be understood by analyzing the contribution of individual molecule through 
their state populations $P_{a}^{\left(n\right)}$. As shown in Fig. \ref{fig:single-plasmon-mode-energy-shift} (d),
the populations of the molecular middle excited states $P_{e}^{\left(n\right)}$ increase while those of the ground states
$P_{g}^{\left(n\right)}$ decrease with increasing $\sigma$. These results reflect that 
 the molecules are on average less affected by the plasmons and thus contribute less to the plasmon excitation.
The features described above indicate that the lasing performance is strongly affected by the inhomogeneous
molecular energy-shift. Finally, we point out that the standard deviation $\sigma$ characterizes energy-shifts due to intra-molecular interactions
and should hence depends on the molecular concentration.

\subsection{Configuration with Two and Three Dipole Plasmon Modes}

\begin{figure}
\begin{centering}
\includegraphics[scale=0.25]{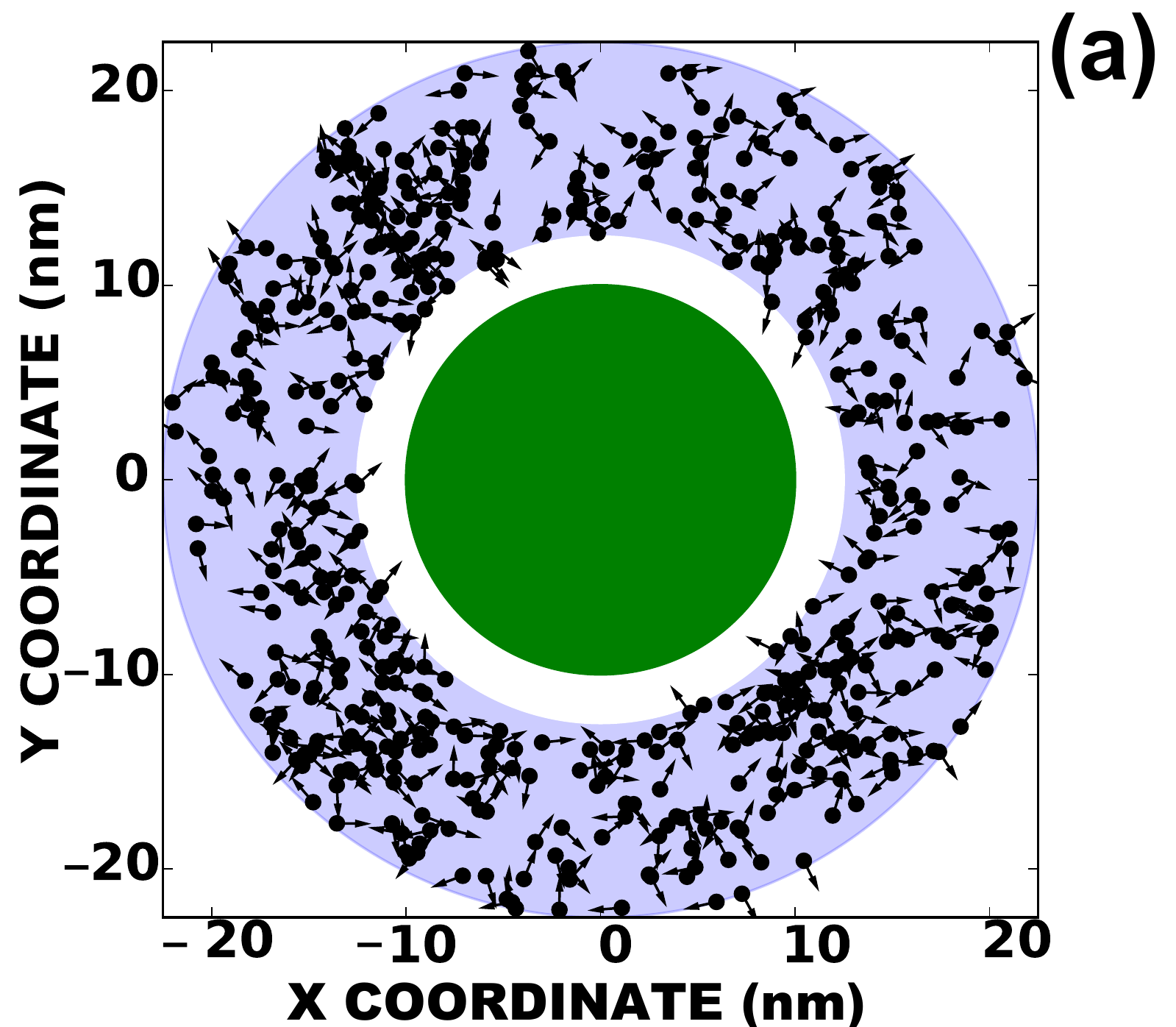}\includegraphics[scale=0.24]{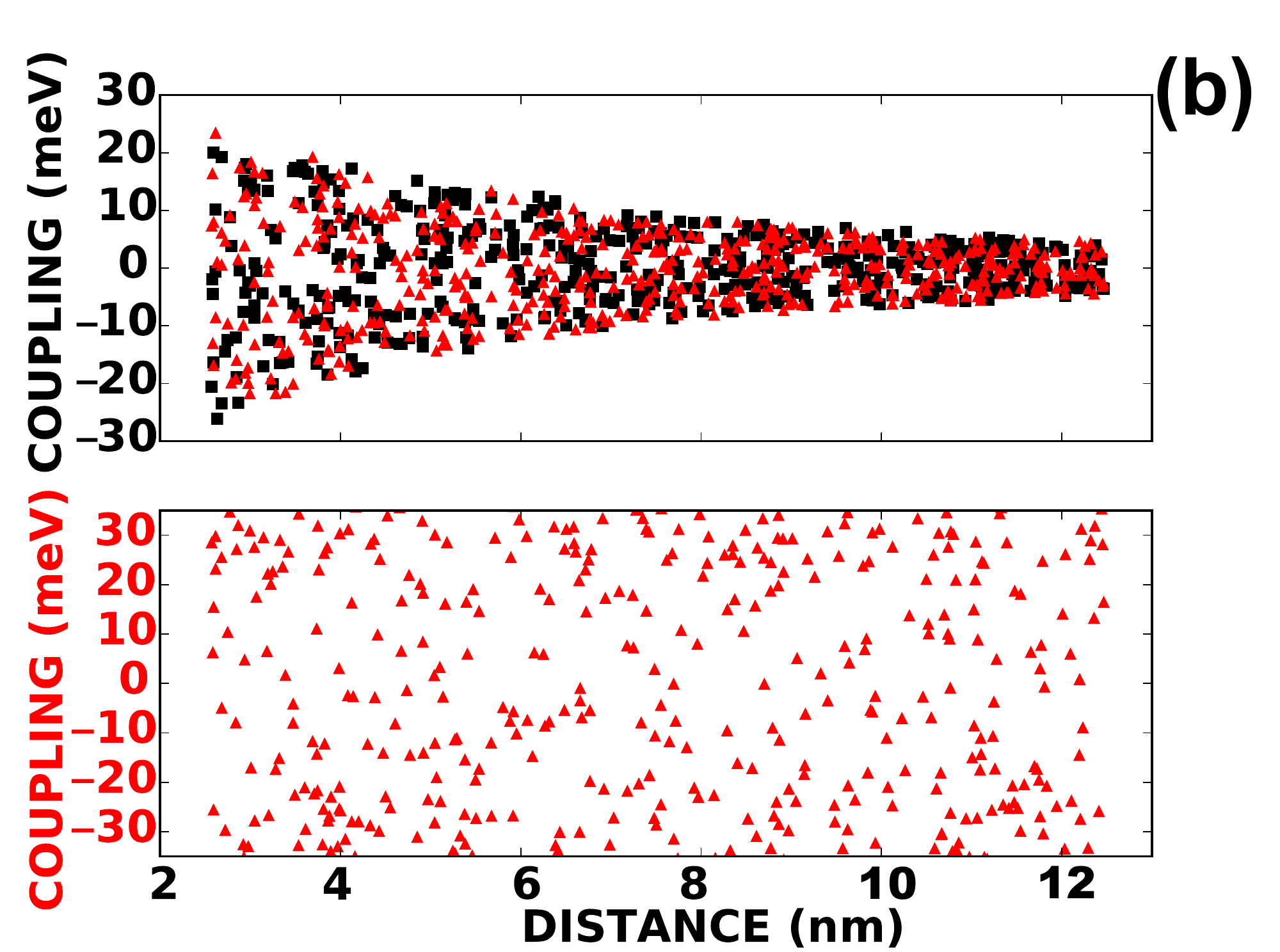}
\par\end{centering}
\begin{centering}
\includegraphics[scale=0.24]{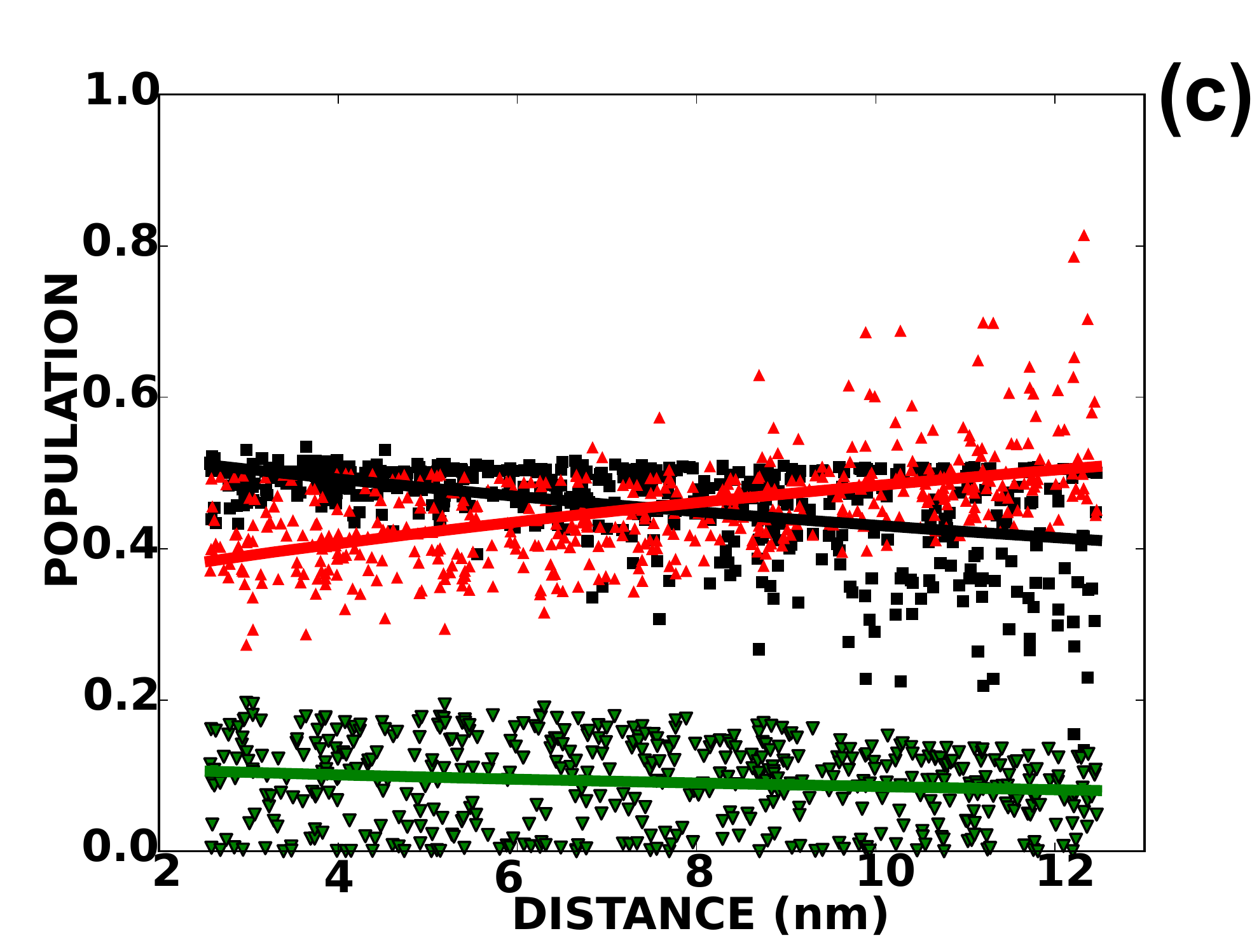}\includegraphics[scale=0.3]{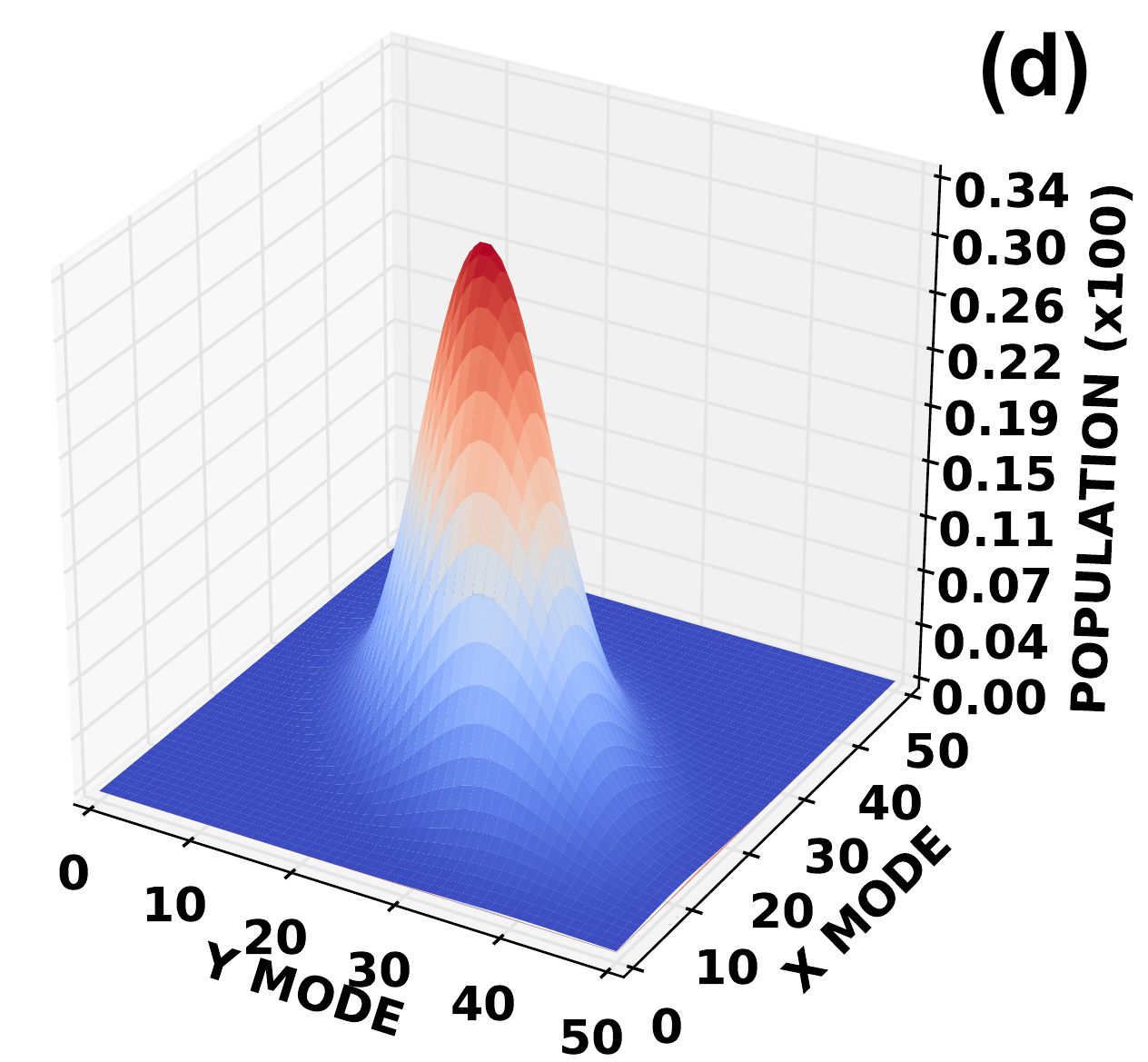}
\par\end{centering}
\caption{\label{fig:double-plasmon-mode}Configuration with two dipole
plasmon modes. Panel (a): the configuration for  $N_{\mathrm{e}}=500$ molecules 
 is similar to Fig.\ref{fig:single-plasmon-mode} (a) except that the molecular transition
 dipole moments orient randomly in the xy-plane and the driving field is along the positive
$x$-axis. Panel (b): the couplings of molecules
at different distances $d^{\left(n\right)}$ to the sphere-surface;
the upper panel shows the coupling with the plasmon x-mode (the black squares),
and with the plasmon y-mode (the red triangles); the lower panel shows
the coupling with the driving field. Panel (c): the population of molecular
states $P_{a}^{\left(n\right)}$ versus $d^{\left(n\right)}$ ; $P_{g}^{\left(n\right)}$
(the black squares), $P_{e}^{\left(n\right)}$ (the red upper triangles),
$P_{f}^{\left(n\right)}$ (the green down triangles); the curves are
exponentially fits to the average population as function of distance. Panel (d): joint population $P_{\mu_{x}\mu_{y}}$ of plasmon Fock states.
Physical parameters are specified in Table \ref{tab:parameters}  in Appendix \ref{sec:parameters}. }
\end{figure}

We now consider the more complex situation where the molecular transition dipole moments  $\mathbf{d}_{ge}^{\left(n\right)}$ 
and  $\mathbf{d}_{gf}^{\left(n\right)}$ orient randomly in the x-y plane, cf. Fig. \ref{fig:double-plasmon-mode}(a).
In this case, the molecules couple with the dipole plasmon $x-$ and $y-$mode simultaneously
with random strength, cf. the upper panel of Fig. \ref{fig:double-plasmon-mode}(b). 
In addition, the driving field coupling becomes also random as shown 
in the lower panel of Fig. \ref{fig:double-plasmon-mode}(b) because it also depends on the orientations.
This implies that the molecules at similar distance to the sphere-surface 
 experience different couplings and this leads to the random population of molecular states 
$P_{a}^{\left(n\right)}$ as displayed in Fig. \ref{fig:double-plasmon-mode}(c). 
However, because the maximum of the molecule-plasmon coupling decreases with increasing
distance to the sphere-surface $d^{\left(n\right)}$, the distance-dependent averaged $P_{a}^{\left(n\right)}$
show a similar behavior as in Fig. \ref{fig:single-plasmon-mode}(d). The co-existence of the 
x- and -y mode is directly illustrated by  the joint plasmon state population 
$P_{\mu_{x}\mu_{y}}$ for a system with $N_{\mathrm{e}}=500$ molecules, cf. Fig. \ref{fig:double-plasmon-mode}(d). 
 Here, to better visualize $P_{\mu_{x}\mu_{y}}$, it is shown as a smooth surface. The population has a maximum around
$\mu_{x}=25$ and $\mu_{y}=25$, which indicates that the both plasmon modes are excited to the same strength. 
In addition, we have also analyzed  the plasmon state populations $P_{\mu_{x}},P_{\mu_{y}}$
, the plasmon mean numbers $N_{x},N_{y}$ and the $g_{x}^{\left(2\right)}\left(0\right)$-
and $g_{y}^{\left(2\right)}\left(0\right)$-function for systems with
increasing molecular density (number of molecules) in Fig. \ref{fig:other-results-two-modes}(a-c) and with increasing 
deviation $\sigma$ of molecular energy shift in Fig. \ref{fig:other-results-two-modes}(d-e) in Appendix
\ref{sec:parameters}. Basically, they show similar features like those in  Fig. \ref{fig:single-plasmon-mode}(c)
and Fig. \ref{fig:single-plasmon-mode-energy-shift} (c) respectively. 

\begin{figure}[t]
\begin{centering}
\includegraphics[scale=0.22]{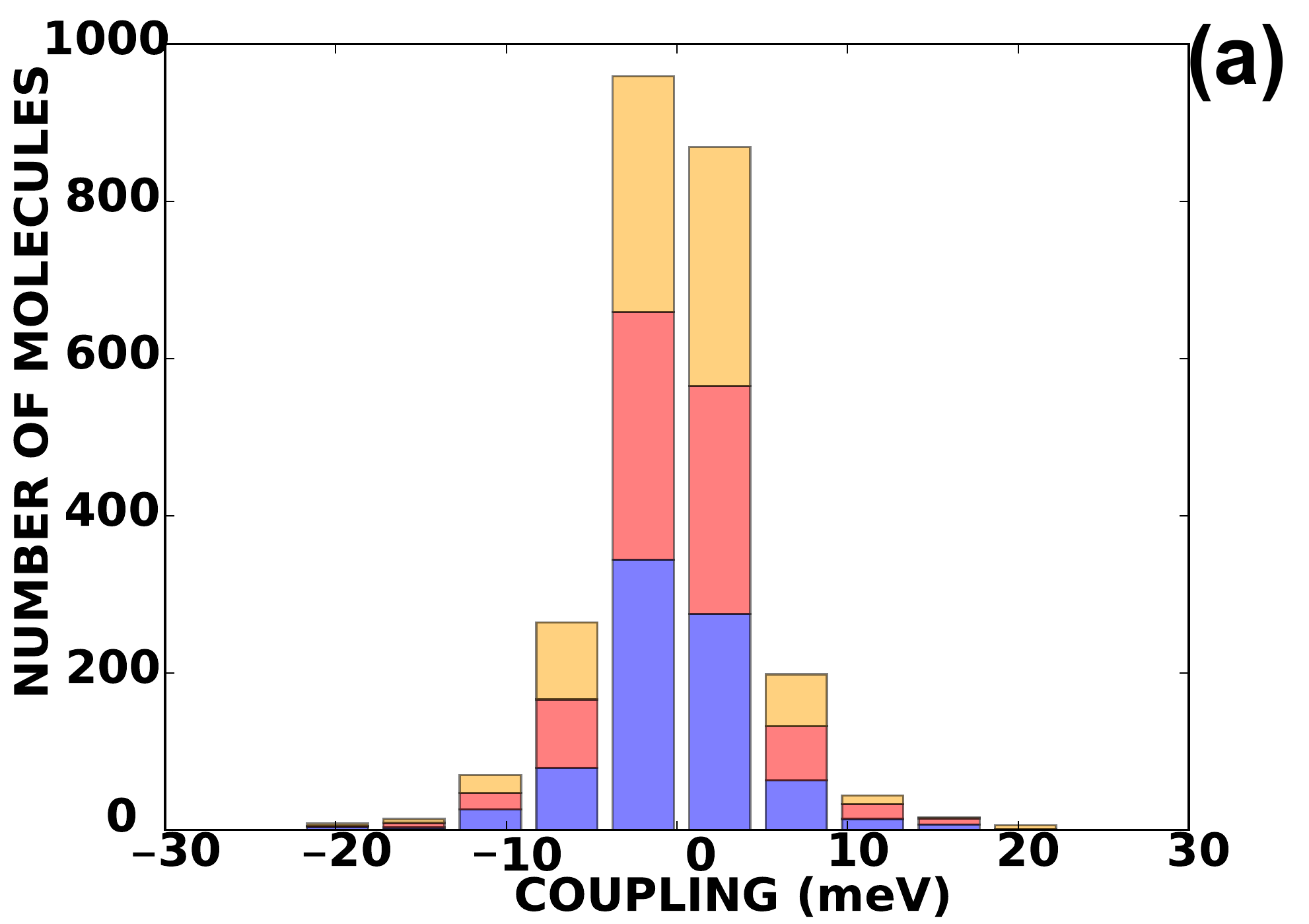}\includegraphics[scale=0.28]{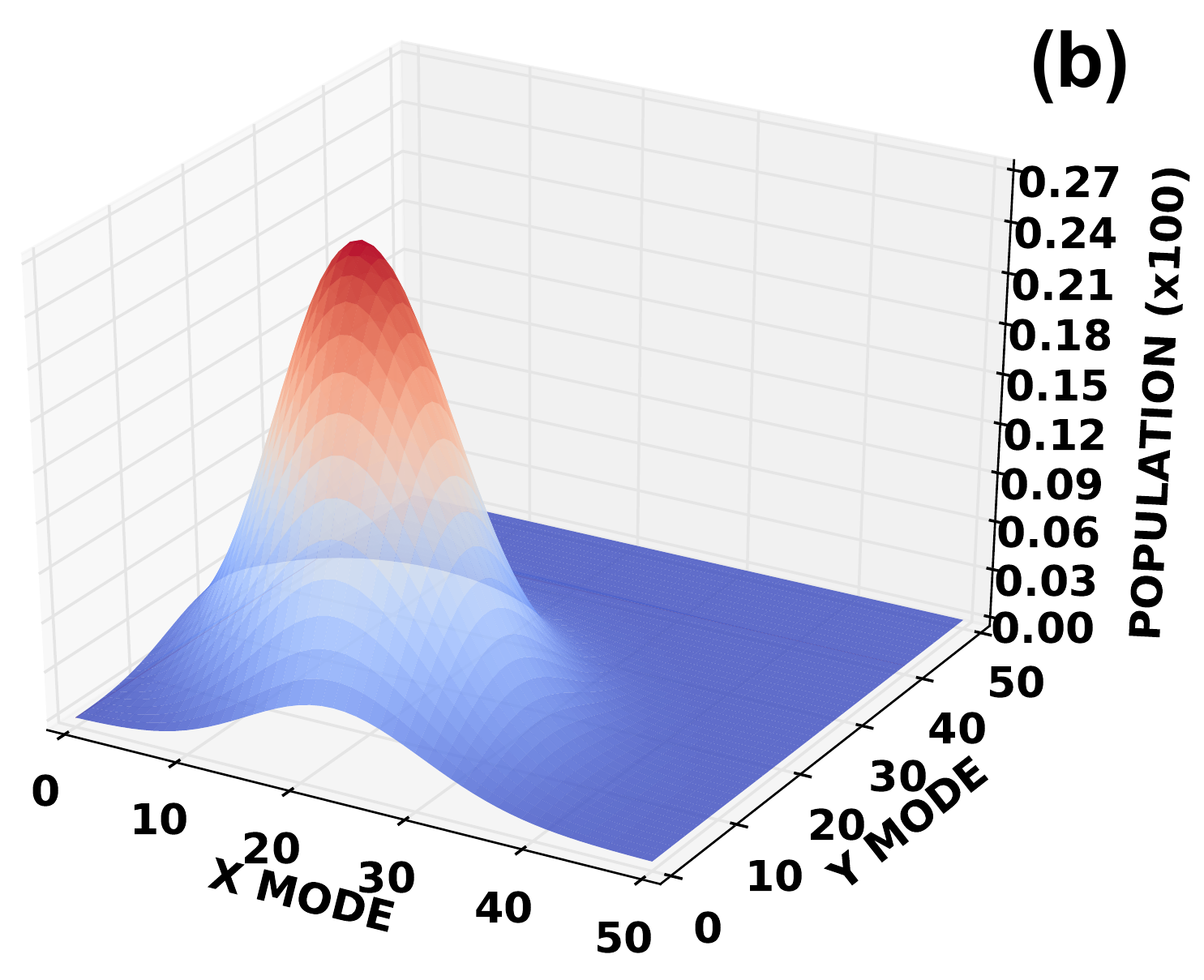}
\par\end{centering}
\caption{\label{fig:three-modes} Configuration shown in Fig. \ref{fig:scheme-three-level}(a) 
with three dipole plasmon modes. Panel
(a): stacked histogram of the molecule-plasmon coupling; blue, red
and orange bars are for x,y,z-mode respectively. Panel (b): the joint
population $P_{\mu_{x}\mu_{y}}$ for a system with $800$ molecules.
Physical parameters are specified the Table \ref{tab:parameters}  in Appendix \ref{sec:parameters}. }
\end{figure}

Finally, let us turn to the realistic configuration of Fig. \ref{fig:scheme-three-level}(a). In this case, 
the randomly distributed molecules in three dimensions couple with the 
three plasmon modes in a similar pattern, cf.  Fig. \ref{fig:three-modes} (a) (see also the  
coupling of individual molecule in Fig. \ref{fig:other-results-three-modes} (a) in Appendix \ref{sec:parameters}). 
This implies that all the plasmon modes will be excited by the molecules in similar way and this 
is reflected by the joint populations $P_{\mu_{x}\mu_{y}}$, $P_{\mu_{y}\mu_{z}}$ and
$P_{\mu_{x}\mu_{z}}$ with a peak around $\left(10,10\right)$ for a system with $N_{\rm e} = 800$ molecules, cf. Fig. \ref{fig:three-modes} (b) (see also Fig.
\ref{fig:other-results-three-modes} (b,c) in Appendix \ref{sec:parameters}). In addition, we also find 
 the increased population $P_{\mu_{x}}$, $P_{\mu_{y}}$, $P_{\mu_{z}}$ of higher plasmon states, 
the increased plasmon mean number $N_{x},N_{y},N_{z}$ as well as 
the reduced $g_{x}^{\left(2\right)}\left(0\right),g_{y}^{\left(2\right)}\left(0\right),g_{z}^{\left(2\right)}\left(0\right)$-functions
 with increasing number of molecules $N_{\mathrm{e}}$ (
cf. Fig. \ref{fig:other-results-three-modes}(d,e,f) in Appendix \ref{sec:parameters}). 

In order to achieve same plasmon excitation per mode, we must double (triple) the number of molecules in the
case with two (three) modes compared to the single mode case.
incidentally, our results show that the polarization of the driving field alone does not cause 
significant asymmetry between the excitation of the three plasmon modes.

\section{Conclusions\label{sec:Conclusions}}

In summary, we have developed a quantum laser theory based on 
reduced density matrix equation and applied it to a plasmonic nano-laser
consisting of a gold nano-sphere and many dye molecules. Our study reveals that the molecular
inhomogeneity and the multi-plasmon modes make strong molecular pumping necessary to compensate strong
plasmon damping and to achieve lasing.  By increasing the molecular density, the plasmon excitation increases, 
but molecular energy-shifts due to inter-molecular interaction may ultimately 
reduce the plasmon excitation.

In this article, we modeled the molecular emitters as three-level systems. However, the procedure illustrated 
can be readily applied to the emitters with arbitrary level structure, which will be necessary 
to study the influence of other intrinsic processes of the emitters on the laser performance. 
For example, by introducing more intermediate molecular vibrational levels, in principle, we can study how the 
intra-molecular vibrational energy redistribution and the temperature of the environment affect the system performance. 
This extended theory may be utilized to analyze the experiments \cite{AYang}, where 
 the varying excitation energy of lattice plasmons due to changing the surrounding material
 affects the dye molecules used by picking up the molecular energy levels resonant to the plasmons. 
This study will not only provide more insights about the interplay of the plasmons and the gain material but may
 also suggest how to optimize the system performance.

\appendix
\newpage

\section{System Parameters and Other Results\label{sec:parameters}}

In Table \ref{tab:parameters}, we list reference parameters
for our simulations. We consider a gold nano-sphere with a radius of
$10$ nm. The corresponding dipole plasmons have an excitation energy
of $\hbar\omega_{\mathrm{pl}}=2.6$ eV, a damping rate $\hbar\gamma_{\mathrm{c}}=100$
meV and an optical transition dipole moment $d_{\mathrm{pl}}=2925$
D. The classical driving field has the photon energy $\hbar\omega_{0}=2.7$
eV and the amplitude $E_{0}=1.2\times10^{8}$ V/m consistent with
the values used in the experiments \cite{WZhou,AYang,AYang-1,XGMeng,YJLu-1,YJLu,RMMa,CYWu,QZhang,KDing}.
The molecules have the transition energy $\hbar\omega_{eg}=2.6$ eV
and the transition dipole moments $d_{gf}=16$ D and $d_{ge}=14.4$
D. The molecular transition energy $\hbar\omega_{fg}=2.7$ eV is off-resonant 
from  the higher multipole plasmons  \cite{YZhang-3}. The decay rate is $\hbar k_{f\to e}^{\left(n\right)}=100$
meV and we assume the other rates $k_{f\to g}^{\left(n\right)}$ and $k_{e\to g}^{\left(n\right)}$
can be ignored compared to the former decay rate.

\begin{figure}
\begin{centering}
\includegraphics[scale=0.23]{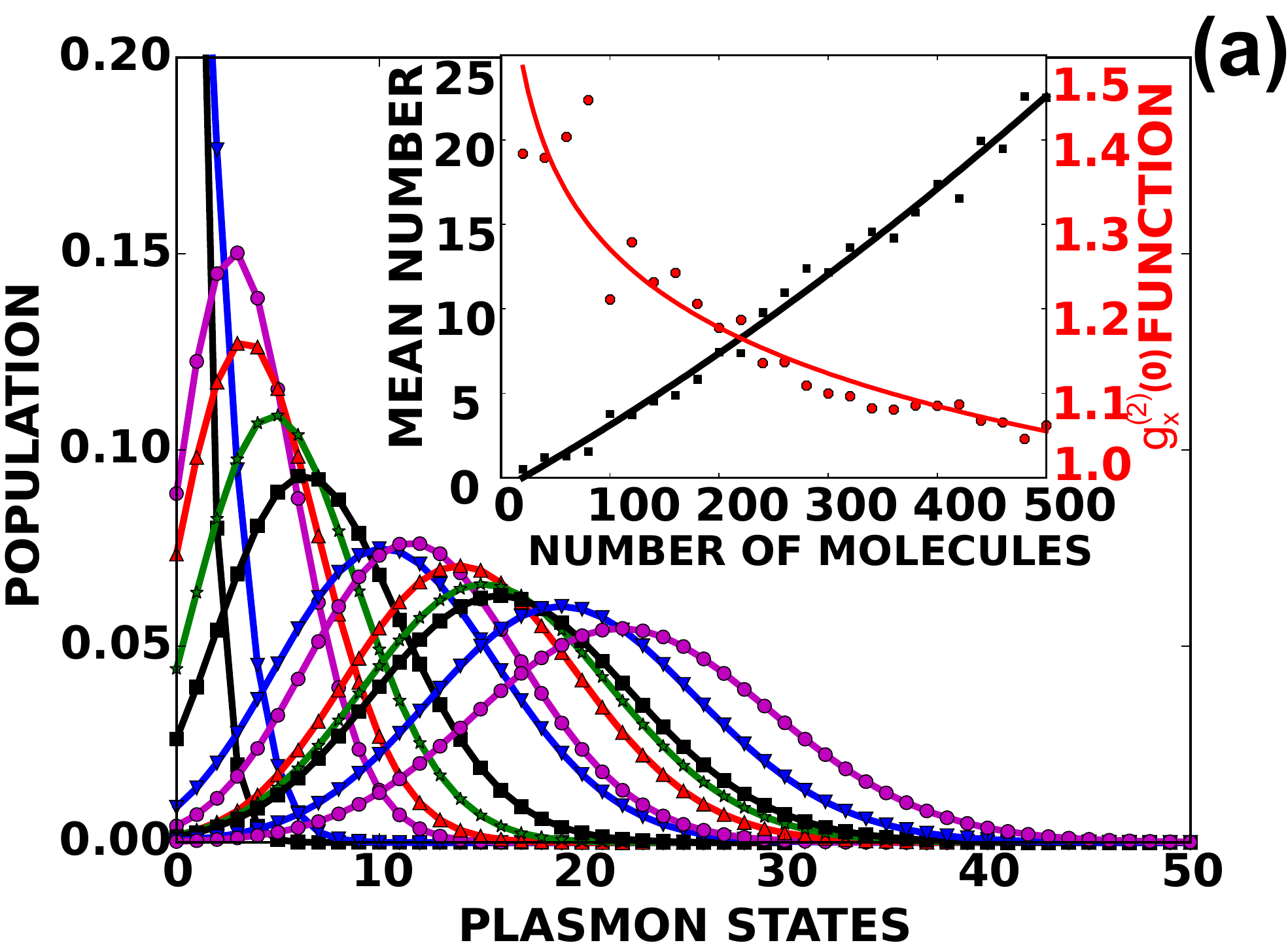}\includegraphics[scale=0.23]{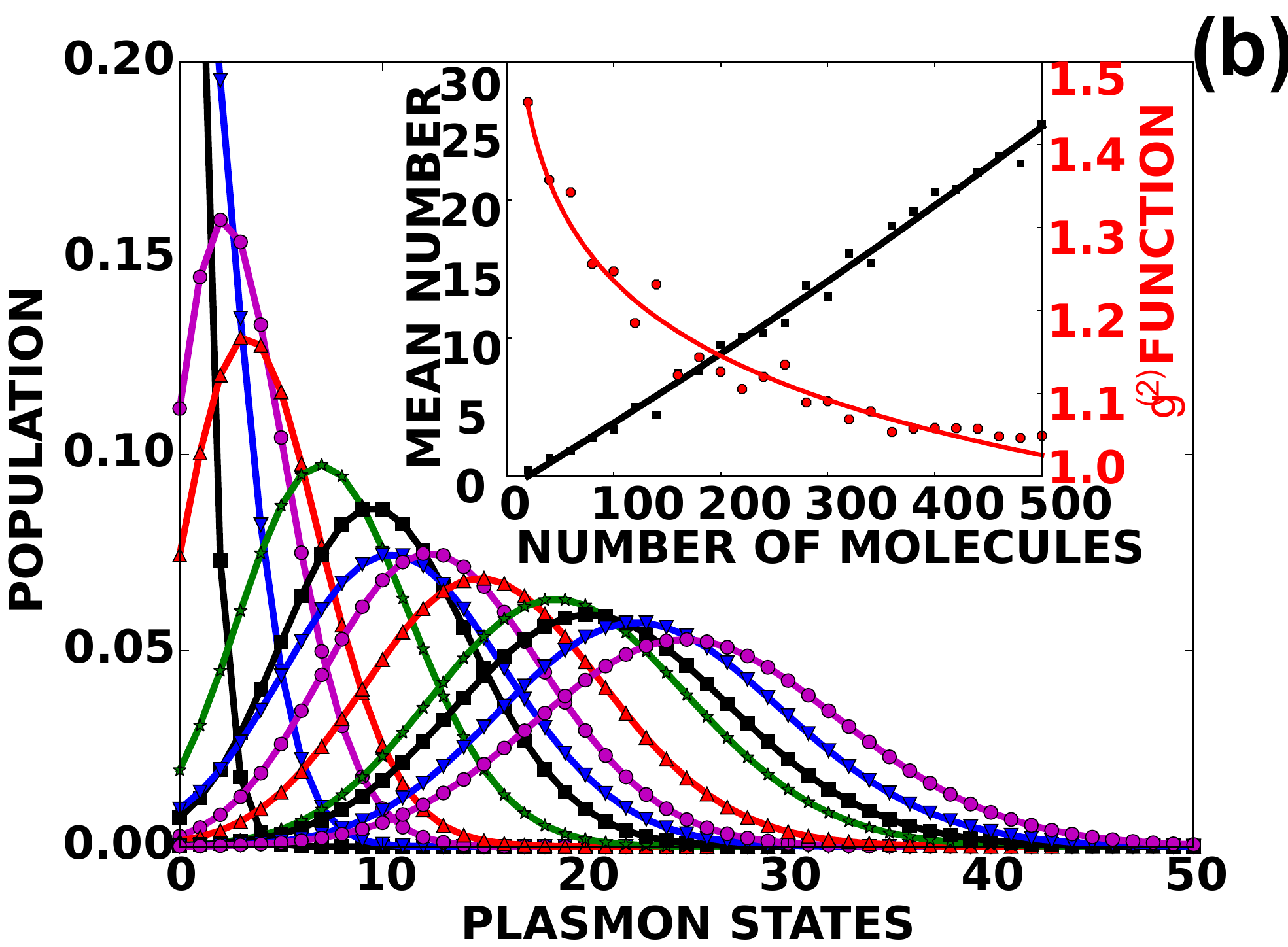}
\par\end{centering}
\begin{centering}
\includegraphics[scale=0.23]{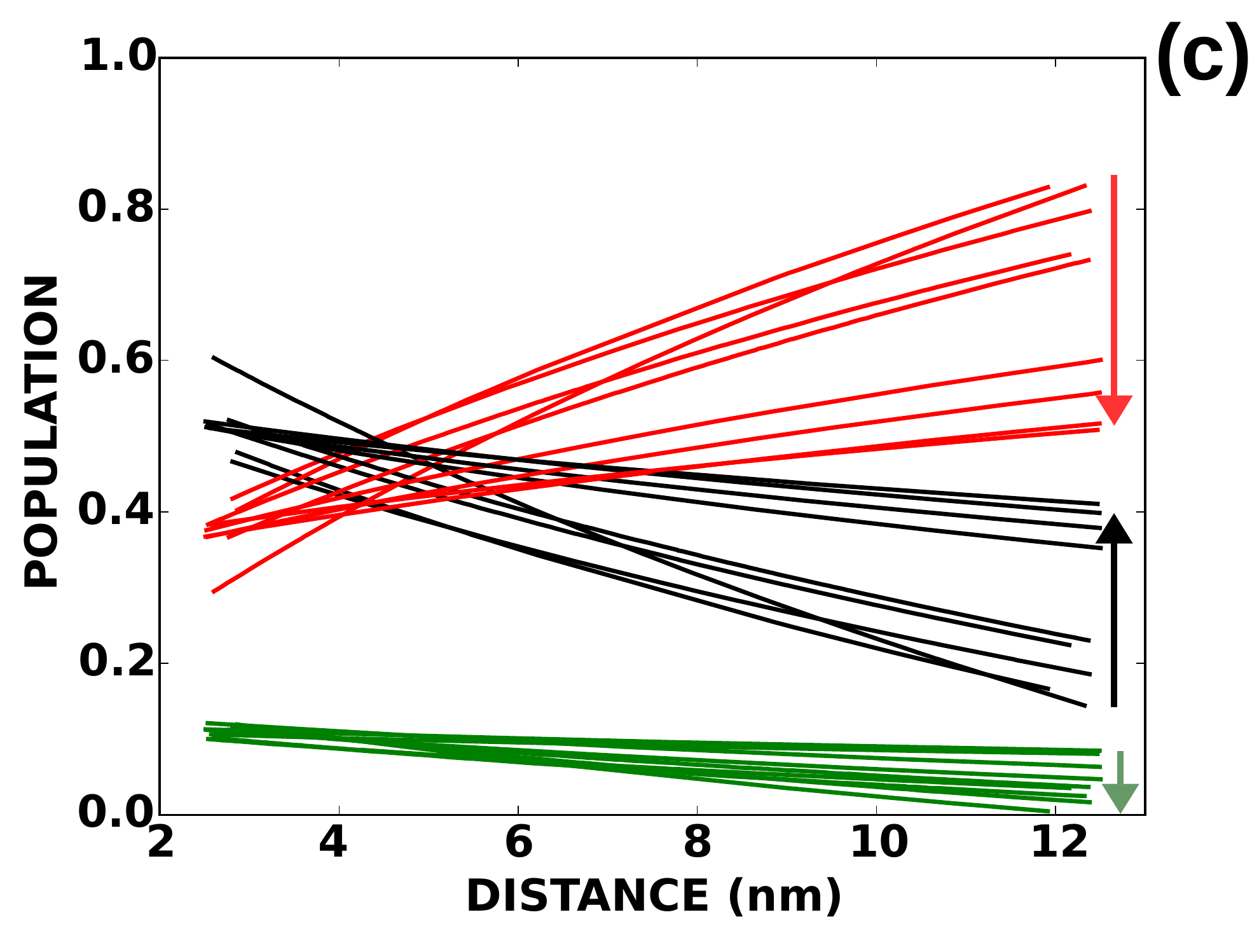}\includegraphics[scale=0.22]{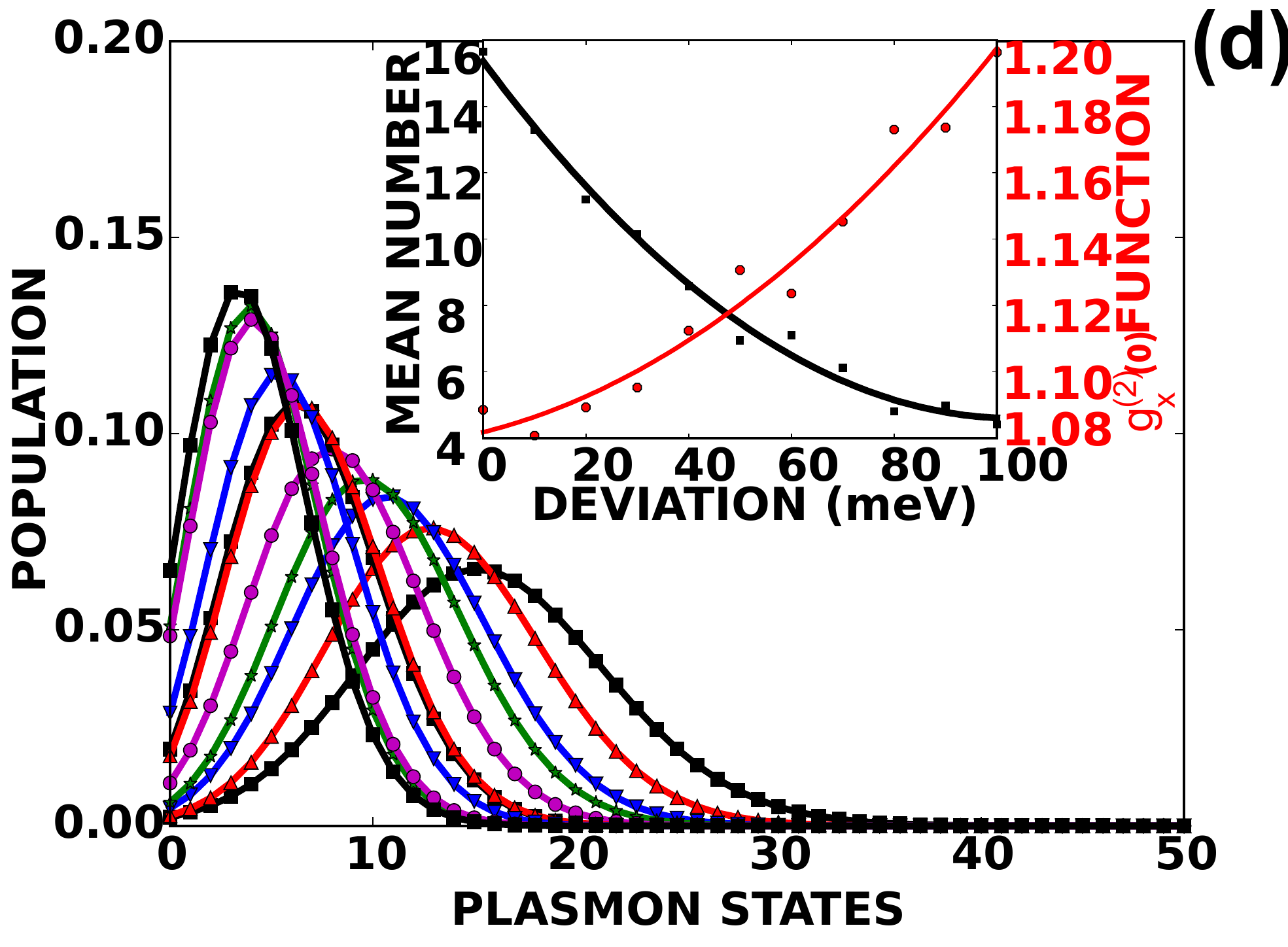}
\par\end{centering}
\begin{centering}
\includegraphics[scale=0.23]{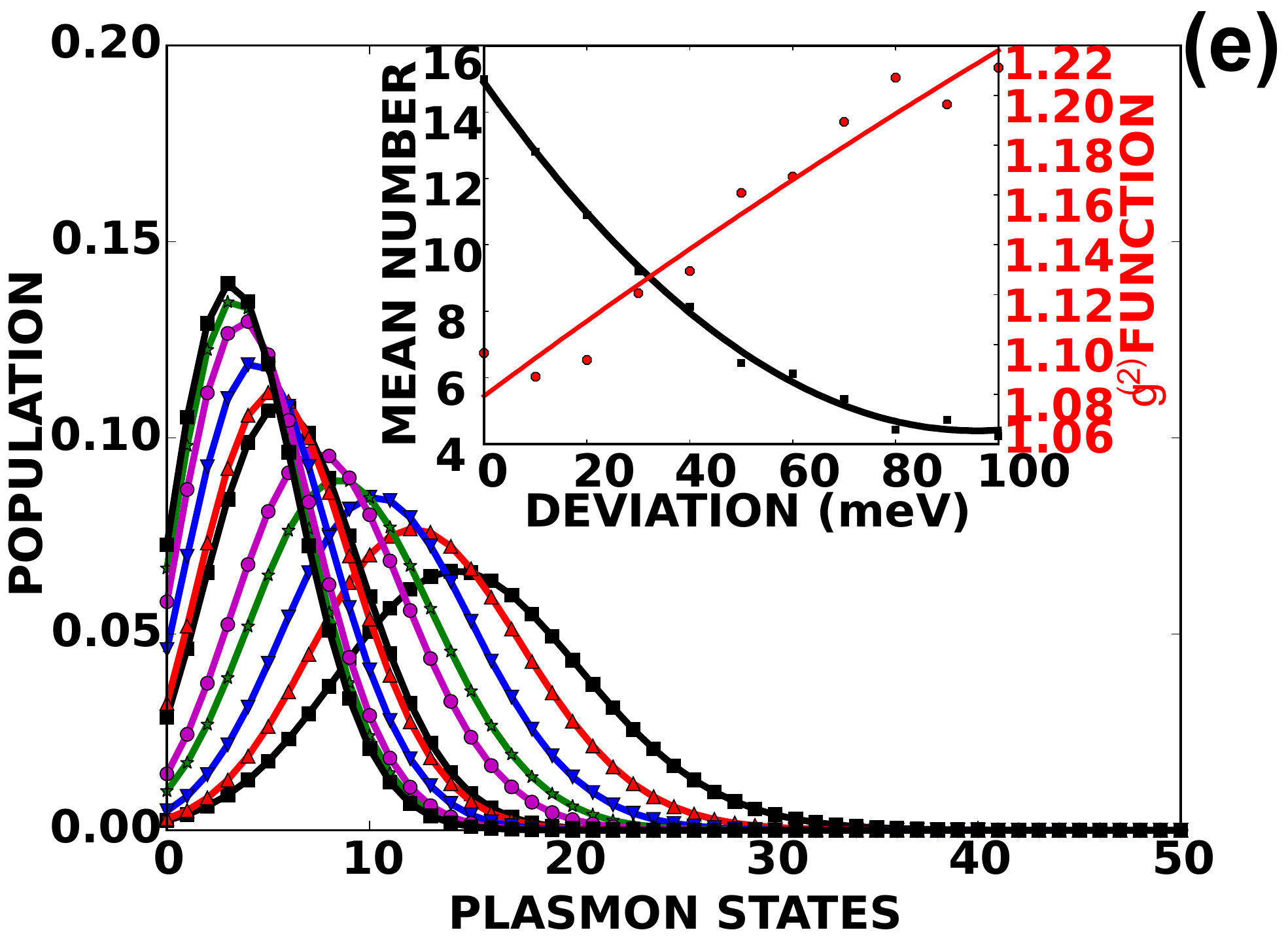}\includegraphics[scale=0.23]{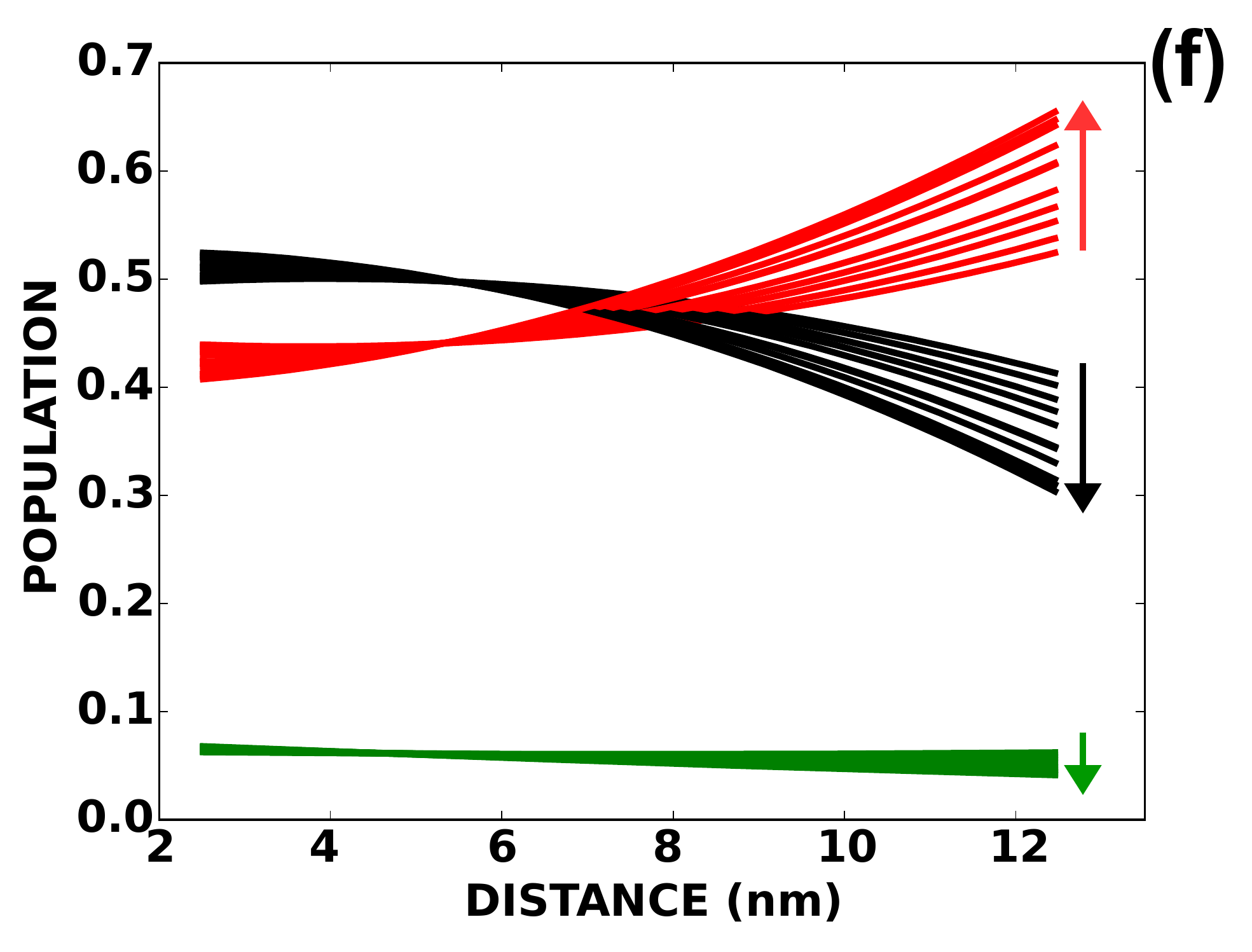}
\par\end{centering}
\caption{\label{fig:other-results-two-modes}Supplemental results to Fig. \ref{fig:double-plasmon-mode}.
Configuration with two plasmon modes. Panel (a) shows the state populations $P_{\mu_{x}}$ of the plasmon x-mode 
with increasing number of molecules $N_{\mathrm{e}}$ from $20$ to $500$ in a step of $40$ (from the left to right curves);
the inset shows the plasmon mean number $N_{x}$ and $g_{x}^{\left(2\right)}\left(0\right)$-function
as functions of $N_{\mathrm{e}}$. Panel (b) shows $P_{\mu_{y}}$, $N_{y}$ and
$g_{y}^{\left(2\right)}\left(0\right)$ as functions of $N_{\mathrm{e}}$. Panel (c) shows fitted populations $P_{a}^{\left(n\right)}$ 
of the molecular states as functions of the molecular distances $d^{\left(n\right)}$ to the
sphere-surface (increasing values of $N_{\mathrm{e}}$  are indicated by the right arrows); 
$P_{g}^{\left(n\right)}$ (the black curves), $P_{e}^{\left(n\right)}$ (the red curves)
and $P_{f}^{\left(n\right)}$ (the blue curves). Panel (d) shows $P_{\mu_{x}}$
,$N_{x}$ and $g_{x}^{\left(2\right)}\left(0\right)$ for increasing deviation 
$\sigma$ of the molecular energy-shift, from $0$ meV to $100$ meV in a step of $10$ meV. Panel (e) shows $P_{\mu_{y}}$ ,$N_{y}$
and $g_{y}^{\left(2\right)}\left(0\right)$ versus $\sigma$. Panel
(f) shows  $P_{a}^{\left(n\right)}$ versus $d^{\left(n\right)}$ with the arrows indicating the increased $\sigma$. Physical parameters
are specified in Table \ref{tab:parameters}. }
\end{figure}

\begin{figure}
\begin{centering}
\includegraphics[scale=0.22]{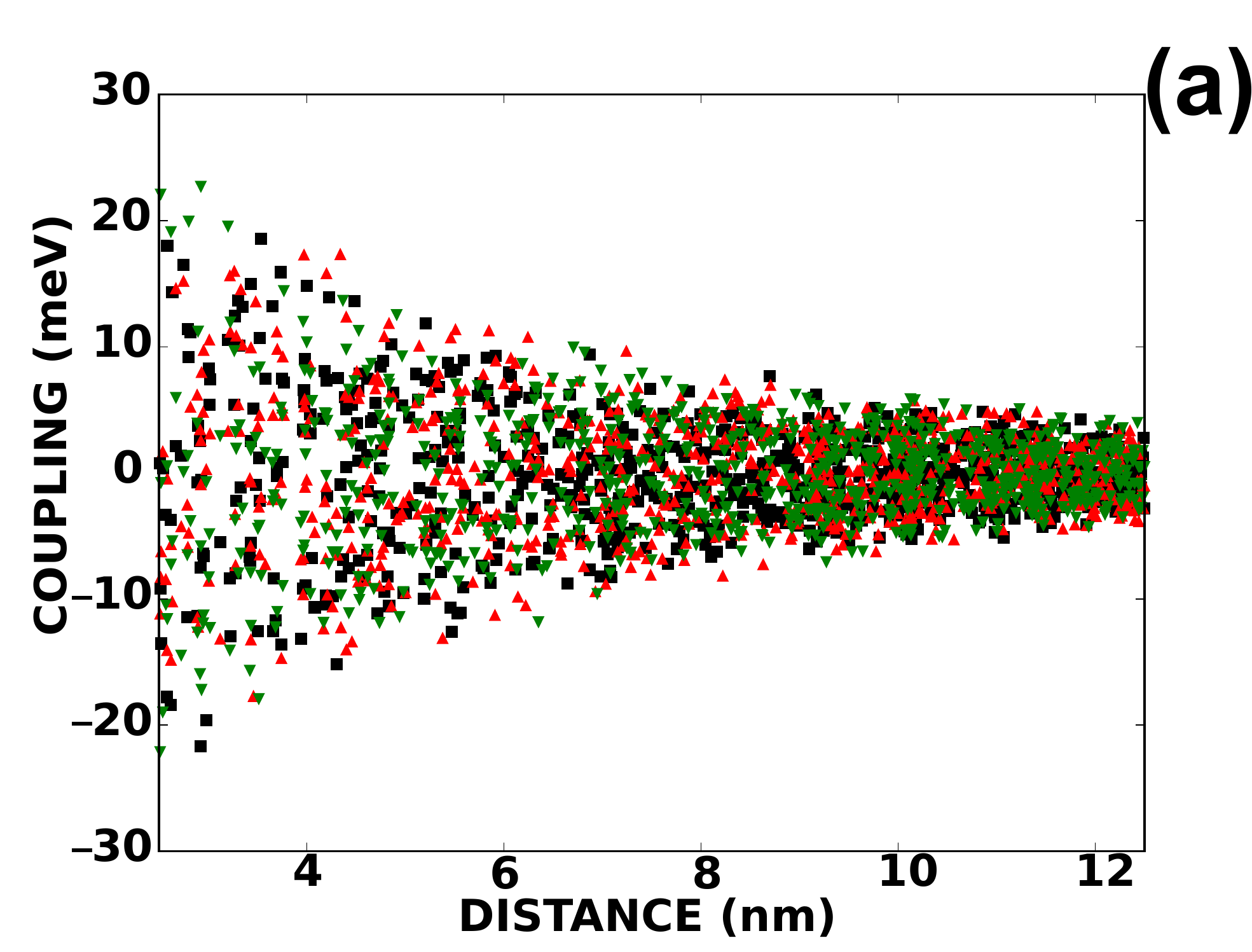}\includegraphics[scale=0.3]{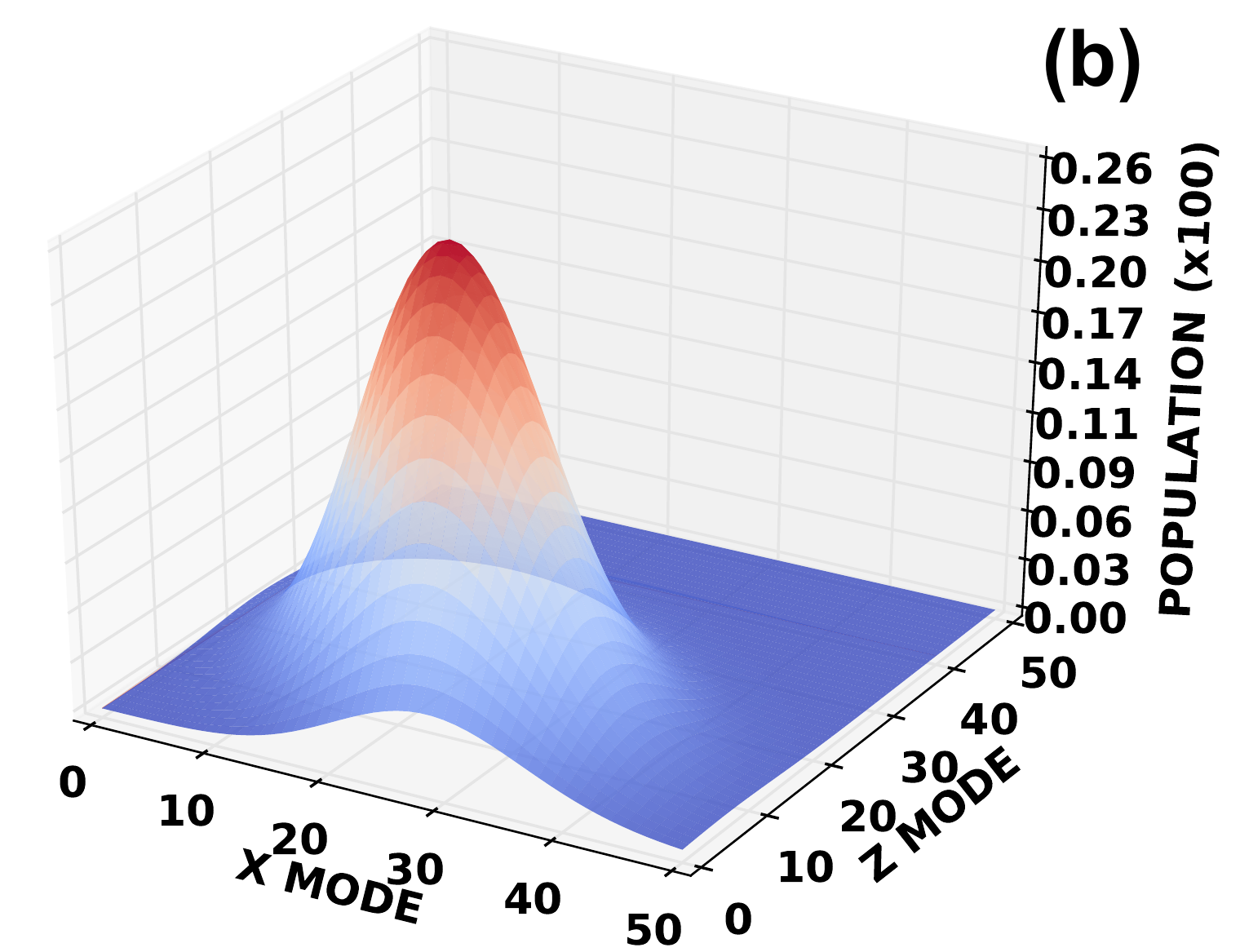}
\par\end{centering}
\begin{centering}
\includegraphics[scale=0.3]{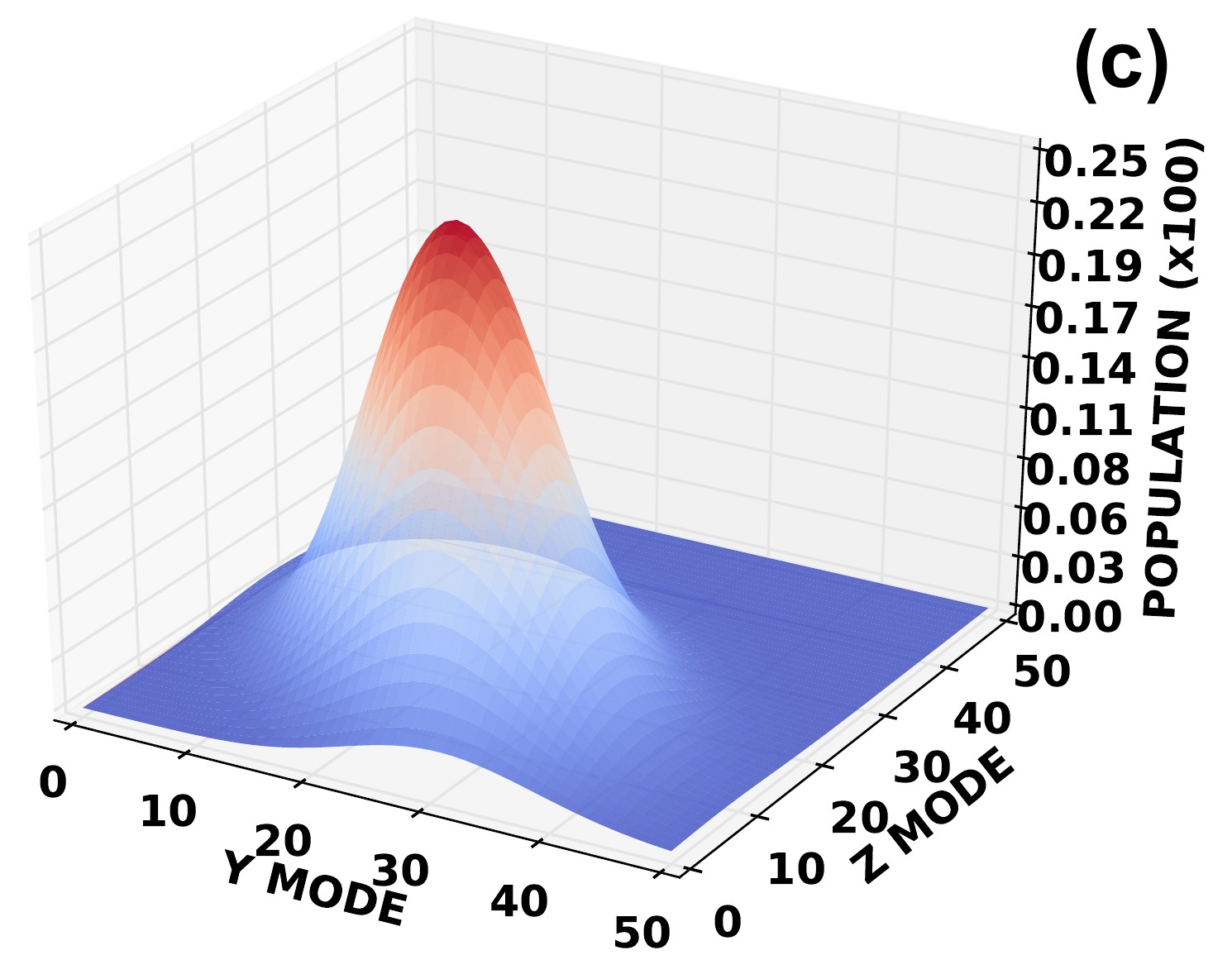}\includegraphics[scale=0.22]{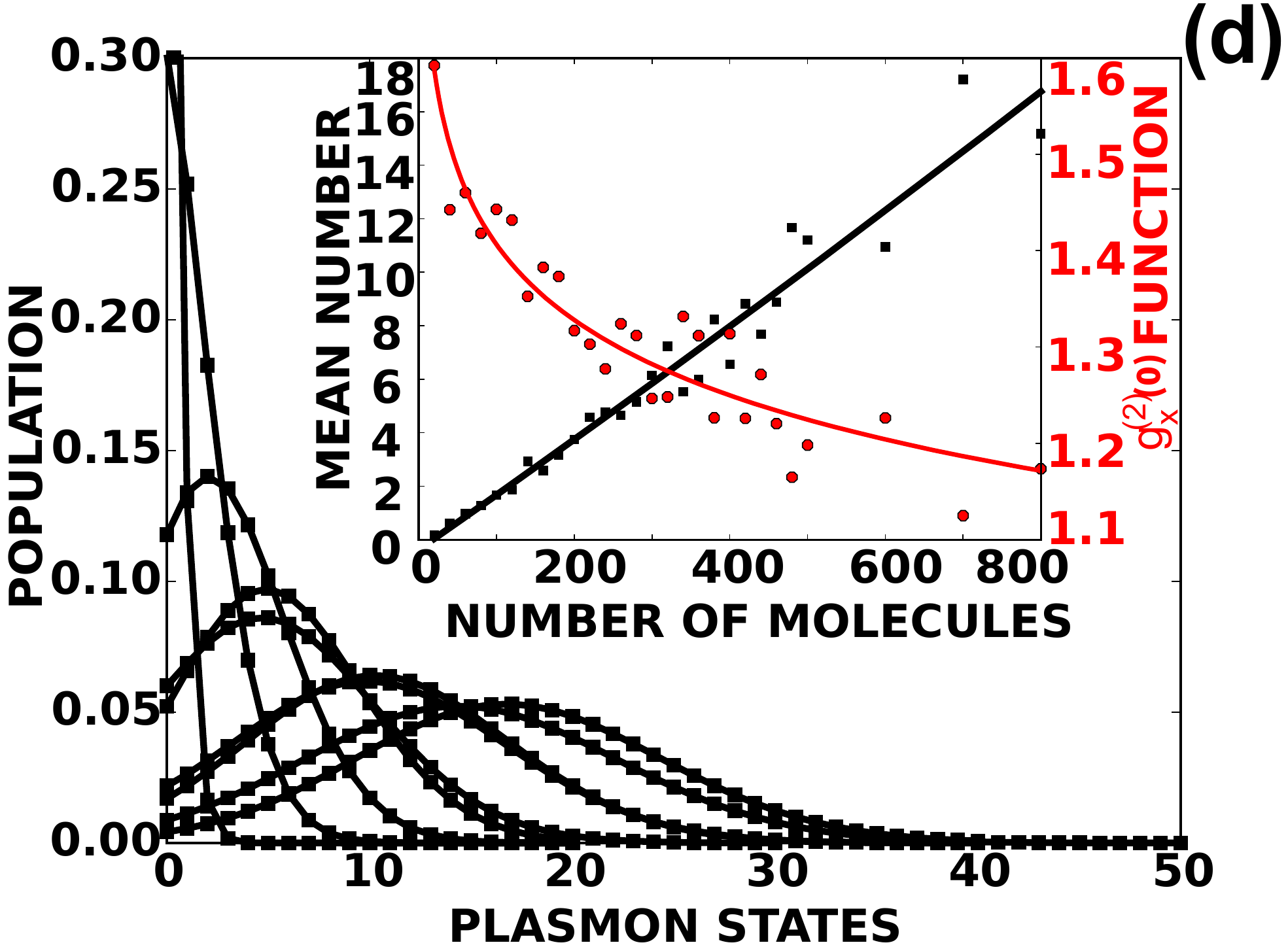}
\par\end{centering}
\begin{centering}
\includegraphics[scale=0.23]{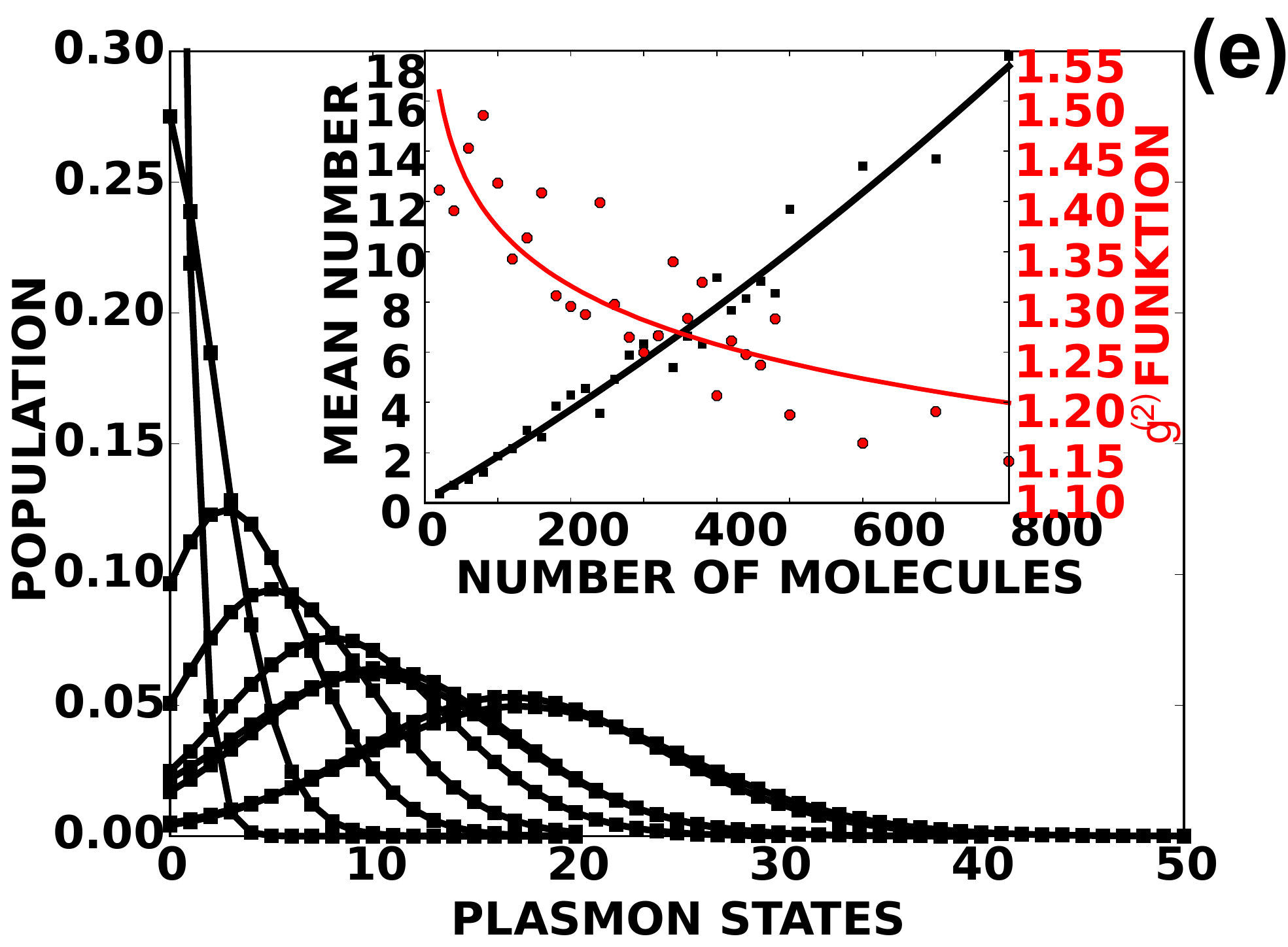}\includegraphics[scale=0.23]{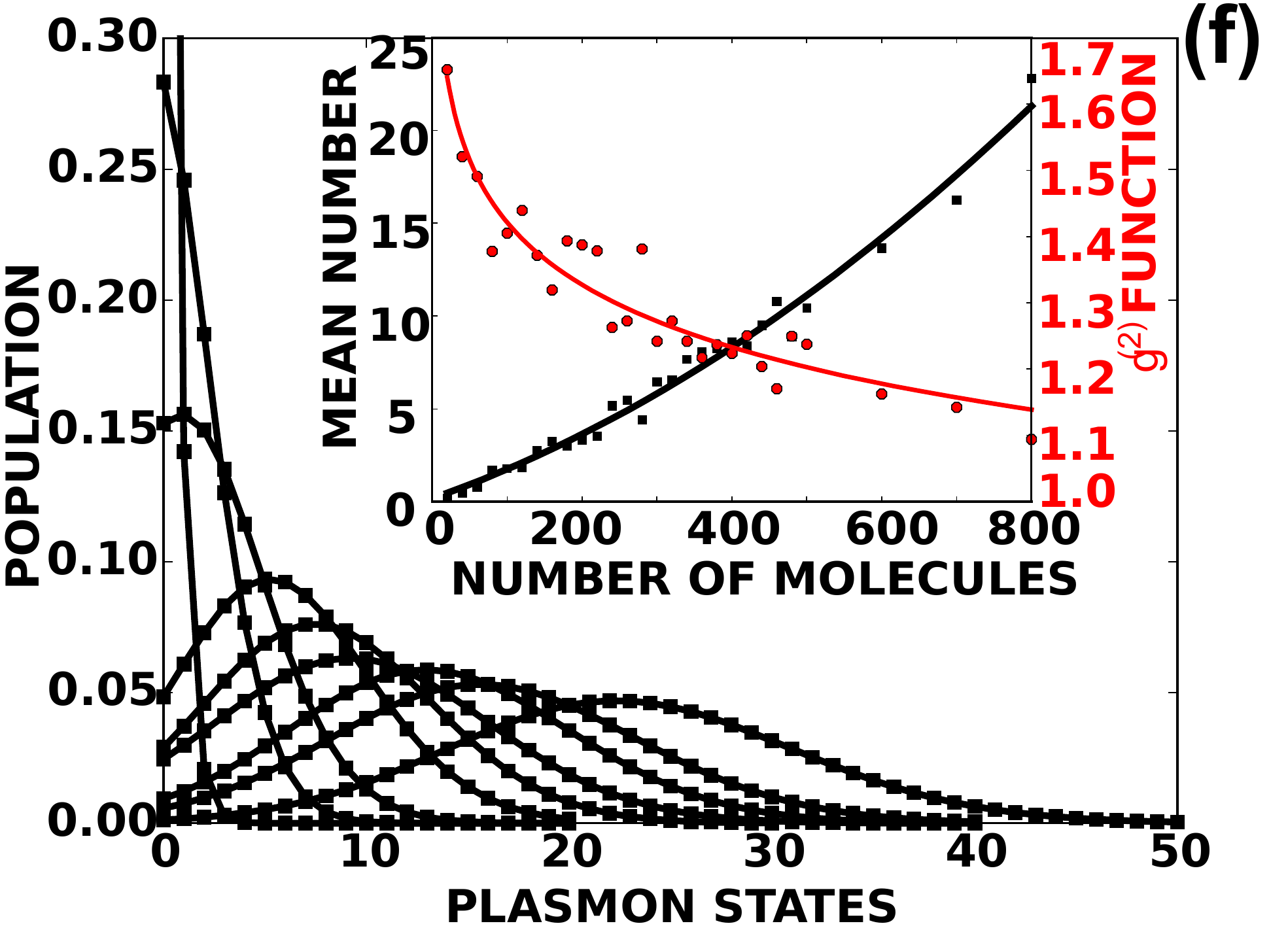}
\par\end{centering}
\caption{\label{fig:other-results-three-modes} Supplemental results to Fig. \ref{fig:three-modes}.
Configuration with three plasmon modes. In panel (a), blue squares,
red upper-triangles and orange down-triangles show the coupling with the dipole plasmon x,y,z-mode for
molecules with different distances to the sphere-surface.
Panel (b,c) show the joint population $P_{\mu_{x}\mu_{z}}$ and $P_{\mu_{y}\mu_{z}}$
for a system with $800$ molecules. Panel (d) shows the reduced plasmon state
population $P_{\mu_{x}}$ for increasing number of molecules $N_{\mathrm{e}}$;
the inset shows the plasmon mean number $N_{x}$ and $g_{x}^{\left(2\right)}(0)$-function
versus $N_{\mathrm{e}}$. Panel (e) is similar to panel (d) but shows
$P_{\mu_{y}}$, $N_{y}$ and $g_{y}^{\left(2\right)}(0)$ of the $y$-mode.
Panel (f) is similar to panel (d) but for $P_{\mu_{z}}$, $N_{z}$ and
$g_{z}^{\left(2\right)}(0)$ of the $z$-mode. Physical parameters are specified in Table \ref{tab:parameters}.}
\end{figure}

In Fig. \ref{fig:other-results-two-modes} and Fig. \ref{fig:other-results-three-modes}, we supplement
the results presented in the main text with numerical results
for system configurations with two and three dipole plasmons.

\begin{table}
\caption{\label{tab:parameters}Physical  parameters (for explanation see text)}

\centering{}%
\begin{tabular}{cc|cc}
\hline
$\hbar\omega_{\mathrm{pl}}$ & $2.6$ eV & $\hbar\omega_{eg}$ & $2.6$ eV\tabularnewline
$\hbar\gamma_{\mathrm{pl}}$ & $100$ meV & $\hbar\omega_{fg}$ & $2.7$ eV\tabularnewline
$d_{\mathrm{pl}}$ & $2925$ D & $d_{gf}$ & $16$ D\tabularnewline
$E_{0}$ & $1.2\times10^{8}$ V/m  & $d_{ge}$ & $14.4$ D\tabularnewline
$\hbar\omega_{0}$ & $2.7$ eV & $\hbar k_{f\to e}$ & $100$ meV \tabularnewline
 &  & others & $0$ meV\tabularnewline
\hline
\end{tabular}
\end{table}

\section{Derivation of Plasmon Reduced Density Matrix Equation\label{sec:derviation-plasmon-rdm}}

In this section, we derive the master equation for the plasmon reduced density matrix
(RDM) $\rho_{\mu\nu}$. From the definition of $\rho_{\mu\nu}$ and
Eq. \eqref{eq:RDO}, we get the following equation:
\begin{align}
 & \frac{\partial}{\partial t}\rho_{\mu\nu}=-\sum_{j=1}^{3}\left(i\omega_{j}\left(\mu_{j}-\nu_{j}\right)+\gamma_{j}\left[\left(\mu_{j}+\nu_{j}\right)/2\right]\right)\rho_{\mu\nu}\nonumber \\
 & +\sum_{j=1}^{3}\gamma_{j}\sqrt{\left(\mu_{j}+1\right)\left(\nu_{j}+1\right)}\rho_{\mu_{j}+1\nu_{j}+1}\nonumber \\
 & -i\sum_{n=1}^{N_{\mathrm{e}}}\sum_{j=1}^{3}v_{ge}^{\left(jn\right)} \Big[ \sqrt{\mu_{j}}\rho_{e\mu_{j}-1,g\nu}^{\left(n\right)}-\sqrt{\nu_{j}}\rho_{g\mu,e\nu_{j}-1}^{\left(n\right)}\nonumber \\
 & -\left(\sqrt{\nu_{j}+1}\rho_{e\mu,g\nu_{j}+1}^{\left(n\right)}-\sqrt{\mu_{j}+1}\rho_{g\mu_{j}+1,e\nu}^{\left(n\right)}\right) \Big].\label{eq:plasmon-rdm}
\end{align}
To simiplify notation, we consider $\rho_{\mu\nu}$ as a reference matrix element and
denote the dependent matrix elements. For example, $\rho_{\mu_{j}+1\nu_{j}+1}$ differ from the reference
element by increasing only the quantum number $\mu_{j}$ and $\nu_{j}$
by one. We see that $\rho_{\mu\nu}$ depends on the molecule-plasmon
correlations: $\rho_{a\mu,b\nu}^{\left(n\right)}\equiv\text{tr}_{\text{S}}\left\{ \hat{\rho}\left(t\right)\left|b_{n}\right\rangle \left\langle a_{n}\right|\times\left|\nu\right\rangle \left\langle \mu\right|\right\} $, more precisely, $\rho_{e\mu_{j}-1,g\nu}^{\left(n\right)}$, where the labels differ 
from the ones of $\rho_{e\mu,g\nu}^{\left(n\right)}$ only by subtracting $\mu_{j}$
by unity.

In the procedure to achieve an equation only for $\rho_{\mu\nu}$,  the most crucial step
is to analyze the equations for $\rho_{a\mu,b\nu}^{\left(n\right)}$ and express them as functions
of $\rho_{\mu\nu}$. In the following, we present the equations for the population-like (coherence-like) correlations $\rho_{a\mu,a\nu}^{\left(n\right)}$ ($\rho_{a\mu,b\nu}^{\left(n\right)}$ with $a\neq b$). 
It turns out that these equations depend on the terms like
$\gamma_{j}\sqrt{\left(\mu_{j}+1\right)\left(v_{j}+1\right)}\rho_{a\mu_{j}+1,b\nu_{j}+1}^{\left(n\right)}$
due to the damping from higher plasmon states. To avoid this dependence,
we carry out the following replacement in all the equations for $\rho_{a\mu,b\nu}^{\left(n\right)}$:
\begin{alignat}{1}
 & -\sum_{j}\Big( \left[i\omega_{\mu\nu}-\gamma_{j}\left(\mu_{j}+\nu_{j}\right)/2\right]\rho_{a\mu,b\nu}^{\left(n\right)}\nonumber \\
 & +\gamma_{j}\sqrt{\left(\mu_{j}+1\right)\left(\nu_{j}+1\right)}\rho_{a\mu_{j}+1,b\nu_{j}+1}^{\left(n\right)}\Big)\to-i\tilde{\omega}_{\mu\nu}\rho_{a\mu,b\nu}^{\left(n\right)}.\label{eq:replacement}
\end{alignat}
Here, we have introduced the complex transition frequency:
\begin{equation}
\tilde{\omega}_{\mu\nu}=\sum_{j} \Big(\omega_{j}\left(\mu_{j}-\nu_{j}\right)-i\gamma_{j}\left[\left(\mu_{j}+\nu_{j}\right)/2-\sqrt{\mu_{j}\nu_{j}}\right] \Big).
\end{equation}

Now, we start with the equations for the population-like correlation $\rho_{a\mu,a\nu}^{\left(n\right)}$ (density matrix elements with same molecular states):
\begin{align}
 & \frac{\partial}{\partial t}\rho_{g\mu,g\nu}^{\left(n\right)}=-i\tilde{\omega}_{\mu\nu}\rho_{g\mu,g\nu}^{\left(n\right)}-\left(k_{g\to f}^{\left(n\right)}+k_{g\to e}^{\left(n\right)}\right)\rho_{g\mu,g\nu}^{\left(n\right)}\nonumber \\
 & +k_{f\to g}^{\left(n\right)}\rho_{f\mu,f\nu}^{\left(n\right)}+k_{e\to g}^{\left(n\right)}\rho_{e\mu,e\nu}^{\left(n\right)}-iv_{gf}^{\left(n\right)}\left(\tilde{\rho}_{f\mu,g\nu}^{\left(n\right)}-\tilde{\rho}_{g\mu,f\nu}^{\left(n\right)}\right)\nonumber \\
 & +i\sum_{j=1}^{3}v_{ge}^{\left(jn\right)}\left(\sqrt{\nu_{j}}\rho_{g\mu,e\nu_{j}-1}^{\left(n\right)}-\sqrt{\mu_{j}}\rho_{e\mu_{j}-1,g\nu}^{\left(n\right)}\right),\label{eq:rhogg}
\end{align}
\begin{align}
 & \frac{\partial}{\partial t}\rho_{f\mu,f\nu}^{\left(n\right)}=-i\tilde{\omega}_{\mu\nu}\rho_{f\mu,f\nu}^{\left(n\right)}-\left(k_{f\to g}^{\left(n\right)}+k_{f\to e}^{\left(n\right)}\right)\rho_{f\mu,f\nu}^{\left(n\right)}\nonumber \\
 & +k_{g\to f}^{\left(n\right)}\rho_{g\mu,g\nu}^{\left(n\right)}+k_{e\to f}^{\left(n\right)}\rho_{e\mu,e\nu}^{\left(n\right)}+iv_{gf}^{\left(n\right)}\left(\tilde{\rho}_{f\mu,g\nu}^{\left(n\right)}-\tilde{\rho}_{g\mu,f\nu}^{\left(n\right)}\right),\label{eq:rhoff}
\end{align}
\begin{align}
 & \frac{\partial}{\partial t}\rho_{e\mu_{j}-1,e\nu_{j}-1}^{\left(n\right)}=-i\tilde{\omega}_{\mu_{j}-1\nu_{j}-1}\rho_{e\mu_{j}-1,e\nu_{j}-1}^{\left(n\right)}+k_{g\to e}^{\left(n\right)}\rho_{g\mu_{j}-1,g\nu_{j}-1}^{\left(n\right)}\nonumber \\
 & -\left(k_{e\to g}^{\left(n\right)}+k_{e\to f}^{\left(n\right)}\right)\rho_{e\mu_{j}-1,e\nu_{j}-1}^{\left(n\right)}+k_{f\to e}^{\left(n\right)}\rho_{f\mu_{j}-1,f\nu_{j}-1}^{\left(n\right)}\nonumber \\
 & +iv_{ge}^{\left(jn\right)}\left(\sqrt{\nu_{j}}\rho_{e\mu_{j}-1,g\nu}^{\left(n\right)}-\sqrt{\mu_{j}}\rho_{g\mu,e\nu_{j}-1}^{\left(n\right)}\right).\label{eq:rhoee}
\end{align}
Then, we present the equations for the coherence-like correlations $\rho_{a\mu,b\nu}^{\left(n\right)}$ (density matrix elements with different molecular states)  appearing in Eqs. \eqref{eq:rhogg}, \eqref{eq:rhoff} and \eqref{eq:rhoee}:
\begin{align}
 & \frac{\partial}{\partial t}\tilde{\rho}_{g\mu,f\nu}^{\left(n\right)}=-i\left(\tilde{\omega}_{gf}^{\left(n\right)}+\tilde{\omega}_{\mu\nu}\right)\tilde{\rho}_{g\mu,f\nu}^{\left(n\right)}\nonumber \\
 & -i\sum_{j=1}^{3}v_{ge}^{\left(jn\right)}\sqrt{\mu_{j}}\tilde{\rho}_{e\mu_{j}-1,f\nu}^{\left(n\right)}-iv_{gf}^{\left(n\right)}\left(\rho_{f\mu,f\nu}^{\left(n\right)}-\rho_{g\mu,g\nu}^{\left(n\right)}\right),\label{eq:rhogf-II}
\end{align}
\begin{align}
 & \frac{\partial}{\partial t}\tilde{\rho}_{f\mu,g\nu}^{\left(n\right)}=i\left(\tilde{\omega}_{gf}^{\left(n\right)*}-\tilde{\omega}_{\mu\nu}\right)\tilde{\rho}_{f\mu,g\nu}^{\left(n\right)}\nonumber \\
 & +i\sum_{j=1}^{3}v_{ge}^{\left(jn\right)}\sqrt{v_{j}}\tilde{\rho}_{f\mu,e\nu_{j}-1}^{\left(n\right)}+iv_{gf}^{\left(n\right)}\left(\rho_{f\mu,f\nu}^{\left(n\right)}-\rho_{g\mu,g\nu}^{\left(n\right)}\right),\label{eq:rhofg-II}
\end{align}
\begin{align}
 & \frac{\partial}{\partial t}\rho_{g\mu,e\nu_{j}-1}^{\left(n\right)}=i\left(\tilde{\omega}_{eg}^{\left(n\right)*}-\tilde{\omega}_{\mu\nu_{j}-1}\right)\rho_{g\mu,e\nu_{j}-1}^{\left(n\right)} -iv_{gf}^{\left(n\right)}\rho_{f\mu,e\nu_{j}-1}^{\left(n\right)} \nonumber \\
 & +iv_{ge}^{\left(jn\right)}\left(\sqrt{\nu_{j}}\rho_{g\mu,g\nu}-\sqrt{\mu_{j}}\rho_{e\mu_{j}-1,e\nu_{j}-1}^{\left(n\right)}\right), \label{eq:rhoge-2}
\end{align}
\begin{align}
 & \frac{\partial}{\partial t}\rho_{e\mu_{j}-1,g\nu}^{\left(n\right)}=-i\left(\tilde{\omega}_{eg}^{\left(n\right)}+\tilde{\omega}_{\mu_{j}-1\nu}\right)\rho_{e\mu_{j}-1,g\nu}^{\left(n\right)} +iv_{gf}^{\left(n\right)}\rho_{e\mu_{j}-1,f\nu}^{\left(n\right)}\nonumber \\
 & +iv_{ge}^{\left(jn\right)}\left(\sqrt{\nu_{j}}\rho_{e\mu_{j}-1,e\nu_{j}-1}^{\left(n\right)}-\sqrt{\mu_{j}}\rho_{g\mu,g\nu}\right),\label{eq:rhoeg-2}
\end{align}
\begin{align}
 & \frac{\partial}{\partial t}\tilde{\rho}_{e\mu_{j}-1,f\nu}^{\left(n\right)}=-i\left(\tilde{\omega}_{ef}^{\left(n\right)}+\tilde{\omega}_{\mu_{j}-1\nu}\right)\tilde{\rho}_{e\mu_{j}-1,f\nu}^{\left(n\right)}\nonumber \\
 & +iv_{gf}^{\left(n\right)}\rho_{e\mu_{j}-1,g\nu}^{\left(n\right)}-iv_{ge}^{\left(jn\right)}\sqrt{\mu_{j}}\rho_{g\mu,f\nu}^{\left(n\right)},\label{eq:rhoef-2}
\end{align}
\begin{align}
 & \frac{\partial}{\partial t}\rho_{f\mu,e\nu_{j}-1}^{\left(n\right)}=i\left(\tilde{\omega}_{ef}^{\left(n\right)*}-\tilde{\omega}_{\mu\nu_{j}-1}\right)\rho_{f\mu,e\nu_{j}-1}^{\left(n\right)}\nonumber \\
 & -iv_{gf}^{\left(n\right)}\tilde{\rho}_{g\mu,e\nu_{j}-1}^{\left(n\right)}+iv_{ge}^{\left(jn\right)}\sqrt{\nu_{j}}\tilde{\rho}_{f\mu,g\nu}^{\left(n\right)}.\label{eq:rhofe-2}
\end{align}
In the above equations, we have introduced the complex transition
frequencies: $\tilde{\omega}_{eg}^{\left(n\right)}=\omega_{eg}^{\left(n\right)}-i\gamma_{eg}^{\left(n\right)}$
with the dephasing rate $\gamma_{eg}^{\left(n\right)} = \left(k_{e\to g}^{\left(n\right)}+k_{e\to f}^{\left(n\right)}+k_{g\to e}^{\left(n\right)}+k_{g\to f}^{\left(n\right)}\right)/2$
and $\tilde{\omega}_{ef}^{\left(n\right)}=\omega_{ef}^{\left(n\right)}+\omega_{0}-i\gamma_{ef}^{\left(n\right)}$
with $\gamma_{ef}^{\left(n\right)}=\left(k_{e\to g}^{\left(n\right)}+k_{e\to f}^{\left(n\right)}+k_{f\to g}^{\left(n\right)}+k_{f\to e}^{\left(n\right)}\right)/2$
as well as $\tilde{\omega}_{gf}^{\left(n\right)}=\omega_{gf}^{\left(n\right)}+\omega_{0}-i\gamma_{gf}^{\left(n\right)}$
with $\gamma_{gf}^{\left(n\right)}\equiv\left(k_{f\to e}^{\left(n\right)}+k_{f\to g}^{\left(n\right)}+k_{g\to e}^{\left(n\right)}+k_{g\to f}^{\left(n\right)}\right)/2$.
In addition, we would like to point out that the pure dephasing rate of the emitters can be readily
included into these dephasing rates. Because of the coupling with
the driving field, we have introduced the following
slowly varying correlations $\tilde{\rho}_{e\mu_{j}-1,g\nu}^{\left(n\right)}\equiv e^{-i\omega_{0}t}\rho_{e\mu_{j}-1,g\nu}^{\left(n\right)}$, $\tilde{\rho}_{g\mu,e\nu_{j}-1}^{\left(n\right)}\equiv e^{i\omega_{0}t}\rho_{g\mu,e\nu_{j}-1}^{\left(n\right)}$, $\tilde{\rho}_{f\mu,g\nu}^{\left(n\right)}\equiv e^{i\omega_{0}t}\rho_{f\mu,g\nu}^{\left(n\right)}$ and
$\tilde{\rho}_{g\mu,f\nu}^{\left(n\right)}\equiv e^{-i\omega_{0}t}\rho_{g\mu,f\nu}^{\left(n\right)}$.

To proceed, we consider the steady-state equations for the coherence-like correlations (density matrix elements with different molecular states). From Eqs. \eqref{eq:rhoge-2}
, \eqref{eq:rhoeg-2},  \eqref{eq:rhoef-2} and \eqref{eq:rhofe-2} we have
\begin{align}
 & \left(\tilde{\omega}_{eg}^{\left(n\right)*}-\tilde{\omega}_{\mu\nu_{j}-1}\right)\rho_{g\mu,e\nu_{j}-1}^{\left(n\right)}=v_{gf}^{\left(n\right)}\rho_{f\mu,e\nu_{j}-1}^{\left(n\right)}\nonumber \\
 & -v_{ge}^{\left(jn\right)}\left(\sqrt{\nu_{j}}\rho_{g\mu,g\nu}-\sqrt{\mu_{j}}\rho_{e\mu_{j}-1,e\nu_{j}-1}^{\left(n\right)}\right),\label{eq:rhoge-s}
\end{align}
\begin{align}
 & \left(\tilde{\omega}_{eg}^{\left(n\right)}+\tilde{\omega}_{\mu_{j}-1\nu}\right)\rho_{e\mu_{j}-1,g\nu}^{\left(n\right)}=v_{gf}^{\left(n\right)}\rho_{e\mu_{j}-1,f\nu}^{\left(n\right)}\nonumber \\
 & -v_{ge}^{\left(jn\right)}\left(\sqrt{\mu_{j}}\rho_{g\mu,g\nu}-\sqrt{\nu_{j}}\rho_{e\mu_{j}-1,e\nu_{j}-1}^{\left(n\right)}\right),\label{eq:rhoeg-s}
\end{align}
\begin{align}
 & \left(\tilde{\omega}_{ef}^{\left(n\right)}+\tilde{\omega}_{\mu_{j}-1\nu}\right)\tilde{\rho}_{e\mu_{j}-1,f\nu}^{\left(n\right)}=v_{gf}^{\left(n\right)}\rho_{e\mu_{j}-1,g\nu}^{\left(n\right)}-v_{ge}^{\left(jn\right)}\sqrt{\mu_{j}}\rho_{g\mu,f\nu}^{\left(n\right)},\label{eq:rhoef-s}
\end{align}
\begin{align}
 & \left(\tilde{\omega}_{ef}^{\left(n\right)*}-\tilde{\omega}_{\mu\nu_{j}-1}\right)\rho_{f\mu,e\nu_{j}-1}^{\left(n\right)}=v_{gf}^{\left(n\right)}\tilde{\rho}_{g\mu,e\nu_{j}-1}^{\left(n\right)}-v_{ge}^{\left(jn\right)}\sqrt{\nu_{j}}\tilde{\rho}_{f\mu,g\nu}^{\left(n\right)}.\label{eq:rhofe-s}
\end{align}
In order to reduce the dependence, we insert Eqs. \eqref{eq:rhoge-s}
and \eqref{eq:rhoeg-s} to Eqs. \eqref{eq:rhoef-s} and \eqref{eq:rhofe-s}
to express $\tilde{\rho}_{e\mu_{j}-1,f\nu}^{\left(n\right)}$ and
$\rho_{f\mu,e\nu_{j}-1}^{\left(n\right)}$ as functions of $\rho_{g\mu,f\nu}^{\left(n\right)}$,
$\rho_{e\mu_{j}-1,e\nu_{j}-1}^{\left(n\right)}$ and $\rho_{g\mu,g\nu}$:
\begin{align}
 & \tilde{\rho}_{e\mu_{j}-1,f\nu}^{\left(n\right)}=-\Xi_{\mu\nu}^{\left(jn\right)}v_{ge}^{\left(jn\right)}\sqrt{\mu_{j}}\rho_{g\mu,f\nu}^{\left(n\right)}\nonumber \\
 & +v_{gf}^{\left(n\right)}\Sigma_{\mu\nu}^{\left(jn\right)}\Xi_{\mu\nu}^{\left(jn\right)}\left(\sqrt{\nu_{j}}\rho_{e\mu_{j}-1,e\nu_{j}-1}^{\left(n\right)}-\sqrt{\mu_{j}}\rho_{g\mu,g\nu}\right),\label{eq:rhoef-s-1}
\end{align}
\begin{align}
 & \rho_{f\mu,e\nu_{j}-1}^{\left(n\right)}=-\tilde{\Xi}_{\mu\nu}^{\left(jn\right)}v_{ge}^{\left(jn\right)}\sqrt{\nu_{j}}\tilde{\rho}_{f\mu,g\nu}^{\left(n\right)}\nonumber \\
 & +v_{gf}^{\left(n\right)}\tilde{\Sigma}_{\mu\nu}^{\left(jn\right)}\tilde{\Xi}_{\mu\nu}^{\left(jn\right)}\left(\sqrt{\mu_{j}}\rho_{e\mu_{j}-1,e\nu_{j}-1}^{\left(n\right)}-\sqrt{\nu_{j}}\rho_{g\mu,g\nu}\right), \label{eq:rhofe-s-1}
\end{align}
with the abbreviations:
\begin{align}
1/\Xi_{\mu\nu}^{\left(jn\right)} & =\tilde{\omega}_{ef}^{\left(n\right)}+\tilde{\omega}_{\mu_{j}-1\nu}- v_{gf}^{\left(n\right)2}/\left(\tilde{\omega}_{eg}^{\left(n\right)}+\tilde{\omega}_{\mu_{j}-1\nu}\right),\\
1/\tilde{\Xi}_{\mu\nu}^{\left(jn\right)} & =\tilde{\omega}_{ef}^{\left(n\right)*}-\tilde{\omega}_{\mu\nu_{j}-1}- v_{gf}^{\left(n\right)2}/\left( \tilde{\omega}_{eg}^{\left(n\right)*}-\tilde{\omega}_{\mu\nu_{j}-1}\right),
\end{align}
\begin{alignat}{1}
 & \Sigma_{\mu\nu}^{\left(jn\right)}=v_{ge}^{\left(jn\right)}/\left(\tilde{\omega}_{eg}^{\left(n\right)}+\tilde{\omega}_{\mu_{j}-1\nu}\right),\\
 & \tilde{\Sigma}_{\mu\nu}^{\left(jn\right)}=v_{ge}^{\left(jn\right)}/\left(\tilde{\omega}_{eg}^{\left(n\right)*}-\tilde{\omega}_{\mu\nu_{j}-1}\right).
\end{alignat}
Then, we consider the steady-state version of Eqs. \eqref{eq:rhogf-II}
and \eqref{eq:rhofg-II}:
\begin{align}
 & \left(\tilde{\omega}_{gf}^{\left(n\right)}+\tilde{\omega}_{\mu\nu}\right)\tilde{\rho}_{g\mu,f\nu}^{\left(n\right)}=v_{gf}^{\left(n\right)}\left(\rho_{g\mu,g\nu}^{\left(n\right)}-\rho_{f\mu,f\nu}^{\left(n\right)}\right)\nonumber \\
 & -\sum_{j=1}^{3}v_{ge}^{\left(jn\right)}\sqrt{\mu_{j}}\tilde{\rho}_{e\mu_{j}-1,f\nu}^{\left(n\right)},\label{eq:rhogf-s}
\end{align}
\begin{align}
 & \left(\tilde{\omega}_{gf}^{\left(n\right)*}-\tilde{\omega}_{\mu\nu}\right)\tilde{\rho}_{f\mu,g\nu}^{\left(n\right)}=v_{gf}^{\left(n\right)}\left(\rho_{g\mu,g\nu}^{\left(n\right)}-\rho_{f\mu,f\nu}^{\left(n\right)}\right)\nonumber \\
 & -\sum_{j=1}^{3}v_{ge}^{\left(jn\right)}\sqrt{v_{j}}\tilde{\rho}_{f\mu,e\nu_{j}-1}^{\left(n\right)}.\label{eq:rhofg-s}
\end{align}
Finally, we insert Eqs.\eqref{eq:rhoef-s-1}
and \eqref{eq:rhofe-s-1} to Eqs.\eqref{eq:rhogf-s} and \eqref{eq:rhofg-s}
to express $\tilde{\rho}_{g\mu,f\nu}^{\left(n\right)}$ and $\tilde{\rho}_{f\mu,g\nu}^{\left(n\right)}$
as functions of the population-like correlations:
\begin{align}
 & \tilde{\rho}_{g\mu,f\nu}^{\left(n\right)}=v_{gf}^{\left(n\right)}\Phi_{\mu\nu}^{\left(n\right)}\left(\rho_{g\mu,g\nu}^{\left(n\right)}-\rho_{f\mu,f\nu}^{\left(n\right)}\right)\nonumber \\
 & -v_{gf}^{\left(n\right)}\Phi_{\mu\nu}^{\left(n\right)}\sum_{j=1}^{3}\Sigma_{\mu\nu}^{\left(jn\right)}\Xi_{\mu\nu}^{\left(jn\right)}v_{ge}^{\left(jn\right)}\left(\sqrt{\mu_{j}\nu_{j}}\rho_{e\mu_{j}-1,e\nu_{j}-1}^{\left(n\right)}-\mu_{j}\rho_{g\mu,g\nu}\right),\label{eq:rhogf-s-1}
\end{align}
\begin{align}
 & \tilde{\rho}_{f\mu,g\nu}^{\left(n\right)}=v_{gf}^{\left(n\right)}\tilde{\Phi}_{\mu\nu}^{\left(n\right)}\left(\rho_{g\mu,g\nu}^{\left(n\right)}-\rho_{f\mu,f\nu}^{\left(n\right)}\right)\nonumber \\
 & -v_{gf}^{\left(n\right)}\tilde{\Phi}_{\mu\nu}^{\left(n\right)}\sum_{j=1}^{3}\tilde{\Sigma}_{\mu\nu}^{\left(jn\right)}\tilde{\Xi}_{\mu\nu}^{\left(jn\right)}v_{ge}^{\left(jn\right)}\left(\sqrt{\mu_{j}v_{j}}\rho_{e\mu_{j}-1,e\nu_{j}-1}^{\left(n\right)}-v_{j}\rho_{g\mu,g\nu}\right), \label{eq:rhofg-s-1}
\end{align}
with the abbreviations
\begin{align}
1/\Phi_{\mu\nu}^{\left(n\right)} & =\tilde{\omega}_{gf}^{\left(n\right)}+\tilde{\omega}_{\mu\nu}-\sum_{j=1}^{3}\Xi_{\mu\nu}^{\left(jn\right)}\mu_{j}v_{ge}^{\left(jn\right)2},\\
1/\tilde{\Phi}_{\mu\nu}^{\left(n\right)} & =\tilde{\omega}_{gf}^{\left(n\right)*}-\tilde{\omega}_{\mu\nu}-\sum_{j=1}^{3}\tilde{\Xi}_{\mu\nu}^{\left(jn\right)}\nu_{j}v_{ge}^{\left(jn\right)2}.
\end{align}
Since $\tilde{\rho}_{e\mu_{j}-1,f\nu}^{\left(n\right)}$ and $\rho_{f\mu,e\nu_{j}-1}^{\left(n\right)}$
depend on $\tilde{\rho}_{g\mu,f\nu}^{\left(n\right)}$ and $\tilde{\rho}_{f\mu,g\nu}^{\left(n\right)}$
through Eqs. \eqref{eq:rhoef-s-1} and \eqref{eq:rhofe-s-1}, we can also
express the former two correlations utilizing Eqs. \eqref{eq:rhogf-s-1} and \eqref{eq:rhofg-s-1} as functions of 
the population-like correlations:
\begin{align}
 & \tilde{\rho}_{e\mu_{j}-1,f\nu}^{\left(n\right)}=v_{gf}^{\left(n\right)}\Sigma_{\mu\nu}^{\left(jn\right)}\Xi_{\mu\nu}^{\left(jn\right)}\sqrt{\nu_{j}}\rho_{e\mu_{j}-1,e\nu_{j}-1}^{\left(n\right)}\nonumber \\
 & +v_{gf}^{\left(n\right)}\sum_{k=1}^{3}\Psi_{\mu\nu}^{\left(jkn\right)}\sqrt{\nu_{k}}\rho_{e\mu_{k}-1,e\nu_{k}-1}^{\left(n\right)}\nonumber \\
 & -v_{gf}^{\left(n\right)} \Big[\Sigma_{\mu\nu}^{\left(jn\right)}\Xi_{\mu\nu}^{\left(jn\right)}\sqrt{\mu_{j}}+\sum_{k=1}^{3}\Psi_{\mu\nu}^{\left(jkn\right)}\sqrt{\mu_{k}}\nonumber \\
 & +v_{ge}^{\left(jn\right)}\Xi_{\mu\nu}^{\left(jn\right)}\Phi_{\mu\nu}^{\left(n\right)}\sqrt{\mu_{j}} \Big]\rho_{g\mu,g\nu}\nonumber \\
 & +v_{gf}^{\left(n\right)}v_{ge}^{\left(jn\right)}\Xi_{\mu\nu}^{\left(jn\right)}\Phi_{\mu\nu}^{\left(n\right)}\sqrt{\mu_{j}}\rho_{f\mu,f\nu}^{\left(n\right)},\label{eq:rhoef}
\end{align}
\begin{align}
 & \rho_{f\mu,e\nu_{j}-1}^{\left(n\right)}=v_{gf}^{\left(n\right)}\tilde{\Sigma}_{\mu\nu}^{\left(jn\right)}\tilde{\Xi}_{\mu\nu}^{\left(jn\right)}\sqrt{\mu_{j}}\rho_{e\mu_{j}-1,e\nu_{j}-1}^{\left(n\right)}\nonumber \\
 & +v_{gf}^{\left(n\right)}\sum_{k=1}^{3}\tilde{\Psi}_{\mu\nu}^{\left(jkn\right)}\sqrt{\mu_{k}}\rho_{e\mu_{k}-1,e\nu_{k}-1}^{\left(n\right)}\nonumber \\
 & -v_{gf}^{\left(n\right)}  \Big[ \tilde{\Sigma}_{\mu\nu}^{\left(jn\right)}\tilde{\Xi}_{\mu\nu}^{\left(jn\right)}\sqrt{\nu_{j}}+\sum_{k=1}^{3}\tilde{\Psi}_{\mu\nu}^{\left(jkn\right)}\sqrt{v_{k}}\nonumber \\
 & +v_{ge}^{\left(jn\right)}\tilde{\Xi}_{\mu\nu}^{\left(jn\right)}\tilde{\Phi}_{\mu\nu}^{\left(n\right)}\sqrt{\nu_{j}}  \Big]\rho_{g\mu,g\nu}\nonumber \\
 & +v_{gf}^{\left(n\right)}v_{ge}^{\left(jn\right)}\tilde{\Xi}_{\mu\nu}^{\left(jn\right)}\tilde{\Phi}_{\mu\nu}^{\left(n\right)}\sqrt{\nu_{j}}\rho_{f\mu,f\nu}^{\left(n\right)},\label{eq:rhofe}
\end{align}
with the abbreviations
\begin{align}
\Psi_{\mu\nu}^{\left(jkn\right)} & =v_{ge}^{\left(jn\right)}v_{ge}^{\left(kn\right)}\Phi_{\mu\nu}^{\left(n\right)}\Xi_{\mu\nu}^{\left(jn\right)}\Xi_{\mu\nu}^{\left(kn\right)}\Sigma_{\mu\nu}^{\left(jn\right)}\sqrt{\mu_{j}\mu_{k}},\\
\tilde{\Psi}_{\mu\nu}^{\left(jkn\right)} & =v_{ge}^{\left(jn\right)}v_{ge}^{\left(kn\right)}\tilde{\Phi}_{\mu\nu}^{\left(n\right)}\tilde{\Xi}_{\mu\nu}^{\left(jn\right)}\tilde{\Xi}_{\mu\nu}^{\left(kn\right)}\tilde{\Sigma}_{\mu\nu}^{\left(jn\right)}\sqrt{\nu_{j}v_{k}}.
\end{align}
Finally, since $\rho_{g\mu,e\nu_{j}-1}^{\left(n\right)}$ and $\rho_{e\mu_{j}-1,g\nu}^{\left(n\right)}$
depend on $\tilde{\rho}_{e\mu_{j}-1,f\nu}^{\left(n\right)}$ and $\rho_{f\mu,e\nu_{j}-1}^{\left(n\right)}$
through Eqs. \eqref{eq:rhoge-s} and \eqref{eq:rhoeg-s}, we can express
them also as functions of the population-like correlations:
\begin{align}
 & \left(\tilde{\omega}_{eg}^{\left(n\right)*}-\tilde{\omega}_{\mu\nu_{j}-1}\right)\rho_{g\mu,e\nu_{j}-1}^{\left(n\right)}=\nonumber \\
 & \left(v_{gf}^{\left(n\right)2}\tilde{\Xi}_{\mu\nu}^{\left(jn\right)}\tilde{\Sigma}_{\mu\nu}^{\left(jn\right)}+v_{ge}^{\left(jn\right)}\right)\sqrt{\mu_{j}}\rho_{e\mu_{j}-1,e\nu_{j}-1}^{\left(n\right)}\nonumber \\
 & +v_{gf}^{\left(n\right)2}\sum_{k=1}^{3}\tilde{\Psi}_{\mu\nu}^{\left(jkn\right)}\sqrt{\mu_{k}}\rho_{e\mu_{k}-1,e\nu_{k}-1}^{\left(n\right)}\nonumber \\
 & -  \Big[ v_{gf}^{\left(n\right)2}\tilde{\Xi}_{\mu\nu}^{\left(jn\right)}\tilde{\Sigma}_{\mu\nu}^{\left(jn\right)}\sqrt{\nu_{j}}+v_{gf}^{\left(n\right)2}\sum_{k=1}^{3}\tilde{\Psi}_{\mu\nu}^{\left(jkn\right)}\sqrt{v_{k}}\nonumber \\
 & +v_{gf}^{\left(n\right)2}v_{ge}^{\left(jn\right)}\tilde{\Xi}_{\mu\nu}^{\left(jn\right)}\tilde{\Phi}_{\mu\nu}^{\left(n\right)}\sqrt{\nu_{j}}+v_{ge}^{\left(jn\right)}\sqrt{\nu_{j}}  \Big]\rho_{g\mu,g\nu}\nonumber \\
 & +v_{gf}^{\left(n\right)2}v_{ge}^{\left(jn\right)}\tilde{\Xi}_{\mu\nu}^{\left(jn\right)}\tilde{\Phi}_{\mu\nu}^{\left(n\right)}\sqrt{\nu_{j}}\rho_{f\mu,f\nu}^{\left(n\right)},\label{eq:rhoge}
\end{align}
\begin{align}
 & \left(\tilde{\omega}_{eg}^{\left(n\right)}+\tilde{\omega}_{\mu_{j}-1\nu}\right)\rho_{e\mu_{j}-1,g\nu}^{\left(n\right)}=\nonumber \\
 & \left(v_{gf}^{\left(n\right)2}\Sigma_{\mu\nu}^{\left(jn\right)}\Xi_{\mu\nu}^{\left(jn\right)}+v_{ge}^{\left(jn\right)}\right)\sqrt{\nu_{j}}\rho_{e\mu_{j}-1,e\nu_{j}-1}^{\left(n\right)}\nonumber \\
 & +v_{gf}^{\left(n\right)2}\sum_{k=1}^{3}\Psi_{\mu\nu}^{\left(jkn\right)}\sqrt{\nu_{k}}\rho_{e\mu_{k}-1,e\nu_{k}-1}^{\left(n\right)}\nonumber \\
 & - \Big[v_{gf}^{\left(n\right)2}\Sigma_{\mu\nu}^{\left(jn\right)}\Xi_{\mu\nu}^{\left(jn\right)}\sqrt{\mu_{j}}+v_{gf}^{\left(n\right)2}\sum_{k=1}^{3}\Psi_{\mu\nu}^{\left(jkn\right)}\sqrt{\mu_{k}}\nonumber \\
 & +v_{gf}^{\left(n\right)2}v_{ge}^{\left(jn\right)}\Xi_{\mu\nu}^{\left(jn\right)}\Phi_{\mu\nu}^{\left(n\right)}\sqrt{\mu_{j}}+v_{ge}^{\left(jn\right)}\sqrt{\mu_{j}} \Big] \rho_{g\mu,g\nu}\nonumber \\
 & +v_{gf}^{\left(n\right)2}v_{ge}^{\left(jn\right)}\Xi_{\mu\nu}^{\left(jn\right)}\Phi_{\mu\nu}^{\left(n\right)}\sqrt{\mu_{j}}\rho_{f\mu,f\nu}^{\left(n\right)}.\label{eq:rhoeg}
\end{align}

Our next step is to obtain equations only for the population-like
correlations. However, before doing so, it is helpful to consider
the following combination of terms appearing in Eqs. \eqref{eq:rhogg} and \eqref{eq:rhoee}: 
\begin{align}
 & iv_{ge}^{\left(jn\right)}\left(\sqrt{\mu_{j}}\rho_{e\mu_{j}-1,g\nu}^{\left(n\right)}-\sqrt{\nu_{j}}\rho_{g\mu,e\nu_{j}-1}^{\left(n\right)}\right)\nonumber \\
 & =\left(d_{\mu\nu}^{\left(jn\right)}+a_{\mu\nu}^{\left(jn\right)}\right)\rho_{e\mu_{j}-1,e\nu_{j}-1}^{\left(n\right)}+\sum_{k=1}^{3}g_{\mu\nu}^{\left(jkn\right)}\rho_{e\mu_{k}-1,e\nu_{k}-1}^{\left(n\right)}\nonumber \\
 & -\left(e_{\mu\nu}^{\left(jn\right)}+k_{\mu\nu}^{\left(jn\right)}+\sum_{k=1}^{3}h_{\mu\nu}^{\left(jkn\right)}+b_{\mu\nu}^{\left(jn\right)}\right)\rho_{g\mu,g\nu}+k_{\mu\nu}^{\left(jn\right)}\rho_{f\mu,f\nu}^{\left(n\right)},\label{eq:rhoeg-rhoge}
\end{align}
\begin{align}
 & iv_{ge}^{\left(jn\right)}\left(\sqrt{\nu_{j}}\rho_{e\mu_{j}-1,g\nu}^{\left(n\right)}-\sqrt{\mu_{j}}\rho_{g\mu,e\nu_{j}-1}^{\left(n\right)}\right)\nonumber \\
 & =\left(f_{\mu\nu}^{\left(jn\right)}+c_{\mu\nu}^{\left(jn\right)}\right)\rho_{e\mu_{j}-1,e\nu_{j}-1}^{\left(n\right)}+\sum_{k=1}^{3}i_{\mu\nu}^{\left(jkn\right)}\rho_{e\mu_{k}-1,e\nu_{k}-1}^{\left(n\right)}\nonumber \\
 & -\left(d_{\mu\nu}^{\left(jn\right)}+l_{\mu\nu}^{\left(jn\right)}+\sum_{k=1}^{3}j_{\mu\nu}^{\left(jkn\right)}+a_{\mu\nu}^{\left(jn\right)}\right)\rho_{g\mu,g\nu}+l_{\mu\nu}^{\left(jn\right)}\rho_{f\mu,f\nu}^{\left(n\right)}, \label{eq:rhoeg-rhoge-1}
\end{align}
with the abbreviations
\begin{alignat}{1}
 & a_{\mu\nu}^{\left(jn\right)}=iv_{ge}^{\left(jn\right)}\left(\Sigma_{\mu\nu}^{\left(jn\right)}-\tilde{\Sigma}_{\mu\nu}^{\left(jn\right)}\right)\sqrt{\mu_{j}\nu_{j}},\\
 & b_{\mu\nu}^{\left(jn\right)}=iv_{ge}^{\left(jn\right)}\left(\Sigma_{\mu\nu}^{\left(jn\right)}\mu_{j}-\tilde{\Sigma}_{\mu\nu}^{\left(jn\right)}\nu_{j}\right),\\
 & c_{\mu\nu}^{\left(jn\right)}=iv_{ge}^{\left(jn\right)}\left(\Sigma_{\mu\nu}^{\left(jn\right)}\nu_{j}-\tilde{\Sigma}_{\mu\nu}^{\left(jn\right)}\mu_{j}\right),\\
 & d_{\mu\nu}^{\left(jn\right)}=iv_{gf}^{\left(n\right)2}\left(\Sigma_{\mu\nu}^{\left(jn\right)2}\Xi_{\mu\nu}^{\left(jn\right)}-\tilde{\Sigma}_{\mu\nu}^{\left(jn\right)2}\tilde{\Xi}_{\mu\nu}^{\left(jn\right)}\right)\sqrt{\mu_{j}\nu_{j}},\\
 & e_{\mu\nu}^{\left(jn\right)}=iv_{gf}^{\left(n\right)2}\left(\Sigma_{\mu\nu}^{\left(jn\right)2}\Xi_{\mu\nu}^{\left(jn\right)}\mu_{j}-\tilde{\Sigma}_{\mu\nu}^{\left(jn\right)2}\tilde{\Xi}_{\mu\nu}^{\left(jn\right)}\nu_{j}\right),\\
 & f_{\mu\nu}^{\left(jn\right)}=iv_{gf}^{\left(n\right)2}\left(\Sigma_{\mu\nu}^{\left(jn\right)2}\Xi_{\mu\nu}^{\left(jn\right)}\nu_{j}-\tilde{\Sigma}_{\mu\nu}^{\left(jn\right)2}\tilde{\Xi}_{\mu\nu}^{\left(jn\right)}\mu_{j}\right),
\end{alignat}
\begin{align}
 & g_{\mu\nu}^{\left(jkn\right)}=iv_{gf}^{\left(n\right)2}\left(\Sigma_{\mu\nu}^{\left(jn\right)}\Psi_{\mu\nu}^{\left(jkn\right)}\sqrt{\mu_{j}\nu_{k}}-\tilde{\Sigma}_{\mu\nu}^{\left(jn\right)}\tilde{\Psi}_{\mu\nu}^{\left(jkn\right)}\sqrt{\nu_{j}\mu_{k}}\right),\\
 & h_{\mu\nu}^{\left(jkn\right)}=iv_{gf}^{\left(n\right)2}\left(\Sigma_{\mu\nu}^{\left(jn\right)}\Psi_{\mu\nu}^{\left(jkn\right)}\sqrt{\mu_{j}\mu_{k}}-\tilde{\Sigma}_{\mu\nu}^{\left(jn\right)}\tilde{\Psi}_{\mu\nu}^{\left(jkn\right)}\sqrt{\nu_{j}\nu_{k}}\right),\\
 & i_{\mu\nu}^{\left(jkn\right)}=iv_{gf}^{\left(n\right)2}\left(\Sigma_{\mu\nu}^{\left(jn\right)}\Psi_{\mu\nu}^{\left(jkn\right)}\sqrt{\nu_{j}\nu_{k}}-\tilde{\Sigma}_{\mu\nu}^{\left(jn\right)}\tilde{\Psi}_{\mu\nu}^{\left(jkn\right)}\sqrt{\mu_{j}\mu_{k}}\right),\\
 & j_{\mu\nu}^{\left(jkn\right)}=iv_{gf}^{\left(n\right)2}\left(\Sigma_{\mu\nu}^{\left(jn\right)}\Psi_{\mu\nu}^{\left(jkn\right)}\sqrt{\nu_{j}\mu_{k}}-\tilde{\Sigma}_{\mu\nu}^{\left(jn\right)}\tilde{\Psi}_{\mu\nu}^{\left(jkn\right)}\sqrt{\mu_{j}\nu_{k}}\right),\\
 & k_{\mu\nu}^{\left(jn\right)}=iv_{gf}^{\left(n\right)2}v_{ge}^{\left(jn\right)}\left(\Sigma_{\mu\nu}^{\left(jn\right)}\Xi_{\mu\nu}^{\left(jn\right)}\Phi_{\mu\nu}^{\left(n\right)}\mu_{j}-\tilde{\Sigma}_{\mu\nu}^{\left(jn\right)}\tilde{\Xi}_{\mu\nu}^{\left(jn\right)}\tilde{\Phi}_{\mu\nu}^{\left(n\right)}\nu_{j}\right),\\
 & l_{\mu\nu}^{\left(jn\right)}=iv_{gf}^{\left(n\right)2}v_{ge}^{\left(jn\right)}\left(\Sigma_{\mu\nu}^{\left(jn\right)}\Xi_{\mu\nu}^{\left(jn\right)}\Phi_{\mu\nu}^{\left(n\right)}-\tilde{\Sigma}_{\mu\nu}^{\left(jn\right)}\tilde{\Xi}_{\mu\nu}^{\left(jn\right)}\tilde{\Phi}_{\mu\nu}^{\left(n\right)}\right)\sqrt{\mu_{j}\nu_{j}}.
\end{align}
We also consider the combination appearing in \eqref{eq:rhoff}:
\begin{align}
 & iv_{gf}^{\left(n\right)}\left(\tilde{\rho}_{f\mu,g\nu}^{\left(n\right)}-\tilde{\rho}_{g\mu,f\nu}^{\left(n\right)}\right)=o_{\mu\nu}^{\left(n\right)}\left(\rho_{f\mu,f\nu}^{\left(n\right)}-\rho_{g\mu,g\nu}^{\left(n\right)}\right)\nonumber \\
 & +\sum_{j=1}^{3}m_{\mu\nu}^{\left(jn\right)}\rho_{e\mu_{j}-1,e\nu_{j}-1}^{\left(n\right)}-\sum_{j=1}^{3}n_{\mu\nu}^{\left(jn\right)}\rho_{g\mu,g\nu}, \label{eq:rhofg-rhogf}
\end{align}
with the abbreviations
\begin{align}
m_{\mu\nu}^{\left(jn\right)} & =iv_{gf}^{\left(n\right)2}v_{ge}^{\left(jn\right)}\left(\Phi_{\mu\nu}^{\left(n\right)}\Sigma_{\mu\nu}^{\left(jn\right)}-\tilde{\Phi}_{\mu\nu}^{\left(n\right)}\tilde{\Sigma}_{\mu\nu}^{\left(jn\right)}\right)\sqrt{\mu_{j}v_{j}},\\
n_{\mu\nu}^{\left(jn\right)} & =iv_{gf}^{\left(n\right)2}v_{ge}^{\left(jn\right)}\left(\Phi_{\mu\nu}^{\left(n\right)}\Sigma_{\mu\nu}^{\left(jn\right)}\mu_{j}-\tilde{\Phi}_{\mu\nu}^{\left(n\right)}\tilde{\Sigma}_{\mu\nu}^{\left(jn\right)}v_{j}\right),\\
o_{\mu\nu}^{\left(n\right)} & =iv_{gf}^{\left(n\right)2}\left(\Phi_{\mu\nu}^{\left(n\right)}-\tilde{\Phi}_{\mu\nu}^{\left(n\right)}\right).
\end{align}
Now, we consider the steady-state version of Eqs. \eqref{eq:rhogg}, \eqref{eq:rhoff} and   \eqref{eq:rhoee}: 
\begin{align}
 & \left(i\tilde{\omega}_{\mu_{j}-1\nu_{j}-1}+k_{f\to e}^{\left(n\right)}+k_{e\to g}^{\left(n\right)}+k_{e\to f}^{\left(n\right)}\right)\rho_{e\mu_{j}-1,e\nu_{j}-1}^{\left(n\right)}=\nonumber \\
 & +\left(k_{g\to e}^{\left(n\right)}-k_{f\to e}^{\left(n\right)}\right)\rho_{g\mu,g\nu}^{\left(n\right)}+k_{f\to e}^{\left(n\right)}\rho_{\mu_{j}-1\nu_{j}-1}\nonumber \\
 & +iv_{ge}^{\left(jn\right)}\left(\sqrt{\nu_{j}}\rho_{e\mu_{j}-1,g\nu}^{\left(n\right)}-\sqrt{\mu_{j}}\rho_{g\mu,e\nu_{j}-1}^{\left(n\right)}\right),\label{eq:rhoee-1}
\end{align}
\begin{align}
 & \left(i\tilde{\omega}_{\mu\nu}+k_{g\to f}^{\left(n\right)}+k_{g\to e}^{\left(n\right)}+k_{e\to g}^{\left(n\right)}\right)\rho_{g\mu,g\nu}^{\left(n\right)}=\nonumber \\
 & \left(k_{f\to g}^{\left(n\right)}-k_{e\to g}^{\left(n\right)}\right)\rho_{f\mu,f\nu}^{\left(n\right)}+k_{e\to g}^{\left(n\right)}\rho_{\mu\nu}\nonumber \\
 & -iv_{gf}^{\left(n\right)}\left(\tilde{\rho}_{f\mu,g\nu}^{\left(n\right)}-\tilde{\rho}_{g\mu,f\nu}^{\left(n\right)}\right)\nonumber \\
 & -i\sum_{j=1}^{3}v_{ge}^{\left(jn\right)}\left(\sqrt{\mu_{j}}\rho_{e\mu_{j}-1,g\nu}^{\left(n\right)}-\sqrt{\nu_{j}}\rho_{g\mu,e\nu_{j}-1}^{\left(n\right)}\right),\label{eq:rhogg-1}
\end{align}
\begin{align}
 & \left(i\tilde{\omega}_{\mu\nu}+k_{f\to g}^{\left(n\right)}+k_{f\to e}^{\left(n\right)}+k_{e\to f}^{\left(n\right)}\right)\rho_{f\mu,f\nu}^{\left(n\right)}=\nonumber \\
 & +\left(k_{g\to f}^{\left(n\right)}-k_{e\to f}^{\left(n\right)}\right)\rho_{g\mu,g\nu}^{\left(n\right)}+k_{e\to f}^{\left(n\right)}\rho_{\mu\nu}\nonumber \\
 & +iv_{gf}^{\left(n\right)}\left(\tilde{\rho}_{f\mu,g\nu}^{\left(n\right)}-\tilde{\rho}_{g\mu,f\nu}^{\left(n\right)}\right).\label{eq:rhoff-1}
\end{align}
In Eq. \eqref{eq:rhoee-1}, we have replaced $\rho_{f\mu_{j}-1,f\nu_{j}-1}^{\left(n\right)}$
by $\rho_{\mu_{j}-1\nu_{j}-1}-\rho_{g\mu_{j}-1,g\nu_{j}-1}^{\left(n\right)}-\rho_{e\mu_{j}-1,e\nu_{j}-1}^{\left(n\right)}$
and then approximated $\rho_{g\mu_{j}-1,g\nu_{j}-1}^{\left(n\right)}$
by $\rho_{g\mu,g\nu}^{\left(n\right)}$. In Eqs. \eqref{eq:rhogg-1}
and \eqref{eq:rhoff-1}, we have replaced $\rho_{e\mu,e\nu}^{\left(n\right)}$
by $\rho_{\mu\nu}-\rho_{g\mu,g\nu}^{\left(n\right)}-\rho_{f\mu,f\nu}^{\left(n\right)}$.
Following this treatment, we get the dependence of the correlations and the plasmon RDM shown in 
Fig. \ref{fig:scheme-three-level} (c) in the main text. To proceed, we insert
Eqs. \eqref{eq:rhoeg-rhoge} and \eqref{eq:rhoeg-rhoge-1} into Eqs.
\eqref{eq:rhoee-1} and \eqref{eq:rhogg-1} and insert Eq.
\eqref{eq:rhofg-rhogf} into Eqs. \eqref{eq:rhogg-1} and \eqref{eq:rhoff-1}
to get closed equations for the population-like correlations:
\begin{align}
 & p_{\mu\nu}^{\left(jn\right)}\rho_{e\mu_{j}-1,e\nu_{j}-1}^{\left(n\right)}-\sum_{k\neq j}i_{\mu\nu}^{\left(jkn\right)}\rho_{e\mu_{k}-1,e\nu_{k}-1}^{\left(n\right)}=\nonumber \\
 & k_{f\to e}^{\left(n\right)}\rho_{\mu_{j}-1\nu_{j}-1}+l_{\mu\nu}^{\left(jn\right)}\rho_{f\mu,f\nu}^{\left(n\right)}\nonumber \\
 & +\left[ k_{g\to e}^{\left(n\right)}-k_{f\to e}^{\left(n\right)}-\left(d_{\mu\nu}^{\left(jn\right)}+l_{\mu\nu}^{\left(jn\right)}+\sum_{k=1}^{3}j_{\mu\nu}^{\left(jkn\right)}+a_{\mu\nu}^{\left(jn\right)}\right)\right]\rho_{g\mu,g\nu}^{\left(n\right)},\label{eq:rhoee-2}
\end{align}
\begin{align}
 & (i\tilde{\omega}_{\mu\nu}+k_{g\to f}^{\left(n\right)}+k_{g\to e}^{\left(n\right)}+k_{e\to g}^{\left(n\right)}-o_{\mu\nu}^{\left(n\right)}-\sum_{j=1}^{3}n_{\mu\nu}^{\left(jn\right)}\nonumber \\
 & -\sum_{j=1}^{3}\left(e_{\mu\nu}^{\left(jn\right)}+k_{\mu\nu}^{\left(jn\right)}+\sum_{k=1}^{3}h_{\mu\nu}^{\left(jkn\right)}+b_{\mu\nu}^{\left(jn\right)}\right))\rho_{g\mu,g\nu}^{\left(n\right)}=\nonumber \\
 & k_{e\to g}^{\left(n\right)}\rho_{\mu\nu}+\left(k_{f\to g}^{\left(n\right)}-k_{e\to g}^{\left(n\right)}-\sum_{j=1}^{3}k_{\mu\nu}^{\left(jn\right)}-o_{\mu\nu}^{\left(n\right)}\right)\rho_{f\mu,f\nu}^{\left(n\right)}\nonumber \\
 & -\sum_{j=1}^{3}\left(d_{\mu\nu}^{\left(jn\right)}+a_{\mu\nu}^{\left(jn\right)}+m_{\mu\nu}^{\left(jn\right)}+\sum_{k=1}^{3}g_{\mu\nu}^{\left(kjn\right)}\right)\rho_{e\mu_{j}-1,e\nu_{j}-1}^{\left(n\right)},\label{eq:rhogg-2}
\end{align}
\begin{align}
 & \left(i\tilde{\omega}_{\mu\nu}+k_{f\to g}^{\left(n\right)}+k_{f\to e}^{\left(n\right)}+k_{e\to f}^{\left(n\right)}-o_{\mu\nu}^{\left(n\right)}\right)\rho_{f\mu,f\nu}^{\left(n\right)}=\nonumber \\
 & k_{e\to f}^{\left(n\right)}\rho_{\mu\nu}+\left(k_{g\to f}^{\left(n\right)}-k_{e\to f}^{\left(n\right)}-o_{\mu\nu}^{\left(n\right)}-\sum_{j=1}^{3}n_{\mu\nu}^{\left(jn\right)}\right)\rho_{g\mu,g\nu}^{\left(n\right)}\nonumber \\
 & +\sum_{j=1}^{3}m_{\mu\nu}^{\left(jn\right)}\rho_{e\mu_{j}-1,e\nu_{j}-1}^{\left(n\right)},\label{eq:rhoff-2}
\end{align}
where  we have introduced
\begin{align}
& p_{\mu\nu}^{\left(jn\right)}=i\tilde{\omega}_{\mu_{j}-1\nu_{j}-1}+k_{f\to e}^{\left(n\right)}+k_{e\to g}^{\left(n\right)}+k_{e\to f}^{\left(n\right)} \nonumber \\
& -\left(f_{\mu\nu}^{\left(jn\right)}+c_{\mu\nu}^{\left(jn\right)}\right)-i_{\mu\nu}^{\left(jjn\right)}
\end{align}
in Eq. \eqref{eq:rhoee-2}.

We notice that Eq. \eqref{eq:rhoee-2} can be rewritten in a matrix-form
and the coefficients before $\rho_{e\mu_{j}-1,e\nu_{j}-1}^{\left(n\right)}$
form a coefficient matrix with the elements $M_{\mu\nu}^{\left(ijn\right)}=\delta_{ij}p_{\mu\nu}^{\left(in\right)}-\left(1-\delta_{ij}\right)i_{\mu\nu}^{\left(ijn\right)}.$
We assume the inverse matrix of the coefficient matrix is $q_{\mu\nu}^{\left(jkn\right)}$
and write the solution of Eq. \eqref{eq:rhoee-2} as
\begin{align}
 & \rho_{e\mu_{j}-1,e\nu_{j}-1}^{\left(n\right)}=\sum_{k=1}^{3}q_{\mu\nu}^{\left(jkn\right)} \Big[ k_{f\to e}^{\left(n\right)}\rho_{\mu_{k}-1\nu_{k}-1}+l_{\mu\nu}^{\left(kn\right)}\rho_{f\mu,f\nu}^{\left(n\right)}\nonumber \\
 & +\left(k_{g\to e}^{\left(n\right)}-k_{f\to e}^{\left(n\right)}-\left(d_{\mu\nu}^{\left(kn\right)}+l_{\mu\nu}^{\left(kn\right)}+\sum_{l=1}^{3}j_{\mu\nu}^{\left(kln\right)}+a_{\mu\nu}^{\left(kn\right)}\right)\right)\rho_{g\mu,g\nu}^{\left(n\right)} \Big].\label{eq:rhoee-3}
\end{align}
We can also rewrite Eqs. \eqref{eq:rhogg-2} and \eqref{eq:rhoff-2}
in a matrix form with the help of Eq. \eqref{eq:rhoee-3}:
\begin{align}
 & \left(\begin{array}{cc}
q_{\mu\nu}^{\left(n\right)} & r_{\mu\nu}^{\left(n\right)}\\
s_{\mu\nu}^{\left(n\right)} & t_{\mu\nu}^{\left(n\right)}
\end{array}\right)\left(\begin{array}{c}
\rho_{g\mu,g\nu}^{\left(n\right)}\\
\rho_{f\mu,f\nu}^{\left(n\right)}
\end{array}\right)=\left(\begin{array}{c}
k_{e\to g}^{\left(n\right)}\\
k_{e\to f}^{\left(n\right)}
\end{array}\right)\rho_{\mu\nu}\nonumber \\
 & +k_{f\to e}^{\left(n\right)}\sum_{k=1}^{3}\left(\begin{array}{c}
-u_{\mu\nu}^{\left(kn\right)}\\
v_{\mu\nu}^{\left(kn\right)}
\end{array}\right)\rho_{\mu_{k}-1\nu_{k}-1}, \label{eq:rhogg-rhoff}
\end{align}
with the abbreviations 
\begin{align}
 & q_{\mu\nu}^{\left(n\right)}=i\tilde{\omega}_{\mu\nu}+k_{g\to f}^{\left(n\right)}+k_{g\to e}^{\left(n\right)}+k_{e\to g}^{\left(n\right)}\nonumber \\
 & -o_{\mu\nu}^{\left(n\right)}-\sum_{j=1}^{3}n_{\mu\nu}^{\left(jn\right)}-\sum_{j=1}^{3}z_{\mu\nu}^{\left(jn\right)}+\sum_{k=1}^{3}u_{\mu\nu}^{\left(kn\right)}y_{\mu\nu}^{\left(kn\right)},\\
 & -r_{\mu\nu}^{\left(n\right)}=k_{f\to g}^{\left(n\right)}-k_{e\to g}^{\left(n\right)}-\sum_{j=1}^{3}k_{\mu\nu}^{\left(jn\right)}-o_{\mu\nu}^{\left(n\right)}-\sum_{k=1}^{3}u_{\mu\nu}^{\left(kn\right)}l_{\mu\nu}^{\left(kn\right)},\\
 & -s_{\mu\nu}^{\left(n\right)}=k_{g\to f}^{\left(n\right)}-k_{e\to f}^{\left(n\right)}-o_{\mu\nu}^{\left(n\right)}-\sum_{j=1}^{3}n_{\mu\nu}^{\left(jn\right)}+\sum_{k=1}^{3}v_{\mu\nu}^{\left(kn\right)}y_{\mu\nu}^{\left(kn\right)},\\
 & t_{\mu\nu}^{\left(n\right)}=i\tilde{\omega}_{\mu\nu}+k_{f\to g}^{\left(n\right)}+k_{f\to e}^{\left(n\right)}+k_{e\to f}^{\left(n\right)}-o_{\mu\nu}^{\left(n\right)}\nonumber \\
 & -\sum_{j=1}^{3}m_{\mu\nu}^{\left(jn\right)}\sum_{k=1}^{3}q_{\mu\nu}^{\left(jkn\right)}l_{\mu\nu}^{\left(kn\right)},\\
 & u_{\mu\nu}^{\left(kn\right)}=\sum_{j=1}^{3}\left(d_{\mu\nu}^{\left(jn\right)}+a_{\mu\nu}^{\left(jn\right)}+m_{\mu\nu}^{\left(jn\right)}+\sum_{l=1}^{3}g_{\mu\nu}^{\left(ljn\right)}\right)q_{\mu\nu}^{\left(jkn\right)},\\
 & v_{\mu\nu}^{\left(kn\right)}=\sum_{j=1}^{3}m_{\mu\nu}^{\left(jn\right)}q_{\mu\nu}^{\left(jkn\right)},\\
 & x_{\mu\nu}^{\left(jkn\right)}=\left(d_{\mu\nu}^{\left(jn\right)}+a_{\mu\nu}^{\left(jn\right)}\right)q_{\mu\nu}^{\left(jkn\right)}+\sum_{l=1}^{3}g_{\mu\nu}^{\left(jln\right)}q_{\mu\nu}^{\left(lkn\right)},\\
 & y_{\mu\nu}^{\left(kn\right)}=k_{g\to e}^{\left(n\right)}-k_{f\to e}^{\left(n\right)}-\left(d_{\mu\nu}^{\left(kn\right)}+l_{\mu\nu}^{\left(kn\right)}+\sum_{o=1}^{3}j_{\mu\nu}^{\left(kon\right)}+a_{\mu\nu}^{\left(kn\right)}\right),\\
 & z_{\mu\nu}^{\left(jn\right)}=e_{\mu\nu}^{\left(jn\right)}+k_{\mu\nu}^{\left(jn\right)}+\sum_{k=1}^{3}h_{\mu\nu}^{\left(jkn\right)}+b_{\mu\nu}^{\left(jn\right)}.
\end{align}
The solution of Eq. \eqref{eq:rhogg-rhoff} is:
\begin{align}
 & \rho_{g\mu,g\nu}^{\left(n\right)}=w_{\mu\nu}^{\left(n\right)}\left(k_{e\to g}^{\left(n\right)}t_{\mu\nu}^{\left(n\right)}-k_{e\to f}^{\left(n\right)}r_{\mu\nu}^{\left(n\right)}\right)\rho_{\mu\nu}\nonumber \\
 & -k_{f\to e}^{\left(n\right)}w_{\mu\nu}^{\left(n\right)}\sum_{k=1}^{3}\left(u_{\mu\nu}^{\left(kn\right)}t_{\mu\nu}^{\left(n\right)}+r_{\mu\nu}^{\left(n\right)}v_{\mu\nu}^{\left(kn\right)}\right)\rho_{\mu_{k}-1\nu_{k}-1},\label{eq:rhogg-3}
\end{align}
\begin{align}
 & \rho_{f\mu,f\nu}^{\left(n\right)}=-w_{\mu\nu}^{\left(n\right)}\left(k_{e\to g}^{\left(n\right)}s_{\mu\nu}^{\left(n\right)}-k_{e\to f}^{\left(n\right)}q_{\mu\nu}^{\left(n\right)}\right)\rho_{\mu\nu}\nonumber \\
 & +w_{\mu\nu}^{\left(n\right)}k_{f\to e}^{\left(n\right)}\sum_{k=1}^{3}\left(u_{\mu\nu}^{\left(kn\right)}s_{\mu\nu}^{\left(n\right)}+q_{\mu\nu}^{\left(n\right)}v_{\mu\nu}^{\left(kn\right)}\right)\rho_{\mu_{k}-1\nu_{k}-1}.\label{eq:rhoff-3}
\end{align}
with 
\begin{equation}
1/w_{\mu\nu}^{\left(n\right)}=q_{\mu\nu}^{\left(n\right)}t_{\mu\nu}^{\left(n\right)}-r_{\mu\nu}^{\left(n\right)}s_{\mu\nu}^{\left(n\right)}.
\end{equation}

In summary, we have expressed  the coherence-like
correlations as functions of the population-like correlations, cf.
Eqs. \eqref{eq:rhoef}, \eqref{eq:rhofe} , \eqref{eq:rhoge} and
\eqref{eq:rhoeg}, and  the population-like correlations
as functions of the plasmon RDM  through Eqs. \eqref{eq:rhoee-3},
\eqref{eq:rhogg-3} and \eqref{eq:rhoff-3}.  By inserting those expressions back into Eq. \eqref{eq:plasmon-rdm},
 we get an explicit, linear equation for the reduced density matrix  $\rho_{\mu\nu}$ of the plasmon modes:
\begin{align}
 & \frac{\partial}{\partial t}\rho_{\mu\nu}=-\sum_{j=1}^{3}\left(i\omega_{j}\left(\mu_{j}-\nu_{j}\right)+\gamma_{j}\left[\left(\mu_{j}+\nu_{j}\right)/2\right]\right)\rho_{\mu\nu}\nonumber \\
 & +\sum_{j=1}^{3}\gamma_{j}\sqrt{\left(\mu_{j}+1\right)\left(\nu_{j}+1\right)}\rho_{\mu_{j}+1\nu_{j}+1}-\sum_{jk=1}^{3}\beta_{\mu\nu}^{\left(jk\right)}\rho_{\mu_{k}-1\nu_{k}-1}\nonumber \\
 & +\sum_{j=1}^{3}\alpha{}_{\mu\nu}^{\left(j\right)}\rho_{\mu\nu}+\sum_{j,k=1}^{3}\tilde{\beta}_{\mu_{j}+1\nu_{j}+1}^{\left(jk\right)}\rho_{\mu_{j}+1\mu_{k}-1\nu_{j}+1\nu_{k}-1}\nonumber \\
 & -\sum_{j=1}^{3}\tilde{\alpha}_{\mu_{j}+1\nu_{j}+1}^{\left(j\right)}\rho_{\mu_{j}+1\nu_{j}+1}.\label{eq:equation-plasmon-rdm}
\end{align}
Here, we have summed the contribution from individual emitter and
introduced the abbreviations $\beta_{\mu\nu}^{\left(jk\right)} \equiv \sum_{n=1}^{N_{\mathrm{e}}}\beta_{\mu\nu}^{\left(jkn\right)},\alpha{}_{\mu\nu}^{\left(j\right)}\equiv\sum_{n=1}^{N_{\mathrm{e}}}\alpha{}_{\mu\nu}^{\left(jn\right)}$
as well as $\tilde{\beta}_{\mu_{j}+1\nu_{j}+1}^{\left(jk\right)} \equiv \sum_{n=1}^{N_{\mathrm{e}}}\tilde{\beta}_{\mu_{j}+1\nu_{j}+1}^{\left(jkn\right)}$,
$\tilde{\alpha}_{\mu_{j}+1\nu_{j}+1}^{\left(j\right)} \equiv \sum_{n=1}^{N_{\mathrm{e}}}\tilde{\alpha}_{\mu_{j}+1\nu_{j}+1}^{\left(jn\right)}$.
The abbreviations are defined as follows:
\begin{align}
 & \alpha{}_{\mu\nu}^{\left(jn\right)}=w_{\mu\nu}^{\left(n\right)} \Big[ \left(\sum_{k=1}^{3}x_{\mu\nu}^{\left(jkn\right)}l_{\mu\nu}^{\left(kn\right)}+k_{\mu\nu}^{\left(jn\right)}\right)\left(k_{e\to g}^{\left(n\right)}s_{\mu\nu}^{\left(n\right)}-k_{e\to f}^{\left(n\right)}q_{\mu\nu}^{\left(n\right)}\right)\nonumber \\
 & -\left(\sum_{k=1}^{3}x_{\mu\nu}^{\left(jkn\right)}y_{\mu\nu}^{\left(kn\right)}-z_{\mu\nu}^{\left(jn\right)}\right)\left(k_{e\to g}^{\left(n\right)}t_{\mu\nu}^{\left(n\right)}-k_{e\to f}^{\left(n\right)}r_{\mu\nu}^{\left(n\right)}\right) \Big],\label{eq:alpha}
\end{align}
\begin{align}
 & \beta_{\mu\nu}^{\left(jkn\right)}=k_{f\to e}^{\left(n\right)} \Big[ w_{\mu\nu}^{\left(n\right)}\left(\sum_{l=1}^{3}x_{\mu\nu}^{\left(jln\right)}l_{\mu\nu}^{\left(ln\right)}+k_{\mu\nu}^{\left(jn\right)}\right)\left(u_{\mu\nu}^{\left(kn\right)}s_{\mu\nu}^{\left(n\right)}+q_{\mu\nu}^{\left(n\right)}v_{\mu\nu}^{\left(kn\right)}\right)\nonumber \\
 & x_{\mu\nu}^{\left(jkn\right)}-w_{\mu\nu}^{\left(n\right)}\left(\sum_{l=1}^{3}x_{\mu\nu}^{\left(jln\right)}y_{\mu\nu}^{\left(ln\right)}-z_{\mu\nu}^{\left(jn\right)}\right)\left(u_{\mu\nu}^{\left(kn\right)}t_{\mu\nu}^{\left(n\right)}+r_{\mu\nu}^{\left(n\right)}v_{\mu\nu}^{\left(kn\right)}\right) \Big],\label{eq:beta}
\end{align}
 and 
\begin{align}
 & \tilde{\alpha}_{\mu\nu}^{\left(jn\right)}=w_{\mu\nu}^{\left(n\right)} \Big[ \left(\sum_{k=1}^{3}\tilde{x}_{\mu\nu}^{\left(jkn\right)}l_{\mu\nu}^{\left(kn\right)}+l_{\mu\nu}^{\left(jn\right)}\right)\left(k_{e\to g}^{\left(n\right)}s_{\mu\nu}^{\left(n\right)}-k_{e\to f}^{\left(n\right)}q_{\mu\nu}^{\left(n\right)}\right)\nonumber \\
 & -\left(\sum_{k=1}^{3}\tilde{x}_{\mu\nu}^{\left(jkn\right)}y_{\mu\nu}^{\left(kn\right)}-\tilde{z}_{\mu\nu}^{\left(jn\right)}\right)\left(k_{e\to g}^{\left(n\right)}t_{\mu\nu}^{\left(n\right)}-k_{e\to f}^{\left(n\right)}r_{\mu\nu}^{\left(n\right)}\right) \Big],
\end{align}
\begin{align}
 & \tilde{\beta}_{\mu\nu}^{\left(jkn\right)}=k_{f\to e}^{\left(n\right)} \Big[ w_{\mu\nu}^{\left(n\right)}\left(\sum_{k=1}^{3}\tilde{x}_{\mu\nu}^{\left(jkn\right)}l_{\mu\nu}^{\left(kn\right)}+l_{\mu\nu}^{\left(jn\right)}\right)\left(u_{\mu\nu}^{\left(kn\right)}s_{\mu\nu}^{\left(n\right)}+q_{\mu\nu}^{\left(n\right)}v_{\mu\nu}^{\left(kn\right)}\right)\nonumber \\
 & +\tilde{x}_{\mu\nu}^{\left(jkn\right)}-w_{\mu\nu}^{\left(n\right)}\left(\sum_{k=1}^{3}\tilde{x}_{\mu\nu}^{\left(jkn\right)}y_{\mu\nu}^{\left(kn\right)}-\tilde{z}_{\mu\nu}^{\left(jn\right)}\right)\left(u_{\mu\nu}^{\left(kn\right)}t_{\mu\nu}^{\left(n\right)}+r_{\mu\nu}^{\left(n\right)}v_{\mu\nu}^{\left(kn\right)}\right) \Big],
\end{align}
with 
\begin{align}
\tilde{x}_{\mu\nu}^{\left(jkn\right)} & =\left(f_{\mu\nu}^{\left(jn\right)}+c_{\mu\nu}^{\left(jn\right)}\right)q_{\mu\nu}^{\left(jkn\right)}+\sum_{l=1}^{3}i_{\mu\nu}^{\left(jln\right)}q_{\mu\nu}^{\left(lkn\right)},\\
\tilde{z}_{\mu\nu}^{\left(jn\right)} & =d_{\mu\nu}^{\left(jn\right)}+l_{\mu\nu}^{\left(jn\right)}+\sum_{k=1}^{3}j_{\mu\nu}^{\left(jkn\right)}+a_{\mu\nu}^{\left(jn\right)}.
\end{align}

\section{Equation for Population of Plasmon Number State and Molecular States\label{sec:collection-population}}

The diagonal elements of the plasmon RDM can be interpreted as the
population $P_{\mu}\equiv\rho_{\mu\mu}$ of plasmon number states  and 
can be obtained by solving Eq. \eqref{eq:EPSP} in the main text. There, $\kappa_{\mu}^{\left(j\right)}\equiv-\sum_{n=1}^{N_{\mathrm{e}}}\alpha{}_{\mu\mu}^{\left(jn\right)}$
and $\eta_{\mu}^{\left(jk\right)}\equiv-\sum_{n=1}^{N_{\mathrm{e}}}\beta_{\mu\mu}^{\left(jkn\right)}$
are the molecule-induced plasmon damping and pumping rate respectively.
Since they only depend on the diagonal elements of $\alpha{}_{\mu\nu}^{\left(jn\right)}$
and $\beta_{\mu\nu}^{\left(jkn\right)}$, they can be given explicitly as:
\begin{align}
 & \eta_{\mu}^{\left(jk\right)}=-\sum_{n=1}^{N_{\mathrm{m}}}k_{f\to e}^{\left(n\right)} \Big[ x_{\mu}^{\left(jkn\right)} \nonumber \\ 
 & + w_{\mu}^{\left(n\right)}\left(\sum_{l=1}^{3}x_{\mu}^{\left(jln\right)}k_{\mu}^{\left(ln\right)}+k_{\mu}^{\left(jn\right)}\right)\left(u_{\mu}^{\left(kn\right)}s_{\mu}^{\left(n\right)}+q_{\mu}^{\left(n\right)}v_{\mu}^{\left(kn\right)}\right)\nonumber \\
 &-w_{\mu}^{\left(n\right)}\left(\sum_{l=1}^{3}x_{\mu}^{\left(jln\right)}y_{\mu}^{\left(ln\right)}-z_{\mu}^{\left(jn\right)}\right)\left(u_{\mu}^{\left(kn\right)}t_{\mu}^{\left(n\right)}+r_{\mu}^{\left(n\right)}v_{\mu}^{\left(kn\right)}\right) \Big],\label{eq:pumping-rates}\\
 & \kappa_{\mu}^{\left(j\right)}=-\sum_{n=1}^{N_{\mathrm{m}}}w_{\mu}^{\left(n\right)}\Big[ \left(\sum_{k=1}^{3}x_{\mu}^{\left(jkn\right)}k_{\mu}^{\left(kn\right)}+k_{\mu}^{\left(jn\right)}\right)\left(k_{e\to g}^{\left(n\right)}s_{\mu}^{\left(n\right)}-k_{e\to f}^{\left(n\right)}q_{\mu}^{\left(n\right)}\right)\nonumber \\
 & -\left(\sum_{k=1}^{3}x_{\mu}^{\left(jkn\right)}y_{\mu}^{\left(kn\right)}-z_{\mu}^{\left(jn\right)}\right)\left(k_{e\to g}^{\left(n\right)}t_{\mu}^{\left(n\right)}-k_{e\to f}^{\left(n\right)}r_{\mu}^{\left(n\right)}\right) \Big].\label{eq:decay-rates}
\end{align}
In the above and also following expressions, all the quantities are
the diagonal elements of the corresponding quantities appearing in
Sec. \ref{sec:derviation-plasmon-rdm}, for example $w_{\mu}^{\left(n\right)}\equiv w_{\mu\mu}^{\left(n\right)}$.
The quantities in Eqs. \eqref{eq:pumping-rates} and \eqref{eq:decay-rates}
have the following expressions:
\begin{align}
 & u_{\mu}^{\left(kn\right)}=\sum_{j=1}^{3}\left(d_{\mu}^{\left(jn\right)}+a_{\mu}^{\left(jn\right)}+m_{\mu}^{\left(jn\right)}+\sum_{l=1}^{3}g_{\mu}^{\left(ljn\right)}\right)q_{\mu}^{\left(jkn\right)},\\
 & v_{\mu}^{\left(kn\right)}=\sum_{j=1}^{3}m_{\mu}^{\left(jn\right)}q_{\mu}^{\left(jkn\right)},\\
 & x_{\mu}^{\left(jkn\right)}=\left(d_{\mu}^{\left(jn\right)}+a_{\mu}^{\left(jn\right)}\right)q_{\mu}^{\left(jkn\right)}+\sum_{l=1}^{3}g_{\mu}^{\left(jln\right)}q_{\mu}^{\left(lkn\right)},\\
 & y_{\mu}^{\left(kn\right)}=k_{g\to e}^{\left(n\right)}-k_{f\to e}^{\left(n\right)}-\left(d_{\mu}^{\left(kn\right)}+k_{\mu}^{\left(kn\right)}+\sum_{o=1}^{3}g_{\mu}^{\left(kon\right)}+a_{\mu}^{\left(kn\right)}\right),\\
 & z_{\mu}^{\left(jn\right)}=d_{\mu}^{\left(jn\right)}+k_{\mu}^{\left(jn\right)}+\sum_{k=1}^{3}g_{\mu}^{\left(jkn\right)}+a_{\mu}^{\left(jn\right)},
\end{align}
\begin{align}
 & q_{\mu}^{\left(n\right)}=k_{g\to f}^{\left(n\right)}+k_{g\to e}^{\left(n\right)}+k_{e\to g}^{\left(n\right)}-o_{\mu}^{\left(n\right)}\nonumber \\
 & -\sum_{j=1}^{3}\left(m_{\mu}^{\left(jn\right)}+z_{\mu}^{\left(jn\right)}-u_{\mu}^{\left(jn\right)}y_{\mu}^{\left(jn\right)}\right),\\
 & r_{\mu}^{\left(n\right)}=-\left(k_{f\to g}^{\left(n\right)}-k_{e\to g}^{\left(n\right)}-o_{\mu}^{\left(n\right)}-\sum_{j=1}^{3}\left(k_{\mu}^{\left(jn\right)}+u_{\mu}^{\left(jn\right)}k_{\mu}^{\left(jn\right)}\right)\right),\\
 & s_{\mu}^{\left(n\right)}=-\left(k_{g\to f}^{\left(n\right)}-k_{e\to f}^{\left(n\right)}-o_{\mu}^{\left(n\right)}-\sum_{j=1}^{3}\left(m_{\mu}^{\left(jn\right)}-v_{\mu}^{\left(jn\right)}y_{\mu}^{\left(jn\right)}\right)\right),\\
 & t_{\mu}^{\left(n\right)}=k_{f\to g}^{\left(n\right)}+k_{f\to e}^{\left(n\right)}+k_{e\to f}^{\left(n\right)}-o_{\mu}^{\left(n\right)}-\sum_{j=1}^{3}m_{\mu}^{\left(jn\right)}\sum_{k=1}^{3}q_{\mu}^{\left(jkn\right)}k_{\mu}^{\left(kn\right)},\\
 & 1/w_{\mu}^{\left(n\right)}=q_{\mu}^{\left(n\right)}t_{\mu}^{\left(n\right)}-r_{\mu}^{\left(n\right)}s_{\mu}^{\left(n\right)},
\end{align}
where we have introduced
\begin{alignat}{1}
 & a_{\mu}^{\left(jn\right)}=-2v_{ge}^{\left(jn\right)}\mathrm{Im}\Sigma_{\mu}^{\left(jn\right)}\mu_{j},\\
 & d_{\mu}^{\left(jn\right)}=-2v_{gf}^{\left(n\right)2}\mathrm{Im}\left(\Sigma_{\mu}^{\left(jn\right)2}\Xi_{\mu}^{\left(jn\right)}\right)\mu_{j},\\
 & g_{\mu}^{\left(jkn\right)}=-2v_{gf}^{\left(n\right)2}\mathrm{Im}\left(\Sigma_{\mu}^{\left(jn\right)}\Psi_{\mu}^{\left(jkn\right)}\right)\sqrt{\mu_{j}\mu_{k}},\nonumber \\
 & k_{\mu}^{\left(jn\right)}=-2v_{gf}^{\left(n\right)2}v_{ge}^{\left(jn\right)}\mathrm{Im}\left(\Sigma_{\mu}^{\left(jn\right)}\Xi_{\mu}^{\left(jn\right)}\Phi_{\mu}^{\left(n\right)}\right)\mu_{j},\\
 & m_{\mu}^{\left(jn\right)}=-2v_{gf}^{\left(n\right)2}v_{ge}^{\left(jn\right)}\mathrm{Im}\left(\Phi_{\mu}^{\left(n\right)}\Sigma_{\mu}^{\left(jn\right)}\right)\mu_{j},\\
 & o_{\mu}^{\left(n\right)}=-2v_{gf}^{\left(n\right)2}\mathrm{Im}\Phi_{\mu}^{\left(n\right)},
\end{alignat}
and 
\begin{equation}
p_{\mu}^{\left(jn\right)}=k_{f\to e}^{\left(n\right)}+k_{e\to g}^{\left(n\right)}+k_{e\to f}^{\left(n\right)}-\left(d_{\mu}^{\left(jn\right)}+a_{\mu}^{\left(jn\right)}\right)-g_{\mu}^{\left(jjn\right)},
\end{equation}
\begin{equation}
q_{\mu}^{\left(jkn\right)}=\mathrm{Inverse}\left(\begin{array}{ccc}
p_{\mu}^{\left(1n\right)} & -g_{\mu}^{\left(12n\right)} & -g_{\mu}^{\left(13n\right)}\\
-g_{\mu}^{\left(21n\right)} & p_{\mu}^{\left(2n\right)} & -g_{\mu}^{\left(23n\right)}\\
-g_{\mu}^{\left(31n\right)} & -g_{\mu}^{\left(32n\right)} & g_{\mu}^{\left(3n\right)}
\end{array}\right),
\end{equation}
as well as 
\begin{align}
 & 1/\Xi_{\mu}^{\left(jn\right)}=\tilde{\omega}_{ef}^{\left(n\right)}+\tilde{\omega}_{\mu_{j}-1\mu}-v_{gf}^{\left(n\right)2}/\left(\tilde{\omega}_{eg}^{\left(n\right)}+\tilde{\omega}_{\mu_{j}-1\mu}\right),\\
 & \Sigma_{\mu}^{\left(jn\right)}=v_{ge}^{\left(jn\right)}/\left(\tilde{\omega}_{eg}^{\left(n\right)}+\tilde{\omega}_{\mu_{j}-1\mu}\right),\\
 & 1/\Phi_{\mu}^{\left(n\right)}=\tilde{\omega}_{gf}^{\left(n\right)}-\sum_{j=1}^{3}\mu_{j}\Xi_{\mu}^{\left(jn\right)}v_{ge}^{\left(jn\right)2},\\
 & \Psi_{\mu}^{\left(jkn\right)}=v_{ge}^{\left(jn\right)}v_{ge}^{\left(kn\right)}\Phi_{\mu}^{\left(n\right)}\Xi_{\mu}^{\left(jn\right)}\Xi_{\mu}^{\left(kn\right)}\Sigma_{\mu}^{\left(jn\right)}\sqrt{\mu_{j}\mu_{k}}.
\end{align}

\begin{figure}
\begin{centering}
\includegraphics[scale=0.55]{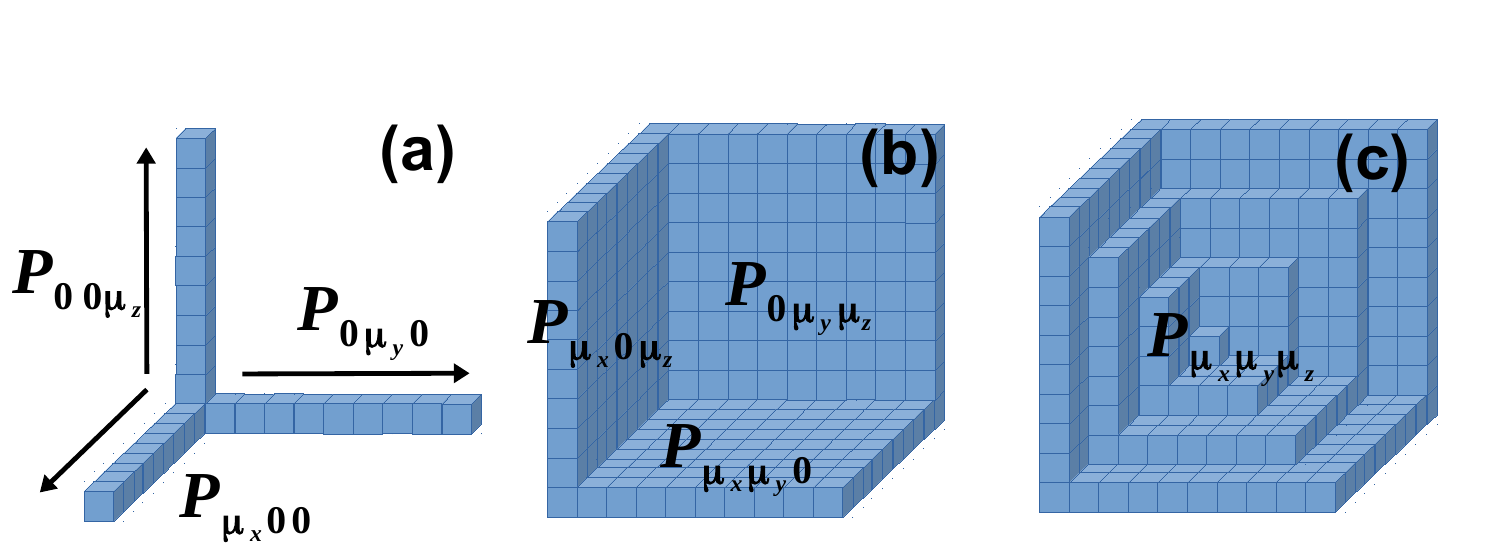}
\par\end{centering}
\caption{\label{fig:procedure}Procedure to calculate the plasmon state population
$P_{\mu_{x}\mu_{y}\mu_{z}}$. Panel (a): the edge elements $P_{\mu_{x}00}$,
$P_{0\mu_{y}0}$ and $P_{00\mu_{z}}$. Panel (b): the surface elements
$P_{\mu_{x}\mu_{y}0}$, $P_{0\mu_{y}\mu_{z}}$ and $P_{\mu_{x}0\mu_{z}}$.
Panel (c): the body elements $P_{\mu_{x}\mu_{y}\mu_{z}}$.}
\end{figure}

In the steady-state of the systems, the time-derivative is zero in Eq. \eqref{eq:EPSP} in the main text and
the resulting equation leads to a recursion relation for the
population, cf. Eq. \eqref{eq:final-recursion-relation} in the main text. In the following,
we explain the procedure to calculate the population with the recursion
relation for three modes (one and two modes follow as special cases). 
First, we assume a fixed value for $P_{000}$ and use it to calculate
the edge elements $P_{\mu_{1}00}$, $P_{0\mu_{2}0}$ and $P_{00\mu_{3}}$
with simplified versions of Eq. \eqref{eq:final-recursion-relation},
cf. Fig. \ref{fig:procedure} (a) :
\begin{alignat}{1}
P_{\mu_{1}00} & =\frac{\eta_{\mu_{1}00}^{\left(11\right)}P_{\mu_{1}-1,00}}{\left(\gamma_{\mathrm{1}}\mu_{1}+\kappa{}_{\mu}^{\left(1\right)}\right)},\label{eq:pmu1}\\
P_{0\mu_{2}0} & =\frac{\eta_{0\mu_{2}0}^{\left(22\right)}P_{0,\mu_{2}-1,0}}{\left(\gamma_{2}\mu_{2}+\kappa{}_{\mu}^{\left(2\right)}\right)},\label{eq:pmu2}\\
P_{00\mu_{3}} & =\frac{\eta_{00\mu_{3}}^{\left(33\right)}P_{00,\mu_{3}-1}}{\left(\gamma_{3}\mu_{3}+\kappa{}_{\mu}^{\left(3\right)}\right)}.\label{eq:pmu3}
\end{alignat}
Secondly, we calculate the surface elements $P_{\mu_{1}\mu_{2}0}$
with the known $P_{\mu_{1}00}$ and $P_{0\mu_{2}0}$ according to
a simplified version of Eq. \eqref{eq:final-recursion-relation},
cf. Fig. \ref{fig:procedure} (b):
\begin{equation}
P_{\mu_{1}\mu_{2}0}=\frac{\sum_{j=1}^{2}\eta_{\mu_{1}\mu_{2}0}^{\left(j1\right)}P_{\mu_{1}-1,\mu_{2},0}+\sum_{j=1}^{2}\eta_{\mu_{1}\mu_{2}0}^{\left(j2\right)}P_{\mu_{1},\mu_{2}-1,0}}{\left(\gamma_{1}\mu_{1}+\kappa{}_{\mu_{1}\mu_{2}0}^{\left(1\right)}\right)+\left(\gamma_{2}\mu_{2}+\kappa{}_{\mu_{1}\mu_{2}0}^{\left(2\right)}\right)}.\label{eq:pmu1mu2}
\end{equation}
Similarly, we can also calculate other surface elements $P_{0\mu_{2}\mu_{3}}$
and $P_{\mu_{1}0\mu_{3}}$ according to simplified versions of Eq.
\eqref{eq:final-recursion-relation}:
\begin{align}
P_{0\mu_{2}\mu_{3}} & =\frac{\sum_{j=2}^{3}\eta_{0\mu_{2}\mu_{3}}^{\left(j2\right)}P_{0,\mu_{2}-1,\mu_{3}}+\sum_{j=2}^{3}\eta_{0\mu_{2}\mu_{3}}^{\left(j3\right)}P_{0\mu_{2}\mu_{3}-1}}{\left(\gamma_{2}\mu_{2}+\kappa{}_{0\mu_{2}\mu_{3}}^{\left(2\right)}\right)+\left(\gamma_{3}\mu_{3}+\kappa{}_{0\mu_{2}\mu_{3}}^{\left(3\right)}\right)},\label{eq:pmu2mu3}\\
P_{\mu_{1}0\mu_{3}} & =\frac{\sum_{j=1,3}\eta_{\mu_{1}0\mu_{3}}^{\left(j1\right)}P_{\mu_{1}-1,0,\mu_{3}}+\sum_{j=1,3}\eta_{\mu_{1}0\mu_{3}}^{\left(j3\right)}P_{\mu_{1}0\mu_{3}-1}}{\left(\gamma_{1}\mu_{1}+\kappa{}_{\mu_{1}0\mu_{3}}^{\left(1\right)}\right)+\left(\gamma_{3}\mu_{3}+\kappa{}_{\mu_{1}0\mu_{3}}^{\left(3\right)}\right)}.\label{eq:pmu1mu3}
\end{align}
Finally, we calculate the body elements $P_{\mu_{1}\mu_{2}\mu_{3}}$
with the known $P_{\mu_{1}\mu_{2}0}$, $P_{0\mu_{2}\mu_{3}}$ and $P_{\mu_{1}0\mu_{3}}$
by applying repeatedly Eq. \eqref{eq:final-recursion-relation}, 
cf. Fig. \ref{fig:procedure} (c). The reason why we can achieve the
above simplified versions of Eq. \eqref{eq:final-recursion-relation}
is that the rates with $\mu_{j}<0$ will vanish.

To calculate the population
of molecular states $P_{g}^{\left(n\right)}=\sum_{\mu}\rho_{g\mu,g\mu}^{\left(n\right)}$,
$P_{f}^{\left(n\right)}=\sum_{\mu}\rho_{f\mu,f\mu}^{\left(n\right)}$
and $P_{e}^{\left(n\right)}=\sum_{\mu}\rho_{e\mu,e\mu}^{\left(n\right)}$, we extract 
$\rho_{g\mu,g\mu}^{\left(n\right)}$,$\rho_{f\mu,f\mu}^{\left(n\right)}$,$\rho_{e\mu,e\mu}^{\left(n\right)}$
from Eqs. \eqref{eq:rhoee-3}, \eqref{eq:rhogg-3}  and \eqref{eq:rhoff-3} and express them as functions 
of the plasmon state population $P_\mu=\rho_{\mu \mu}$. 
\begin{align}
 & \rho_{g\mu,g\mu}^{\left(n\right)}=w_{\mu}^{\left(n\right)}\left(k_{e\to g}^{\left(n\right)}t_{\mu}^{\left(n\right)}-k_{e\to f}^{\left(n\right)}r_{\mu}^{\left(n\right)}\right)P_{\mu}\nonumber \\
 & -k_{f\to e}^{\left(n\right)}w_{\mu}^{\left(n\right)}\sum_{k=1}^{3}\left(u_{\mu}^{\left(kn\right)}t_{\mu}^{\left(n\right)}+r_{\mu}^{\left(n\right)}v_{\mu}^{\left(kn\right)}\right)P_{\mu_{k}-1},\label{eq:rhogg-4}
\end{align}
\begin{align}
 & \rho_{f\mu,f\mu}^{\left(n\right)}=-w_{\mu}^{\left(n\right)}\left(k_{e\to g}^{\left(n\right)}s_{\mu}^{\left(n\right)}-k_{e\to f}^{\left(n\right)}q_{\mu}^{\left(n\right)}\right)P_{\mu}\nonumber \\
 & +w_{\mu}^{\left(n\right)}k_{f\to e}^{\left(n\right)}\sum_{k=1}^{3}\left(u_{\mu}^{\left(kn\right)}s_{\mu}^{\left(n\right)}+q_{\mu}^{\left(n\right)}v_{\mu}^{\left(kn\right)}\right)P_{\mu_{k}-1},\label{eq:rhoff-4}
\end{align}
\begin{align}
 & \rho_{e\mu_{j}-1,e\mu_{j}-1}^{\left(n\right)}=\sum_{k=1}^{3}q_{\mu}^{\left(jkn\right)} \Big[ k_{f\to e}^{\left(n\right)}P_{\mu_{k}-1}+k_{\mu}^{\left(kn\right)}P_{f\mu}^{\left(n\right)}\nonumber \\
 & +\left(k_{g\to e}^{\left(n\right)}-k_{f\to e}^{\left(n\right)}-\left(d_{\mu}^{\left(kn\right)}+k_{\mu}^{\left(kn\right)}+\sum_{l=1}^{3}g_{\mu}^{\left(kln\right)}+a_{\mu}^{\left(kn\right)}\right)\right)P_{g\mu}^{\left(n\right)} \Big].\label{eq:rhoee-4}
\end{align}


\begin{thebibliography}{10}
\bibitem{ThreeLevels} The three molecular states do not need to be electronic
states. They can also be the relevant electronic-vibrational states in dye molecules.
\bibitem{NFO} Although the gold nano-sphere can locally enhance
an off-resonant incoming field, the enhancement is often very
weak. Therefore, here, we assume an homogeneous driving field
 in the small region of the nano-laser.
\end{thebibliography}
\end{document}